\documentclass[12pt]{article}
\usepackage{jheppub}
\pdfoutput=1
\usepackage{amssymb}
\usepackage{graphicx}
\usepackage{amsmath}
\usepackage{multirow}
\usepackage{verbatim}
\usepackage{rotating}
\usepackage{graphics}
\usepackage{pstricks}
\usepackage{bm}% bold math
\usepackage{amsfonts}
\usepackage{bbm}
\usepackage[FIGTOPCAP,scriptsize,tight]{subfigure}

\usepackage[textsize=scriptsize,backgroundcolor=red!70,linecolor=red]{todonotes}
\usepackage{paralist}
\usepackage{arydshln}
\newcommand{\nue}{{\nu_e}}

\newcommand{\be}{\begin{eqnarray}}
\newcommand{\beq}{\begin{equation}}
\newcommand{\eeq}{\end{equation}}
\newcommand{\ee}{\end{eqnarray}}

\newcommand{\bmp}{\noindent\begin{minipage}{16cm}}
\newcommand{\emp}{\end{minipage}\vskip 7mm} % 7mm untightened

\newcommand{\kmc}{\ensuremath{\text{km}^3}}

\newcommand{\ie}{\textit{i.e.}}

\newcommand{\dm}{\ensuremath{\text{DM}}}

\newcommand{\chisq}{\ensuremath{\chi^{2}}}

\newcommand{\tdm}{\ensuremath{\tau_{\dm}}}
\newcommand{\mdm}{\ensuremath{m_{\dm}}}
\newcommand{\ndm}{\ensuremath{N_{\dm}}}
\newcommand{\nast}{\ensuremath{N_{\rm astro}}}
\newcommand{\PhiDM}{\ensuremath{\Phi_{\dm}}}
\newcommand{\PhiAst}{\ensuremath{\Phi_\text{astro}}}
\newcommand{\phia}{\ensuremath{\phi_{\rm astro}}}

\newcommand{\like}{\ensuremath{\mathcal{L}}}
\newcommand{\Emin}{\ensuremath{{E_{\rm min}}}}
\newcommand{\Emax}{\ensuremath{{E_{\rm max}}}}
\newcommand{\pdf}{\ensuremath{\mathcal{P}}}
\newcommand{\hhatn}{\ensuremath{\skew{3}\widehat{\widehat{N}}}}
\newcommand{\hhatnu}{\ensuremath{\skew{1}\widehat{\widehat{\nu}}}}

\preprint{{\tt IFIC/17-19}}

\title{Probing decaying heavy dark matter with the 4-year IceCube HESE data}
 
\author[a]{Atri Bhattacharya,}
\author[b]{Arman Esmaili,}
\author[c]{Sergio Palomares-Ruiz,} 
\author[d,e]{Ina Sarcevic.}

\affiliation[a]{Space sciences, Technologies and Astrophysics Research (STAR) Institute,
                Universit\'{e} de Li\`{e}ge, B\^{a}t.~B5a, 4000 Li\`{e}ge,
                Belgium}
\affiliation[b]{Departamento de F\'{\i}sica, Pontif\'{\i}cia Universidade Cat\'olica do Rio de Janeiro, C. P. 38071, 22452- 970, Rio de Janeiro, Brazil}
\affiliation[c]{Instituto de F\'{\i}sica Corpuscular (IFIC),
                CSIC-Universitat de Val\`{e}ncia,  
	              Apartado de Correos 22085, E-46071 Valencia, Spain}
\affiliation[d]{Department of Physics, University of Arizona, 1118 E.\ 4th St.\ Tucson, AZ 85704}
\affiliation[e]{Department of Astronomy, University  of Arizona, 933 N.\ Cherry Ave., Tucson, AZ 85721}

\emailAdd{a.bhattacharya@ulg.ac.be}
\emailAdd{arman@puc-rio.br}
\emailAdd{sergio.palomares.ruiz@ific.uv.es}
\emailAdd{ina@physics.arizona.edu}

\date{\today}

\abstract{
After the first four years of data taking, the IceCube neutrino telescope has observed 54 high-energy starting events (HESE) with deposited energies between 20~TeV and 2~PeV. The background from atmospheric muons and neutrinos is expected to be of about 20 events, all below 100~TeV, thus pointing towards the astrophysical origin of about 8 events per year in that data set. However, their precise origin remains unknown. Here, we perform a detailed analysis of this event sample (considering simultaneously the energy, hemisphere and topology of the events) by assuming two contributions for the signal events: an isotropic power-law flux and a flux from decaying heavy dark matter. We fit the mass and lifetime of the dark matter and the normalization and spectral index of an isotropic power-law flux, for various decay channels of dark matter. We find that a significant contribution from dark matter decay is always slightly favored, either to explain the excess below 100~TeV, as in the case of decays to quarks or, as in the case of neutrino channels, to explain the three multi-PeV events. Also, we consider the possibility to interpret all the data by dark matter decays only, considering various  combinations of two decay channels. We show that the decaying dark matter scenario provides a better fit to HESE data than the isotropic power-law flux.
}

\begin{document}
\maketitle

\section{Introduction}
During the last years, the \kmc\ neutrino telescope IceCube has been steadily accumulating high-energy neutrino signals, leading to the detection of neutrinos at PeV energies~\cite{Aartsen:2013bka}. Being the first detection of neutrinos at such high energies, there is considerable speculation regarding the source of the neutrino flux that gave rise to the observed events. Standard explanations suggest neutrinos being produced in the same hadronuclear or photohadronic interactions that lead to the production of cosmic rays inside high-energy astrophysical sources, such as active galactic nuclei~\cite{Kalashev:2013vba, Stecker:2013fxa, Murase:2014foa, Tjus:2014dna, Krauss:2014tna, Dermer:2014vaa, Tavecchio:2014iza, Sahu:2014fua, Kalashev:2014vya, Tavecchio:2014eia, Kimura:2014jba, Petropoulou:2015upa, Resconi:2016ggj}, star-forming galaxies or regions in our own galaxy~\cite{Murase:2013rfa, Tamborra:2014xia, Anchordoqui:2014yva, Chang:2014hua, Bartos:2015xpa, Chakraborty:2016mvc, Yoast-Hull:2017gaj}, gamma-ray bursts~\cite{Cholis:2012kq, Liu:2012pf, Murase:2013ffa, Razzaque:2013dsa, Fraija:2013cha, Petropoulou:2014lja, Dado:2014mea, Razzaque:2014ola, Tamborra:2015qza, Tamborra:2015fzv, Senno:2015tsn}, hypernova and supernova remnants~\cite{Fox:2013oza, Murase:2013ffa, Liu:2013wia, Bhattacharya:2014sta, Chakraborty:2015sta, Tamborra:2015fzv, Senno:2015tsn}, the galactic halo~\cite{Taylor:2014hya}, galaxy clusters~\cite{Zandanel:2014pva}, microquasars~\cite{Anchordoqui:2014rca}, neutron stars mergers~\cite{Gao:2013rxa}, tidal disruption events~\cite{Wang:2015mmh, Lunardini:2016xwi}, transient sources~\cite{Kistler:2017lep} and from a more general perspective, relating the neutrino flux to the cosmic-ray spectrum~\cite{Kistler:2013my, Gupta:2013xfa, Anchordoqui:2013qsi, Neronov:2013lza, Liu:2013wia, Joshi:2013aua, Katz:2013ooa, Fang:2014uja, Kachelriess:2014oma, Dado:2014mea, Winter:2014pya, Anchordoqui:2014pca, Guo:2014laa, Fang:2017zjf, Kachelriess:2017tvs}. Since such cosmic rays have been detected at energies beyond hundreds of PeV, the kinematics of the corresponding interactions which suggest the production of neutrinos as final products of the interaction chain at energies about an order of magnitude or two lower, could point to extragalactic high-energy objects as being the sources for the dominant part of the IceCube high-energy neutrino flux, which would thus be isotropically distributed. Indeed, the observed distribution is consistent with an isotropic flux~\cite{Aartsen:2013jdh, Bai:2013nga, Aartsen:2014gkd, Esmaili:2014rma, Troitsky:2015cnk, Chianese:2016opp, Palladino:2016zoe, Vincent:2016nut, Denton:2017csz} (however, see also Ref.~\cite{Neronov:2015osa}) and the typical diffuse neutrino energy spectrum expected is assumed to be described by an uniform and unbroken power-law flux. Given that oscillation probabilities are averaged out during propagation~\cite{Learned:1994wg} and the standard cosmic production of neutrinos results from pion decays at the sources, a flavor ratio at the Earth of $(1 : 1 : 1)_\oplus$ for both neutrinos and antineutrinos is usually assumed  (however, see Refs.~\cite{Mena:2014sja, Xu:2014via, Palomares-Ruiz:2014zra, Palomares-Ruiz:2015mka, Vincent:2015woa, Watanabe:2014qua, Palladino:2015zua, Aartsen:2015ivb, Bustamante:2015waa, Shoemaker:2015qul, Arguelles:2015dca, Aartsen:2015knd, Vincent:2016nut, Brdar:2016thq} for analyses of the flavor composition of IceCube neutrinos).

However, if interpreted as a single power-law flux, there is presently considerable tension between different sets of IceCube data; in particular, the 4-year high-energy starting events (HESE) with deposited energies in the range of about $20 ~{\rm TeV}$--$2~{\rm PeV}$~\cite{Aartsen:2013jdh, Aartsen:2014gkd, Aartsen:2015zva} and the upgoing muon track event data collected over 6 years and sensitive to neutrino energies above $\sim 200$~TeV~\cite{Aartsen:2015rwa, Aartsen:2016xlq}. The HESE correspond to those signals which have a starting vertex contained within the IceCube detector volume. These include tracks arising from charged-current interactions of $\nu_\mu$ and $\bar{\nu}_\mu$ (and a small fraction of $\nu_\tau$ and $\bar{\nu}_\tau$) with nuclei in the detector volume, as well as cascades produced from charged-current interactions of $\nu_e$, $\bar{\nu}_e$, $\nu_\tau$ and $\bar{\nu}_\tau$ and neutral-current interactions of all flavors. The 4-year IceCube HESE data, with a threshold of 10~TeV, contains 54 events, with $\sim$21 of them expected to be associated to atmospheric muons and neutrinos. The non-atmospheric signal is consistent with a power-law flux with a steeply falling nature, with an spectral index of $\gamma_{\rm HESE} = 2.58 \pm 0.25$~\cite{Aartsen:2015zva}. On the other hand, a complementary measurement by IceCube searches for up-going muon tracks passing through the detector, having been produced in charged-current interactions of (mainly) $\nu_\mu$ and $\bar{\nu}_\mu$ with nuclei either in the Earth's rock before entering the detector or within the IceCube volume~\cite{Aartsen:2015rwa, Aartsen:2016xlq}. By considering events whose interaction vertex can be outside the detector, the effective area significantly increases. However, this necessarily restricts the observation to the North hemisphere to avoid the enormous background from atmospheric muons. Data collected over six years point to a much flatter flux than the HESE result, with an spectral index of $\gamma_{\mu \rm N} = 2.13 \pm 0.13$~\cite{Aartsen:2016xlq}. In addition, one up-going muon track was detected with deposited energy of $2.6 \pm 0.3$~PeV, which results in a reconstructed median muon energy of $4.5 \pm 1.2$~PeV~\cite{Aartsen:2016xlq}, with a median expected muon neutrino energy of $8.7$~PeV~\cite{Kistler:2016ask, Aartsen:2016xlq}. In turn, this poses further tension to the two results, as the through-going muons require a flux that extends to multi-PeV energies, whereas the HESE analysis finds a small preference for a cut-off at a few PeV due to the lack of events in the sample~\cite{Aartsen:2015zva, Anchordoqui:2016ewn}. As a result, the best-fit spectral indices obtained in these two analyses are inconsistent with each other at more than $3\sigma$~\cite{Aartsen:2016xlq}. In addition, using two years of data, an analysis of an optimized cascade selection, including some which are partially contained in the detector volume, resulted in a spectral index of $\gamma_{\rm C} = 2.67^{+0.12}_{-0.13}$~\cite{Aartsen:2015zva}, in good agreement with the HESE result (dominated by cascades).

Furthermore, when fitting with a uniform power-law flux, the HESE analysis has issues of its own. As mentioned before, the best-fit spectral index resulting from this analysis currently settles at $\gamma_{\rm HESE} = 2.58$, assuming equal flavor ratios and a range in deposited energies of [60~TeV--3~PeV]~\cite{Aartsen:2015zva} (see also Ref.~\cite{Vincent:2016nut} for an analysis of several assumptions). However, the corresponding power-law form and its normalization gives event rates at PeV energies that are barely consistent, or even in tension, with the $1\sigma$ range of observed signals. With a somewhat harder spectral index, more consistent with the PeV HESE and the through-going muon-track analysis, the corresponding power-law spectrum predicts event rates poorly matching with the events with deposited energies $\lesssim 100$~TeV. Specifically, in this case, there is an excess of events below $\sim 100$~TeV, which cannot be explained by the corresponding power-law flux.

One explanation that has been proposed to explain the apparent inconsistencies with a single power-law fit is to introduce a break in the flux beyond some energy threshold~\cite{Palomares-Ruiz:2015mka, Aartsen:2015zva, Anchordoqui:2016ewn}, rather than assuming a uniform index throughout all the energies. On another hand, it is likely that different astrophysical sources would contribute to the neutrino flux and thus, it is reasonable to also consider different spectral features and anisotropies in its angular distribution. Therefore, another phenomenological approach is to consider a two-component astrophysical flux~\cite{Chen:2014gxa, Aartsen:2015knd, Palladino:2016zoe, Vincent:2016nut}.

In addition to conventional sources, a possible contribution from new physics scenarios has also been considered, either to interpret all or part of the IceCube high-energy events, such as modifications to fundamental physics~\cite{Borriello:2013ala, Alikhanov:2014uja, Anchordoqui:2014hua, Stecker:2014xja, Aeikens:2014yga, Stecker:2014oxa, Lai:2017bbl}, dark matter (DM) decays or annihilations~\cite{Anisimov:2008gg, Feldstein:2013kka, Esmaili:2013gha, Bai:2013nga, Ema:2013nda, Bhattacharya:2014vwa, Zavala:2014dla, Higaki:2014dwa, Ema:2014ufa, Rott:2014kfa, Esmaili:2014rma, Fong:2014bsa, Daikoku:2015vsa, Murase:2015gea, Esmaili:2015xpa, Aisati:2015vma, Roland:2015yoa, Anchordoqui:2015lqa, Boucenna:2015tra, Ko:2015nma, Troitsky:2015cnk, EsmailiTaklimi:2016bbx, Esmaili:2016swq, Chianese:2016opp, Dev:2016uxj, Fiorentin:2016avj, Dev:2016qbd, DiBari:2016guw, Chianese:2016smc, Chianese:2016kpu, Kuznetsov:2016fjt, Cohen:2016uyg, Borah:2017xgm, Hiroshima:2017hmy}, and existence of new particles or interactions~\cite{Barger:2013pla, Akay:2014tga, Ioka:2014kca, Ng:2014pca, Ibe:2014pja, Bhattacharya:2014yha, Blum:2014ewa, Araki:2014ona, Akay:2014qka, Illana:2014bda, Cherry:2014xra, Kopp:2015bfa, Davis:2015rza, DiFranzo:2015qea, Araki:2015mya, Dey:2015eaa, Ema:2016zzu, Bhattacharya:2016tma, Arguelles:2017atb, Chauhan:2017ndd}. In this work, using the 4-year HESE data, we revisit the possibility to explain the IceCube spectrum with decaying DM, either as the only source of high-energy neutrinos or as an additional contribution to the astrophysical (isotropic) power-law flux.

The DM interpretation of the 2-year IceCube HESE dataset~\cite{Aartsen:2013jdh} (considering the energy spectrum of the events in this dataset) has been studied in Ref.~\cite{Esmaili:2013gha}, finding a mildly better agreement with the data for the decaying DM scenario than for the isotropic power-law flux. The same dataset has been analyzed in Ref.~\cite{Bhattacharya:2014vwa} assuming both contributions, from DM decays and from an isotropic astrophysical power-law flux. In a more detailed analysis, considering the 3-year IceCube HESE dataset~\cite{Aartsen:2014gkd}, a mild preference for DM decay has been found in both the angular and energy distributions of the events~\cite{Esmaili:2014rma}. Here, we consider the recent 4-year IceCube HESE data, which contains a combination of muon track as well as shower events, providing an opportunity to compare with and update previous analyses. In particular, we perform a simultaneous likelihood analysis of the topology and energy distributions, including also the hemisphere of the events. We consider three possible scenarios as the origin of IceCube neutrinos: isotropic unbroken power-law flux, decaying DM, or a combination of both. We allow all signal parameters to vary, both the normalization and the spectral index for the astrophysical flux, as well as the DM mass, its decay lifetime and branching ratio for several two-body decay channels. In this way, we explore the parameter space of several combinations of fluxes from an astrophysical source and DM decays that gives the best fit to the observed event spectrum. First, we do not impose any restriction on the parameter space and then we also study the case when a hard cut on the astrophysical index is imposed, more in agreement with theoretical expectations. Finally, we also study the possibility to explain the data with only DM decays into two channels. In order to compute the event distributions, we follow the procedure described in Refs.~\cite{Palomares-Ruiz:2015mka, Vincent:2016nut}.

The paper is organized as follows. In Section~\ref{sec:flux} we describe the expected flux of neutrinos from decaying DM, considering both the galactic and extragalactic contributions. Section~\ref{sec:stat} is devoted to the description of the statistical methods used in the analyses. The results are presented in Sections~\ref{sec:fit-results}, \ref{sec:prior-tracks} and \ref{sec:multi-channel}. In Section~\ref{sec:fit-results} we show the best-fit points and some event spectra for the scenario with contributions from a flux from DM decays and from an astrophysical power-law flux. Also, the preferred regions in the parameter space are depicted for some selected cases. The correlations among the four parameters are presented in Appendix~\ref{sec:appA} for a couple of illustrative cases. We also compute limits on DM decay lifetimes for all the decay channels under consideration, shown in Appendix~\ref{sec:appB}. In Section~\ref{sec:prior-tracks} we reanalyze the data by assuming a prior on the spectral index of the astrophysical flux, motivated by theoretical expectations and the though-going muon track dataset. In Section~\ref{sec:multi-channel} we consider the possibility to interpret all the HESE dataset by DM decaying into two channels with a branching ratio treated as an additional free parameter. Various choices of decay channels are studied. We discuss our results and draw our conclusions in Section~\ref{sec:conc}. 

\pagebreak

%%%%%%%%%%%%%%%%%%%%%%%%%%%
%%%%%%%%%%%%%%%%%%%%%%%%%%%
\section{Neutrino flux from DM decay\label{sec:flux}}
%%%%%%%%%%%%%%%%%%%%%%%%%%%
%%%%%%%%%%%%%%%%%%%%%%%%%%%

For a general cosmic neutrino flux, given that propagation distances are much larger than oscillation lengths (even for galactic sources), neutrino oscillations get averaged out. Therefore, the fluxes at Earth of neutrinos of different flavors are independent of the propagation distance and are given by
\begin{equation}
\left. \frac{d \Phi_{\nu_\alpha}}{d E_\nu} \right|_\oplus = \sum_\beta \, P_{\alpha \beta} \, \frac{d \Phi_{\nu_\beta}}{d E_\nu} = \sum_{\beta } \sum_i |U_{\alpha i}|^2 \, |U_{\beta i}|^2 \, \frac{d \Phi_{\nu_\beta}}{d E_\nu} ~, 
\end{equation}
where $U$ is the neutrino mixing matrix and $P_{\alpha \beta}$ represents the oscillation probability of flavor neutrino $\nu_\alpha$ into $\nu_\beta$, after all terms depending on mass squared differences are averaged out. 

For the case of DM decays, the neutrino flux has two contributions: 1) an extragalactic component, $\Phi_{\rm EG}$, which originates from the decay of DM particles in halos at all redshifts, and thus it is isotropic; 2) a galactic contribution, $\Phi_{\rm G}$, that comes from DM decays within our galactic halo and it is anisotropic, i.e., it follows the galactic morphology. Thus, the total $\nu_\alpha+\bar{\nu}_\alpha$ flux (with $\alpha=e, \mu$ and $\tau$) can be written, without including neutrino oscillations, as~\cite{Esmaili:2012us, Esmaili:2014rma} 
\begin{equation}\label{eq:flux}
\frac{d \Phi_{\dm,\nu_\alpha}}{d E_\nu} = \frac{d \Phi_{{\rm G},\nu_\alpha}}{d E_\nu} + \frac{d \Phi_{{\rm EG},\nu_\alpha}}{d E_\nu} ~,  
\end{equation}  
where  the isotropic extragalactic flux is given by
\begin{equation}
\frac{d \Phi_{{\rm EG},\nu_\alpha}}{d E_\nu} (E_\nu) =  \frac{\Omega_{\rm DM}\rho_c}{4\pi m_{\rm DM}\tau_{\rm DM}} \int_0^\infty dz\, \frac{1}{H(z)}\frac{dN_{\nu_\alpha}}{dE_\nu}\left[ (1+z)E_\nu \right]~,
\end{equation}
where $m_{\rm DM}$ and $\tau_{\rm DM}$ are the DM mass and lifetime, respectively, $dN_{\nu_\alpha}/dE_\nu$ is the spectrum of $\nu_\alpha+\bar{\nu}_\alpha$ from DM decays, $z$ is the redshift of the emitted neutrinos and $\rho_{c} = 5.6 \times 10^{-6}\text{ GeV cm}^{-3}$ represents the critical density of the Universe. The Hubble function is $H(z) = \sqrt{\Omega_{\Lambda} + \Omega_\text{m}(1+z)^{3}}$, and using the results from the most recent Planck data~\cite{Ade:2013zuv}, we take $\Omega_\Lambda = 0.6825$, $\Omega_\text{m} = 0.3175$, $\Omega_{\dm} = 0.2685$ and $H_{0} = 67.1 \text{ km}\text{ s}^{-1}\text{Mpc}^{-1}$.

The galactic contribution is given by
\begin{equation}
\frac{d \Phi_{{\rm G},\nu_\alpha}}{d E_\nu} (E_\nu,b,l)= \frac{1}{4\pi m_{\rm DM}\tau_{\rm DM}}\frac{dN_{\nu_\alpha}}{dE_\nu} \int_0^\infty \rho\left[ r\left(s,b,l\right)\right]\, ds ~,  
\end{equation}
where $r(s,b,l)=\sqrt{s^2+R_\odot^2-2sR_\odot\cos b\cos l}$ is the distance to the galactic center and depends on the galactic latitude and longitude, $b$ and $l$, in galactic coordinates (or equivalently, on the declination, $\delta$, and right ascension, RA, in the equatorial coordinates), and $R_\odot=8.5$~kpc is the Sun's distance to the galactic center.
$\rho(r)$ is the DM radial density profile of our Galaxy, which we assume to be of Navarro-Frenk-White type~\cite{Navarro:1995iw, Navarro:1996gj}, given by
\begin{equation}
\rho(r) = \frac{\rho_0}{\left(r/r_s\right) \, \left(1+r/r_s\right)^2}~,
\end{equation}     
with $r_s = 20$~kpc and $\rho_0 = 0.33$~GeV~${\rm cm}^{-3}$, \ie, $\rho(R_\odot) = 0.38$~GeV~${\rm cm}^{-3}$.

\begin{figure}[t]
	\subfigure{\includegraphics[width=0.5\textwidth]{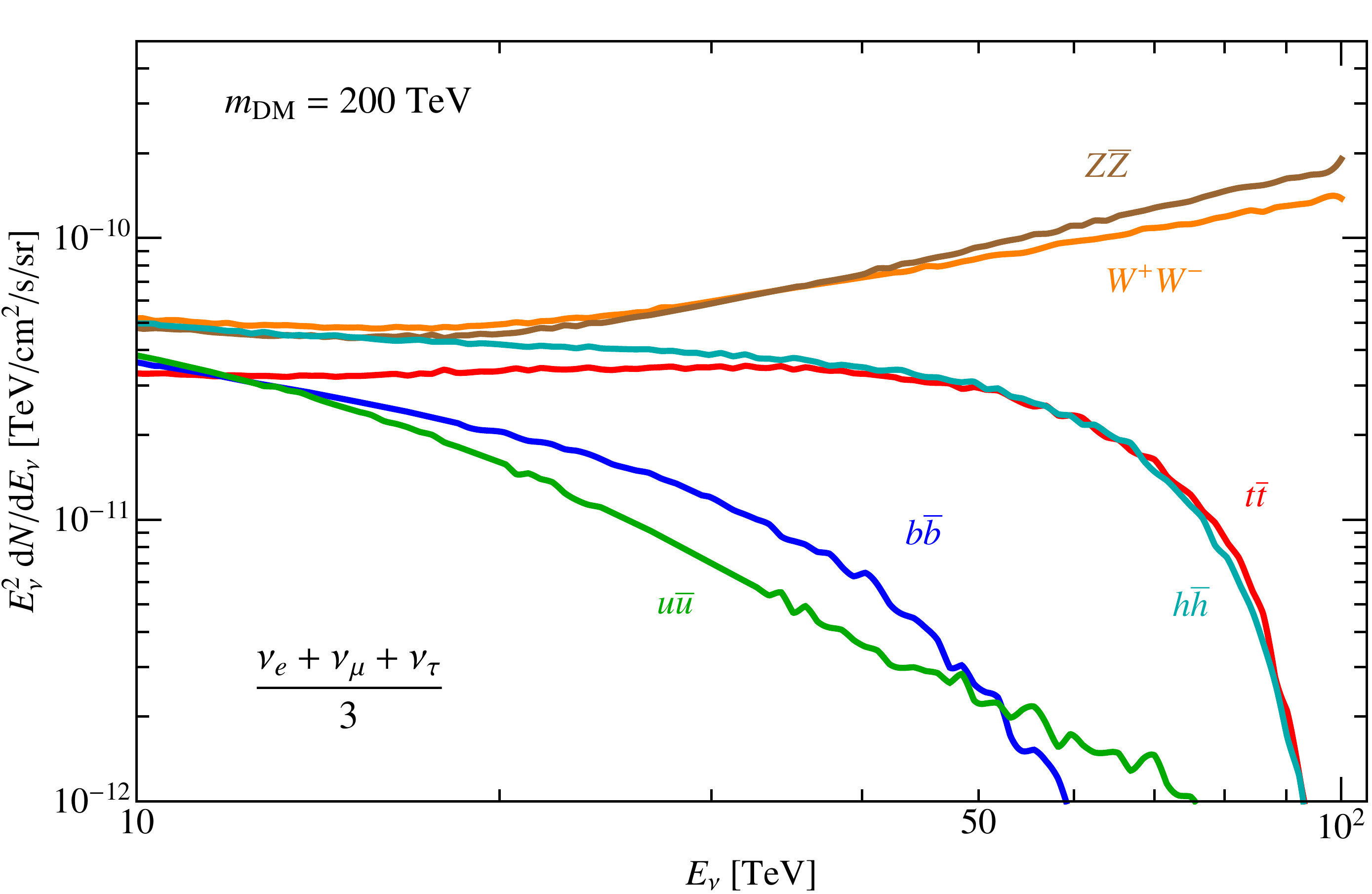}}\quad
	\subfigure{\includegraphics[width=0.5\textwidth]{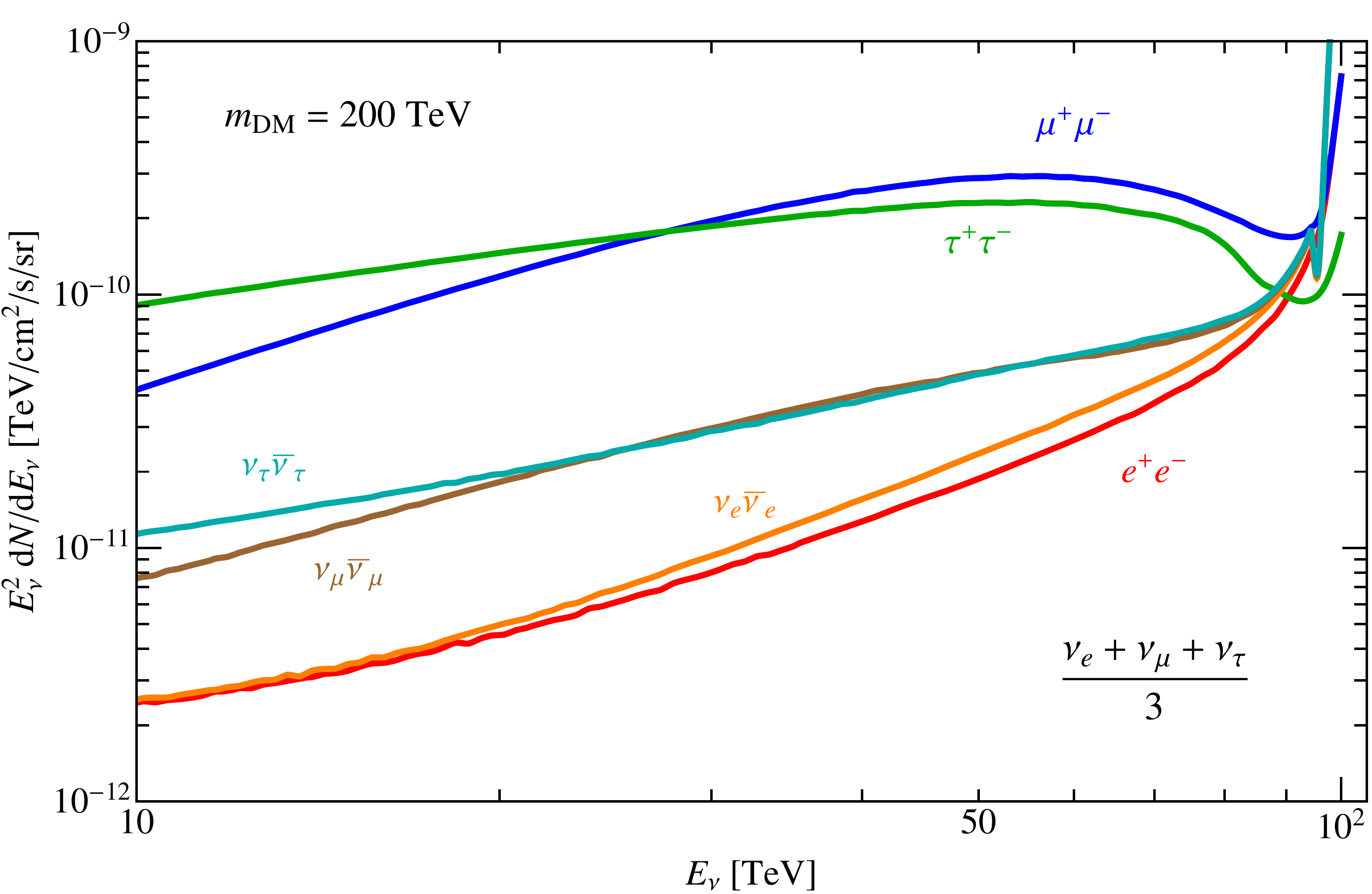}}\\
	\subfigure{\includegraphics[width=0.5\textwidth]{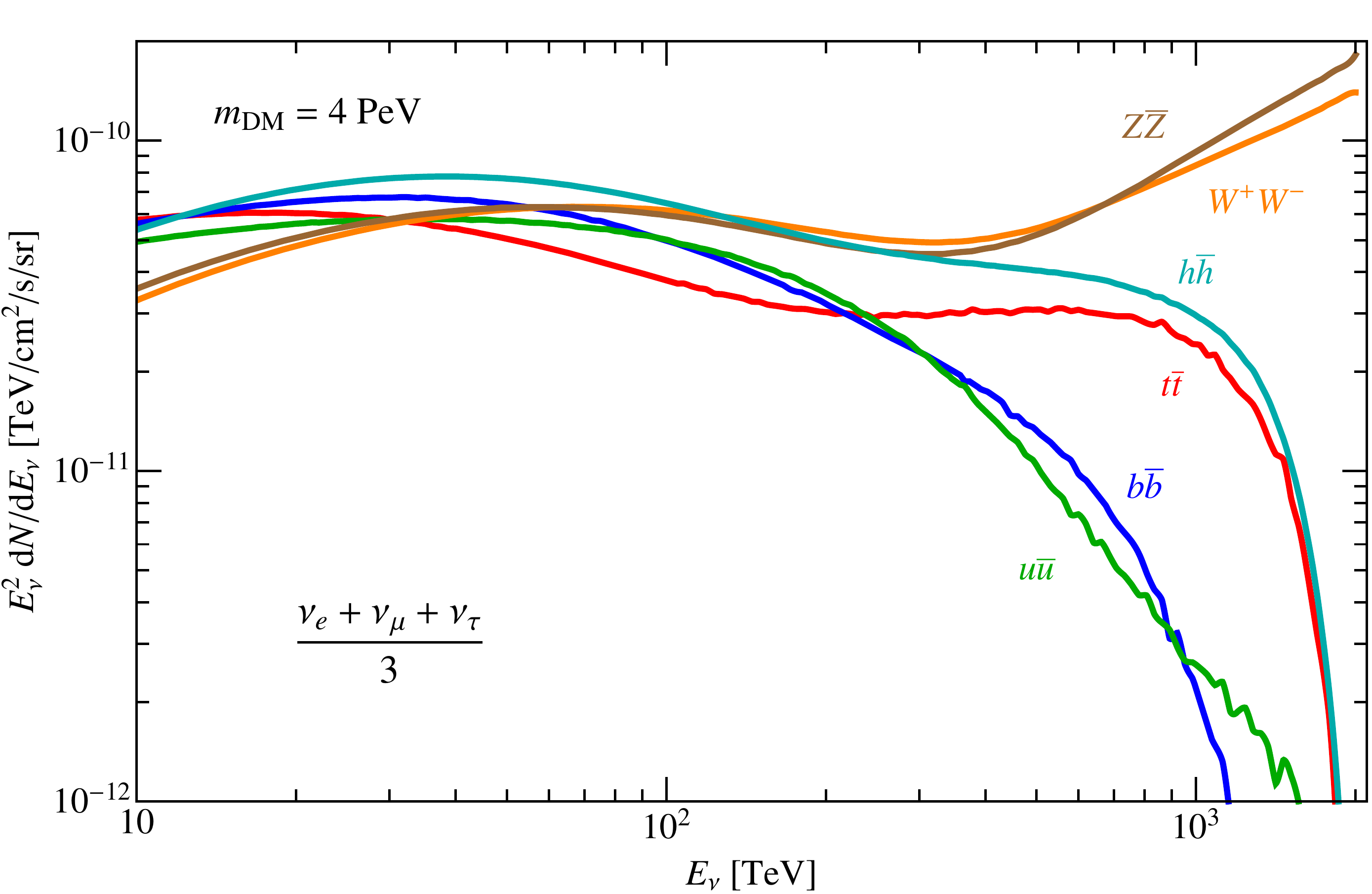}}\quad
	\subfigure{\includegraphics[width=0.5\textwidth]{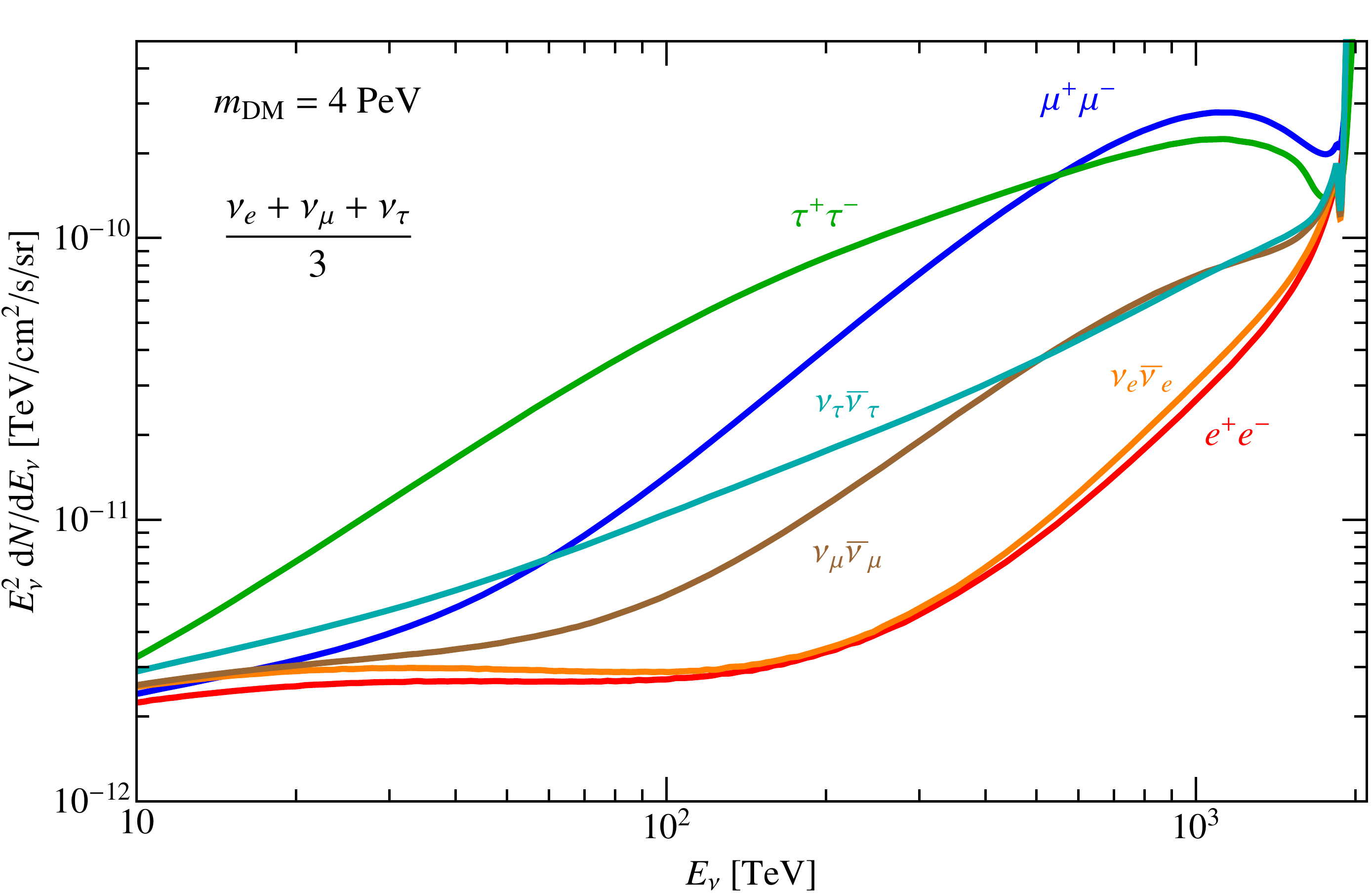}}
	\caption{\label{fig:flux} 
		All-sky averaged neutrino plus antineutrino flux (averaged over flavors, i.e., $(\nu_e+\nu_\mu+\nu_\tau)/3$) from DM decays into various two-body channels and for two DM masses, $m_{\rm DM}=200$~TeV (top panels) and $m_{\rm DM}=4$~PeV (bottom panels). For all panels, $\tau_{\rm DM}=10^{27}$~s. Note that the average over neutrino flavors results in fluxes which are identical with or without neutrino oscillations.
		}
\end{figure}

In order to compute the neutrino spectrum of flavor $\alpha$ from DM decays into different final state two-body channels, $dN_{\nu_\alpha}/dE_\nu$, we use the event generator \verb+PYTHIA+ 8.2~\cite{Sjostrand:2014zea}, which includes the weak gauge bosons radiation corrections~\cite{Christiansen:2014kba}. In Figure~\ref{fig:flux} we show the expected flux of neutrinos (averaged over the neutrino flavors and averaged over all directions) at Earth for two DM masses $m_{\rm DM}=200$~TeV and 4~PeV, with $\tau_{\rm DM}=10^{27}$~s. As can be seen, all the neutrino channels (${\rm DM} \to \nu_\alpha \, \bar{\nu}_\alpha$), and to some extent the charged lepton channels (${\rm DM} \to e^+ \, e^-$, $\mu^+ \, \mu^-$ and $\tau^+ \, \tau^-$), share the pronounced peak at an energy equal to $m_{\rm DM}/2$. This feature results in an important contribution to the IceCube event spectrum in the bins immediately below $m_{\rm DM}/2$. However, electroweak corrections also populate lower energies and so, finally, the event spectrum is expected to be wider than that from a monochromatic flux. The weak gauge boson channels (${\rm DM}\to W^+ \, W^-$ and $Z \, Z$) also exhibit a mild peak at $m_{\rm DM}/2$ (mainly coming from the direct neutrino production from their decays). On the other hand, in the case of ``hadronic" channels (${\rm DM}\to$ quarks) and decays to Higgs bosons, the neutrino flux is much softer and does not present a peak, spreading over several decades of energy. For these channels, the contribution of DM decays to the event rate at IceCube covers several bins of energy significantly below $m_{\rm DM}/2$. Based on these considerations, the different decay channels can be classified into two clear groups: a) soft channels (\dm\ decays into quarks or Higgs bosons) and b) hard channels (\dm\ decays into charged and neutral leptons). Decays into gauge bosons can be classified as being intermediate to these two categories.

Finally, note that the galactic and extragalactic contributions are of the same order, while the peak of the flux at energies $\lesssim m_{\rm DM}/2$ in the case of leptonic channels comes from the galactic contribution. The extragalactic flux is smoothed due to the redshifting of the energy from DM decays occurring at different times.

%%%%%%%%%%%%%%%%%%%%%%%%%%%%%%%
%%%%%%%%%%%%%%%%%%%%%%%%%%%%%%%
\section{\label{sec:stat}Statistical analysis of the 4-year HESE data}
%%%%%%%%%%%%%%%%%%%%%%%%%%%%%%%
%%%%%%%%%%%%%%%%%%%%%%%%%%%%%%%

After 1347~days of data taking, 54 events with electromagnetic (EM)-equivalent deposited energies in the range $\sim [20~{\rm TeV}$--$2~{\rm PeV}]$ were detected by the IceCube neutrino observatory,\footnote{Note that one of these events is a pair of coincident muon tracks whose energy and direction cannot be unambiguously assigned, and we have not included it in our analysis.} divided into muon tracks or showers, depending on the way the energy is deposited in the detector. In addition, the direction of each event can be also determined, with an uncertainty that depends on the event topology ($\lesssim 1^\circ$ for tracks and $\sim 10^\circ$--$20^\circ$ for showers). In this work, we consider the HESE sample in two EM-equivalent deposited energy intervals: all 53 events for [10~TeV--10~PeV] and 32 events for [60~TeV--10~PeV]. With the increased low-energy threshold, the effect of the background is reduced; however, the statistics also becomes weaker. 

At these energies, the sources of background are atmospheric muons and neutrinos, produced as secondaries in cosmic-ray interactions with the nuclei of the atmosphere. Given that the best fit value for the contribution from prompt atmospheric neutrinos is zero, we only consider conventional atmospheric neutrinos. For the two energy intervals we consider in this work, the corresponding numbers for 4 years (1347 days) are: $N_\mu =12.6$ and $N_\nu = 9.0$ for [10~TeV--10~PeV]; and $N_\mu =0.6$ and $N_\nu = 3.3$ for [60~TeV--10~PeV]~\cite{Aartsen:2014gkd, Aartsen:2015zva}. In this work we fix the expected number of background events. While this definitely has no significant impact on the analysis only involving events above 60~TeV, given the similarity of the results for both energy intervals, we expect the results to stay qualitatively unchanged when including uncertainties in the background expectations.

In addition to the background, in part of this work we consider two components of the neutrino flux at Earth: neutrinos from DM decays and an isotropic power-law flux. The former was described in the previous section and the latter is parameterized as
\begin{equation}
\left. \frac{d \Phi_{{\rm astro}, \nu_\alpha}}{d E_\nu} \right|_\oplus = \phia \, \left(\frac{E_\nu}{\rm 100~TeV}\right)^{-\gamma} ~, 
\end{equation}
where $\gamma$ is the spectral index and $\phia$ is the flux normalization (per flavor), which  will be given in units of $10^{-18} \, {\rm GeV}^{-1} \, {\rm cm}^{-2} \, {\rm s}^{-1} \, {\rm sr}^{-1}$. In what follows we consider this flux to be mainly originate from the usual hadronic production mechanism at neutrino sources, which results in the canonical homogeneous flavor ratios at Earth, $(1 : 1 : 1)_\oplus$, for both neutrinos and antineutrinos.

Therefore, the combined flux from an astrophysical power-law and single-channel DM decays can be represented in terms of four fundamental parameters: the spectral index $\gamma$ and normalization $\phia$ of the astrophysical flux and the lifetime $\tau_{\rm DM}$ and DM mass $m_{\rm DM}$ governing the neutrino flux from DM decays into channel $c$,
\begin{equation}
\label{eq:tot-flux}
\frac{d \Phi^c}{d E_\nu} (E_\nu; \tau_{\rm DM}, \mdm, \phia, \gamma) = 
\frac{d\PhiDM^c}{d E_\nu} (E_\nu; \tdm, \mdm) + \frac{d \PhiAst}{d E_\nu} (E_\nu; \phia, \gamma) ~.
\end{equation}
However, in the analysis, it is more convenient to trade the power-law normalization and the DM lifetime by the number of astrophysical and DM events, $N_{\rm astro}$ and $N_{\rm DM}$, respectively. Therefore, the four parameters entering most of our fits are: $\boldsymbol{\theta} = \{N_{\rm DM}, m_{\rm DM}, N_{\rm astro}, \gamma\}$, which are then converted back into $\{\tau_{\rm DM}, \mdm, \phia, \gamma\}$ to present the results. 

We also separately consider the possibility that the signal comes only from DM decays via two dominant channels. In these cases, our analyses involve three free parameters, $\boldsymbol{\theta}_{2c} = \{N_{\rm DM}, m_{\rm DM}, \textrm{BR}\}$, where $\textrm{BR}$ is the branching ratio into one of the channels. Analogously, we convert $N_{\rm DM}$ into $\tau_{\rm DM}$. In what follows, we describe the statistical analyses for the cases involving both an astrophysical and a single-channel decaying DM components, although it is straightforward to adapt them to the two-channel decaying DM cases.
 
In order to compute the event spectra of the signal (the astrophysical power-law flux and DM decays) and of the background contributions, we closely follow the approach of Ref.~\cite{Vincent:2016nut}, which updates and improves the detailed description provided in Ref.~\cite{Palomares-Ruiz:2015mka}. It includes modeling of the deposited EM-equivalent energy as a function of the true neutrino energy for different processes, modeling of the effective target mass as a function of the deposited energy, and the computation of the rates of electromagnetic and hadronic showers, as well as of muon tracks, as a function of their energy. We also compute the effects of neutrino attenuation in the Earth for all neutrino fluxes (atmospheric background, astrophysical and from DM decays), including the contribution from secondary neutrinos from $\nu_\tau$ regeneration.\footnote{The flux of secondary neutrinos from $\nu_\tau$ regeneration in the Earth is especially relevant for hard incoming $\nu_\tau$ fluxes and thus, it represents a significant contribution for the case of the DM signal.} We model the energy spectra of the atmospheric muon background and also add the veto for downgoing atmospheric neutrinos. We refer the reader to these two works~\cite{Palomares-Ruiz:2015mka, Vincent:2016nut} for details.

Following Ref.~\cite{Vincent:2016nut}, we perform an unbinned extended maximum likelihood analysis, using the EM-equivalent deposited energy, event topology and hemisphere of origin of the 53 events (or 32, depending on the energy interval) in the 4-year HESE sample. Each observed event $i$ is identified by $\{E_{\rm dep,i}, H_i, T_i\}$, which refer to EM-equivalent deposited energy ($E_{\mathrm{dep},i}$); direction ($H_i =$ upgoing or downgoing); and topology ($T_i =$ track or shower). The full likelihood, for a given DM decay channel $c$, is given by:

\begin{equation}
\like^c (\boldsymbol{\theta}) = \frac{e^{-N_{\rm DM} - N_{\rm astro} - N_\nu - N_\mu}}{N_{\rm obs}!}  \,  \prod_{i = 1}^{N_{\rm obs}} \like_i^c (\boldsymbol{\theta}) ~,
\end{equation}
where $N_{\rm obs}$ is the total number of observed events. The likelihood for each event $i$, for a given DM decay channel $c$, is given by:
\begin{equation}
\like_i^c (\boldsymbol{\theta}) =N_{\rm DM} \, \pdf_{{\rm DM}, i}^c(m_{\rm DM}) + N_{\rm astro} \, \pdf_{{\rm astro}, i} (\gamma) + N_\nu \, \pdf_{\nu, i} + N_\mu \, \pdf_{\mu, i}~,
\label{eq:fullLike}
\end{equation}
where $\pdf_{f, i}$ is the probability density of event $i$ corresponding to source $f = \{ {\rm DM}, {\rm astro}, \nu, \mu \}$, which is normalized to one for each source. 
 
The probability density of a given event $i$ that corresponds to a flux of neutrinos from DM decays into channel $c$ with direction $H_i$ to be produced with EM-equivalent deposited energy $E_{\textrm{dep}, i}$ and topology $T_i$ can be written as
\begin{equation}
\pdf_{{\rm DM},i}^c (\mdm)= \frac{1}{\sum_{\ell, H', T'} \int_\Emin^\Emax dE_{\textrm{dep}} \, \frac{d\left(N_{\rm DM}^c\right)_{\ell, H'}^{T'}}{dE_{\textrm{dep}} }} \, \sum_\ell \frac{d\left(N_{\rm DM}^c\right)_{\ell, H_i}^{T_i}}{dE_{\textrm{dep},i}}   ~.
\label{eq:PDFDM}
\end{equation}
where $\Emin$ and $\Emax$ are the minimum and maximum EM-equivalent deposited energies considered in a given analysis, and the sum in the denominator goes over the three neutrino flavors, $\ell = \{e, \mu, \tau\}$, the direction, $H' =$\{upgoing, downgoing\}, and the event topology, $T'=\{\textrm{track, shower}\}$. The event spectrum $d\left(N_{\dm}^c\right)_{\ell, H}^{T}/dE_{\textrm{dep}}$ results from the sum of all the partial contributions (including secondary neutrinos from $\nu_\tau$ regeneration in the Earth) from different processes to topology $T$ from neutrinos and antineutrinos of flavor $\ell$ and with direction $H$, for a given neutrino flux at Earth $d\PhiDM^c/d E_\nu$. The probability densities corresponding to any other source of events are analogously defined (for more details, see Refs.~\cite{Palomares-Ruiz:2015mka, Vincent:2016nut}).

In order to compute two-dimensional parameter correlations, for each DM decay channel $c$, we define the following test statistic,
\begin{equation}
{\rm TS}_{2\rm D}^c ({\boldsymbol{\theta_{\rm test}}}) = 
- 2 \, \ln \frac{\like^c (\boldsymbol{\theta}_{\rm test}, \boldsymbol{\hhatnu}(\boldsymbol{\theta_{\rm test}}))}{\like^c (\boldsymbol{\widehat{\theta}})} ~,
\label{eq:TS2D}
\end{equation}
where $\boldsymbol{\theta}_{\rm test}$ indicates the pair of parameters to evaluate and $\boldsymbol{\nu}$ the other two remaining ones, taken as nuisance parameters. The single-hat quantity $\boldsymbol{\widehat{\theta}}$ corresponds to the values of the four parameters that maximize the likelihood (the global best fit) and the double-hat quantity $\hhatnu(\boldsymbol{\theta_{\rm test}})$ corresponds to the values of $\boldsymbol{\nu}$ that maximize $\like^c$ for the specified pair $\boldsymbol{\theta_{\rm test}}$.  We also impose positivity on the values of the number of events from DM decays and astrophysical neutrinos, i.e., $N_{\rm DM} \geq 0$ and $N_{\rm astro} \geq 0$. We assume ${\rm TS}_{2\rm D}^c$ asymptotically follows a $\chi^2$ distribution with two degrees of freedom. However, note that this approximation fails for very small number of events, either $N_{\rm DM}$ or $\nast$, as the other associated parameter ($\mdm$ or $\gamma$) is not specified in the limit of $N_{\rm DM} = 0$ or $\nast = 0$~\cite{Davies:1977, Davies:1987}. For instance, for $\boldsymbol{\theta_{\rm test}} = \{\ndm, \mdm\}$, in the $\ndm = 0$ limit, $\mdm$ is undefined under the hypothesis of only astrophysical (and background) events, and hence, part of the fluctuations of ${\rm TS}_{2\rm D}^c$ are unrelated to the difference of the two hypotheses tested (DM versus no DM).
Consequently, ${\rm TS}_{2\rm D}^c$ is expected to follow a distribution with a larger expectation value than that of a $\chi^2$ distribution with two degrees of freedom. With this word of caution in mind, we proceed with the approximation.
 
Finally, we also compute limits on the DM lifetime for different channels for each DM mass, $\mdm$. To do so, we define a new test statistic, given by
\begin{equation}
 {\rm TS}_{\rm lim}^c (N_{\rm DM}; \mdm) = \begin{cases}
- 2 \, \ln \frac{\like^c (N_{\rm DM}, \hhatn_{\rm astro} (N_{\rm DM}),  \widehat{\widehat{\gamma}} (N_{\rm DM}); \mdm)}{\like^c (\boldsymbol{\widehat{\theta}})} \, , \hspace{1cm} N_{\rm DM} \geq \widehat{N}_{\rm DM} \\[2ex] 
0 \, , \hspace{7cm} N_{\rm DM} < \widehat{N}_{\rm DM}
\end{cases}
 \label{eq:TSlim}
\end{equation}
where, again, we impose $N_{\rm DM} \geq 0$ and $N_{\rm astro} \geq 0$. In practice, the results are obtained using the fact that ${\rm TS}_{\rm lim}^c (N_{\rm DM}; \mdm)$ is asymptotically distributed as half a delta function at zero plus half a $\chi^2$ distribution for one degree of freedom~\cite{Cowan:2010js}, i.e., ${\rm TS}_{\rm lim}^c (N_{\rm DM}; \mdm) = 1.642$ (2.706) corresponds to a bound at 90\%~CL (95\%~CL), for the given mass $\mdm$ and decay channel $c$.

%%%%%%%%%%%%%%%%%%%%%%%%%
%%%%%%%%%%%%%%%%%%%%%%%%%
\section{\label{sec:fit-results}Results: DM decays plus isotropic astrophysical power-law flux}
%%%%%%%%%%%%%%%%%%%%%%%%%
%%%%%%%%%%%%%%%%%%%%%%%%%

As previously discussed, we analyze the IceCube 4-year HESE data assuming an astrophysical (isotropic and unbroken) power-law spectrum plus a signal from heavy DM decays. We do so for different two-body final states from DM decays. We show the best-fit values for the four parameters we fit $\boldsymbol{\theta} = \{\ndm, \mdm, \nast, \gamma\}$ (we also provide the physical parameters $\tau_{\rm DM}$ and $\phia$) in Table~\ref{tab:fits-noprior10} for the EM-equivalent deposited energy interval [10~TeV--10~PeV] and in Table~\ref{tab:fits-noprior60} for [60~TeV--10~PeV], which corresponds to the same lower energy threshold used in the official IceCube HESE analyses. 

In general, note that the best fit for the DM mass for quark channels span more than an order of magnitude (from $\mdm \sim 500$~TeV for decays into $u \, \bar{u}$ to $\mdm \sim 11$~PeV for decays into $t \, \bar{t}$), while decays into the gauge and Higgs bosons, charged leptons, and neutrinos best fit the data with DM masses in a narrower range, $\mdm \sim 4\text{--}8$~PeV. The former tend to better explain the low-energy excess in the HESE sample, whereas the latter help to explain the PeV events (gauge boson channels also partly contribute to events at $\sim 100$~TeV). Except for DM decays into $u \, \bar{u}$ and $b \, \bar{b}$ (and $h \, h$ when the threshold is set at 60~TeV), the best fit for the astrophysical index points to a very soft spectrum ($\gamma > 3$), hard to explain with standard acceleration mechanisms. However, for the few cases with harder astrophysical flux ($\gamma < 2.5$), the corresponding DM lifetime is inevitably too low ($\tau_{\rm DM} \lesssim 10^{27}$~s), and in tension with constraints from gamma-ray observations (see below).

\begin{table}[t]
	\caption{Best-fit values for $\boldsymbol{\theta} = \{\ndm (\tau_{\rm DM}), \mdm, \nast (\phia), \gamma\}$, where $\phia$ is given in units of $10^{-18}~{\rm GeV}^{-1}~{\rm cm}^{-2}~{\rm s}^{-1}~{\rm sr}^{-1}$. The EM-equivalent deposited energy interval is [10~TeV--10~PeV].}  	  	
	\begin{center}
		\begin{tabular}{c|cc|cc}
			\hline
			Decay channel & $ \ndm (\tau_{\rm DM} [10^{28}~{\rm s}])$  &  $ \mdm $ [TeV] &
			$\nast (\phia) $  & $\gamma$ \\
			\hline
			$u \, \bar{u}$                &       14.6 (0.033) &        521 &       22.2 (1.4) &      2.48 \\
			$b \, \bar{b}$                &       21.2 (0.082) &       1040 &       14.7 (0.73) &      2.29 \\
			$t \, \bar{t}$                &       18.1 (0.59) &      11167 &       18.4 (1.6) &      3.64 \\
			$W^{+} \, W^{-}$              &       11.9 (1.5) &       4864 &       24.7 (2.2) &      3.43 \\
			$Z \, Z$                      &       11.1 (1.6) &       4811 &       25.5 (2.3) &      3.40 \\
			$h \, h$                      &       18.5 (0.86) &       8729 &       18.1 (1.5) &      3.69 \\
			$e^{+} \, e^{-}$              &        4.6 (1.3) &       4131 &       31.9 (2.8) &      3.20 \\
			$\mu^{+} \, \mu^{-}$          &        5.8 (5.5) &       6513 &       30.9 (2.7) &      3.26 \\
			$\tau^{+} \, \tau^{-}$        &        7.1 (4.8) &       6836 &       29.6 (2.6) &      3.30 \\
			$\nu_e \, \bar{\nu}_e$        &        3.6 (2.7) &       4048 &       32.6 (2.8) &      3.16 \\
			$\nu_\mu \, \bar{\nu}_\mu$    &        6.0 (2.6) &       4151 &       30.8 (2.7) &      3.27 \\
			$\nu_\tau \, \bar{\nu}_\tau$  &        6.4 (2.4) &       4132 &       30.3 (2.7) &      3.29 \\
			\hline
		\end{tabular}
	\end{center}	 
	\label{tab:fits-noprior10}
\end{table}

\begin{table}[t]	
	\caption{Same as Table~\ref{tab:fits-noprior10}, but for the EM-equivalent deposited energy interval [60~TeV--10~PeV].}  	  	
	\begin{center}
		\begin{tabular}{c|cc|cc}
			\hline
			Decay channel & $ \ndm (\tau_{\rm DM} [10^{28}~{\rm s}])$  &  $ \mdm $ [TeV] &
			$\nast (\phia) $  & $\gamma$ \\
			\hline
			$u \, \bar{u}$                &       10.2 (0.021) &        522 &       16.6 (1.2) &      2.42 \\
			$b \, \bar{b}$                &       12.9 (0.089) &       1066 &       13.8 (0.83) &      2.32 \\
			$t \, \bar{t}$                &       16.1 (0.58) &      11134 &       10.7 (1.9) &      3.91 \\
			$W^{+} \, W^{-}$              &       11.3 (1.4) &       4860 &       15.5 (2.5) &      3.66 \\
			$Z \, Z$                     &       10.5 (1.6) &       4800 &       16.3 (2.6) &      3.61 \\
			$h \, h$                     &       13.6 (0.17) &        606 &       13.2 (0.76) &      2.29 \\
			$e^{+} \, e^{-}$              &        5.0 (1.2) &       4116 &       21.9 (3.2) &      3.33 \\
			$\mu^{+} \, \mu^{-}$          &        6.3 (5.0) &       6437 &       20.7 (3.2) &      3.46 \\
			$\tau^{+} \, \tau^{-}$        &        7.6 (4.4) &       6749 &       19.3 (3.0) &      3.53 \\
			$\nu_e \, \bar{\nu}_e$        &        3.7 (2.6) &       4041 &       22.7 (3.2) &      3.24 \\
			$\nu_\mu \, \bar{\nu}_\mu$    &        6.4 (2.4) &       4133 &       20.6 (3.2) &      3.48 \\
			$\nu_\tau \, \bar{\nu}_\tau$  &        6.7 (2.3) &       4117 &       20.1 (3.1) &      3.50 \\
			\hline
		\end{tabular}
	\end{center}
	\label{tab:fits-noprior60}	
\end{table}

Comparing the two EM-equivalent deposited energy intervals, we find that the best-fit values of the four parameters to be quite similar. Nevertheless, the astrophysical power-law index obtained using events above 60~TeV is slightly larger, indicating that the astrophysical flux prefers a steeper (softer) flux in this case. Between the two selected energy intervals, the only channel that has different best-fit values is the $h \, h$ channel. For this channel and the analysis with the 10~TeV threshold, there is a local maximum in the likelihood around $\mdm \sim 600$~TeV and $\gamma \sim 2.3$, although the global maximum is located at $\mdm \sim 8.7$~PeV and $\gamma \sim 3.7$. However, that local maximum develops into the global maximum when the analysis is performed with the low-energy threshold at 60~TeV. Hence, the difference in the best fits obtained for this channel. Although not illustrated here, for the [60~TeV--10~PeV] interval, we find that there is also a near-degenerate maximum around $\mdm \sim 8.8$~PeV and $\gamma \sim 4.1$.

%%%%%%%%%%%%%%%%%%%%%%%%%%%%%%
%%%%%%%%%%%%%%%%%%%%%%%%%%%%%%
\subsection{Event rates}
\label{subsec:events-noprior}
%%%%%%%%%%%%%%%%%%%%%%%%%%%%%%
%%%%%%%%%%%%%%%%%%%%%%%%%%%%%%

\begin{figure}[t]
	\begin{center}
	  \subfigcapskip=-4pt
		\subfigure[{\scriptsize DM $\to b \, \bar{b}$}]{\includegraphics[width=0.49\linewidth]{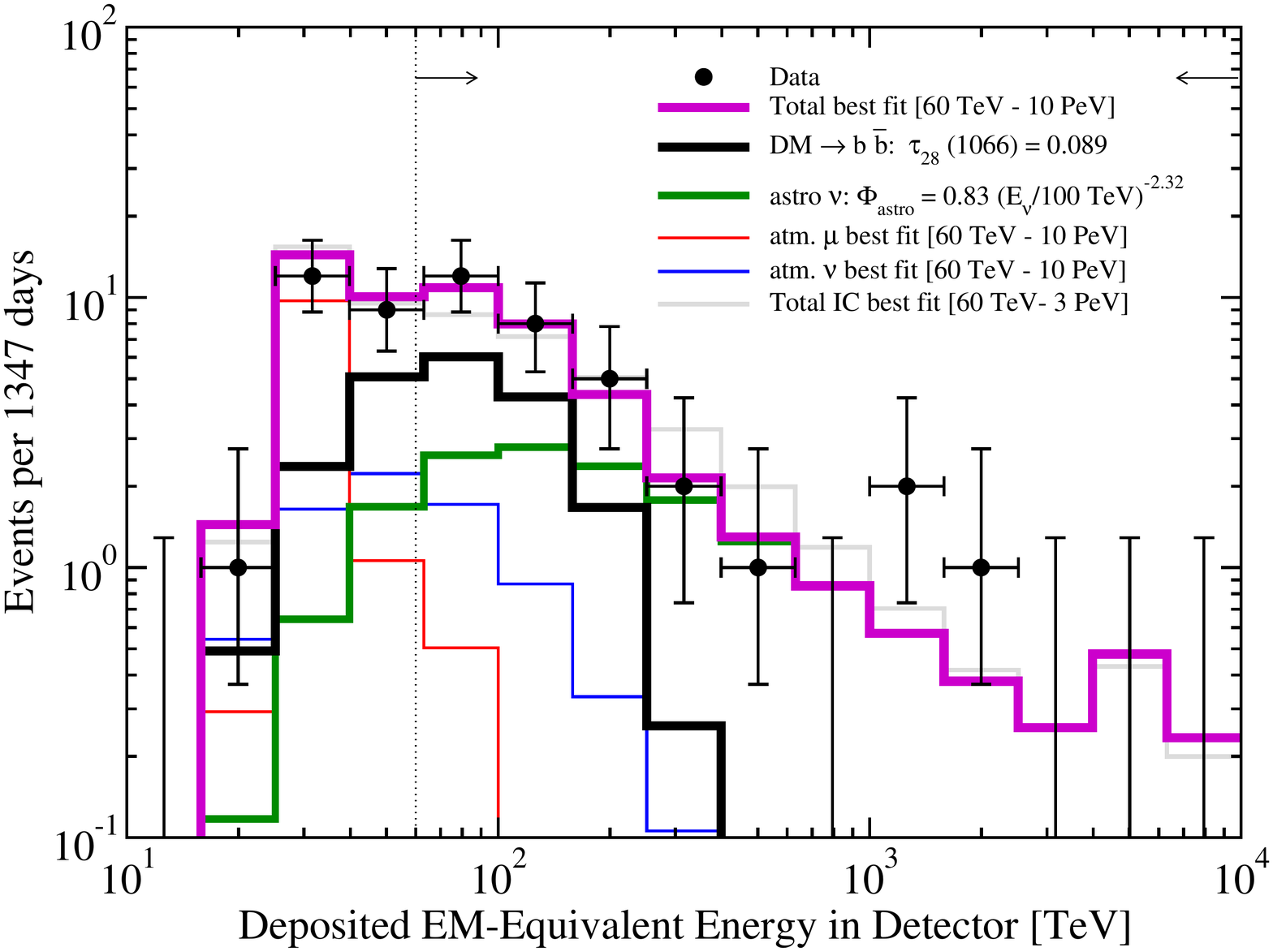}}
		\subfigure[{\scriptsize DM $\to W^{+} \, W^{-}$}]{\includegraphics[width=0.49\linewidth]{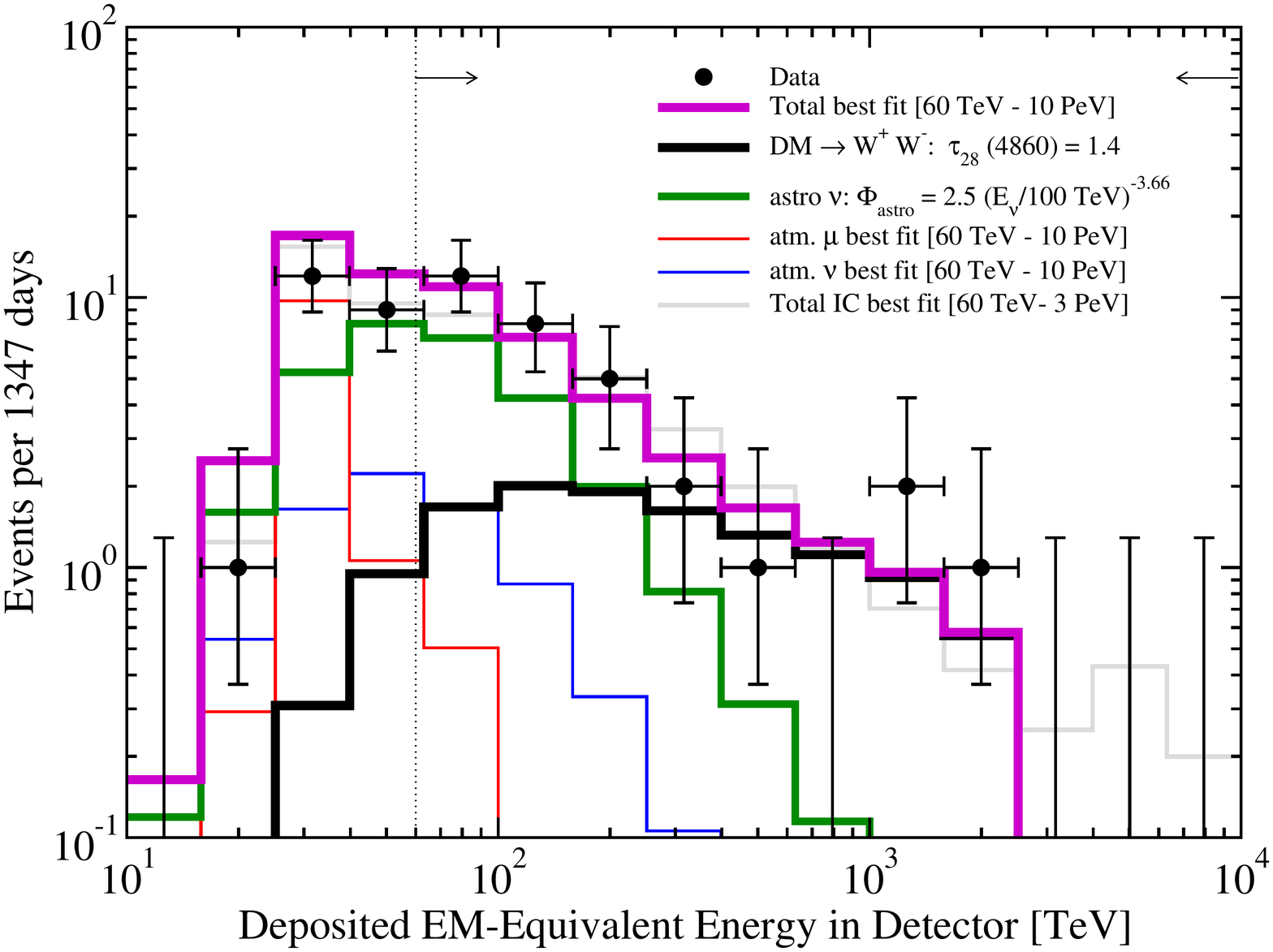}} \\
		\vspace{-0.25cm}
		\subfigure[{\scriptsize DM $\to \mu^-  \, \mu^+$}]{\includegraphics[width=0.49\linewidth]{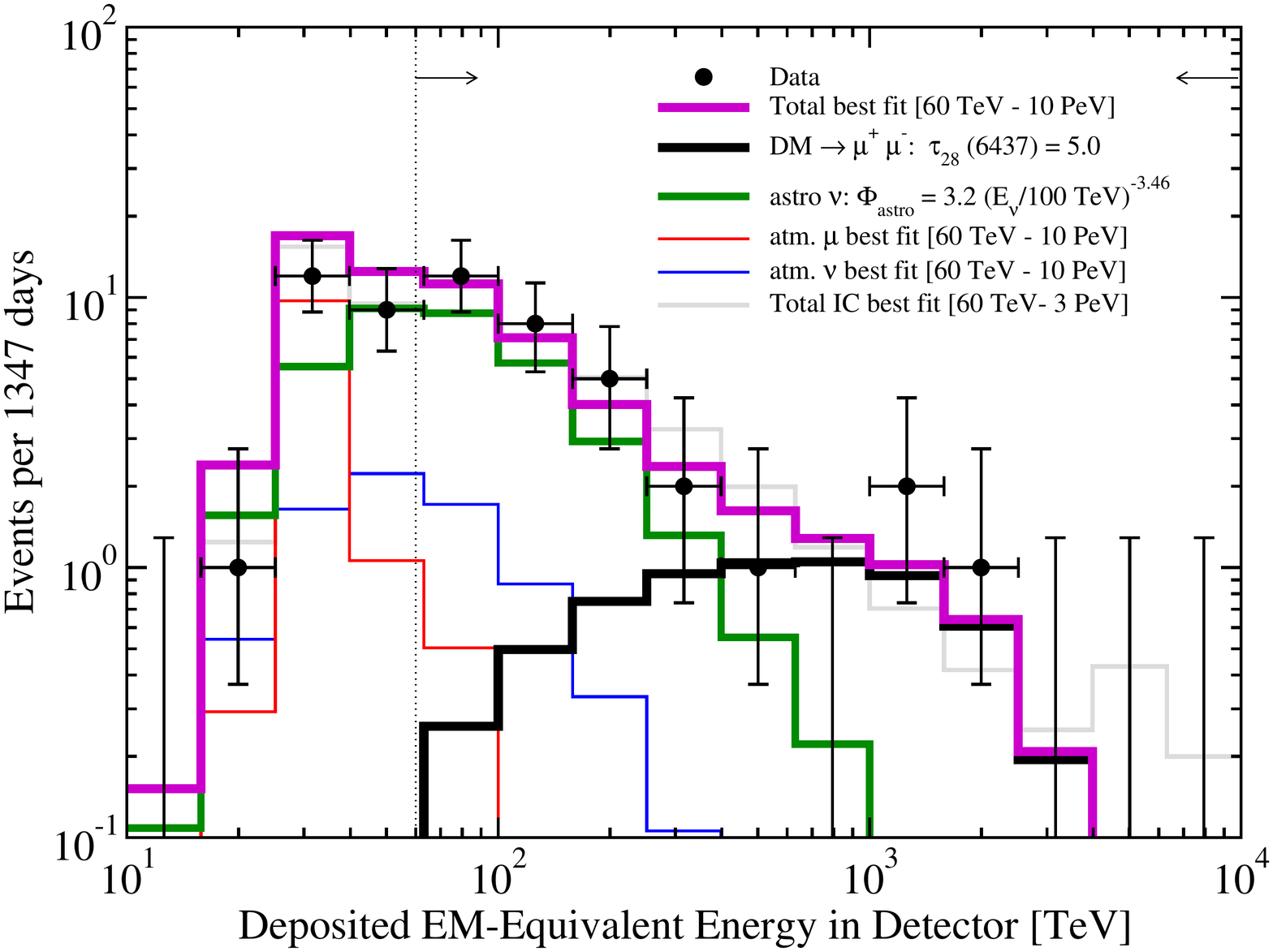}}
		\subfigure[{\scriptsize DM $\to \nu_e \, \bar{\nu}_e$}]{\includegraphics[width=0.49\linewidth]{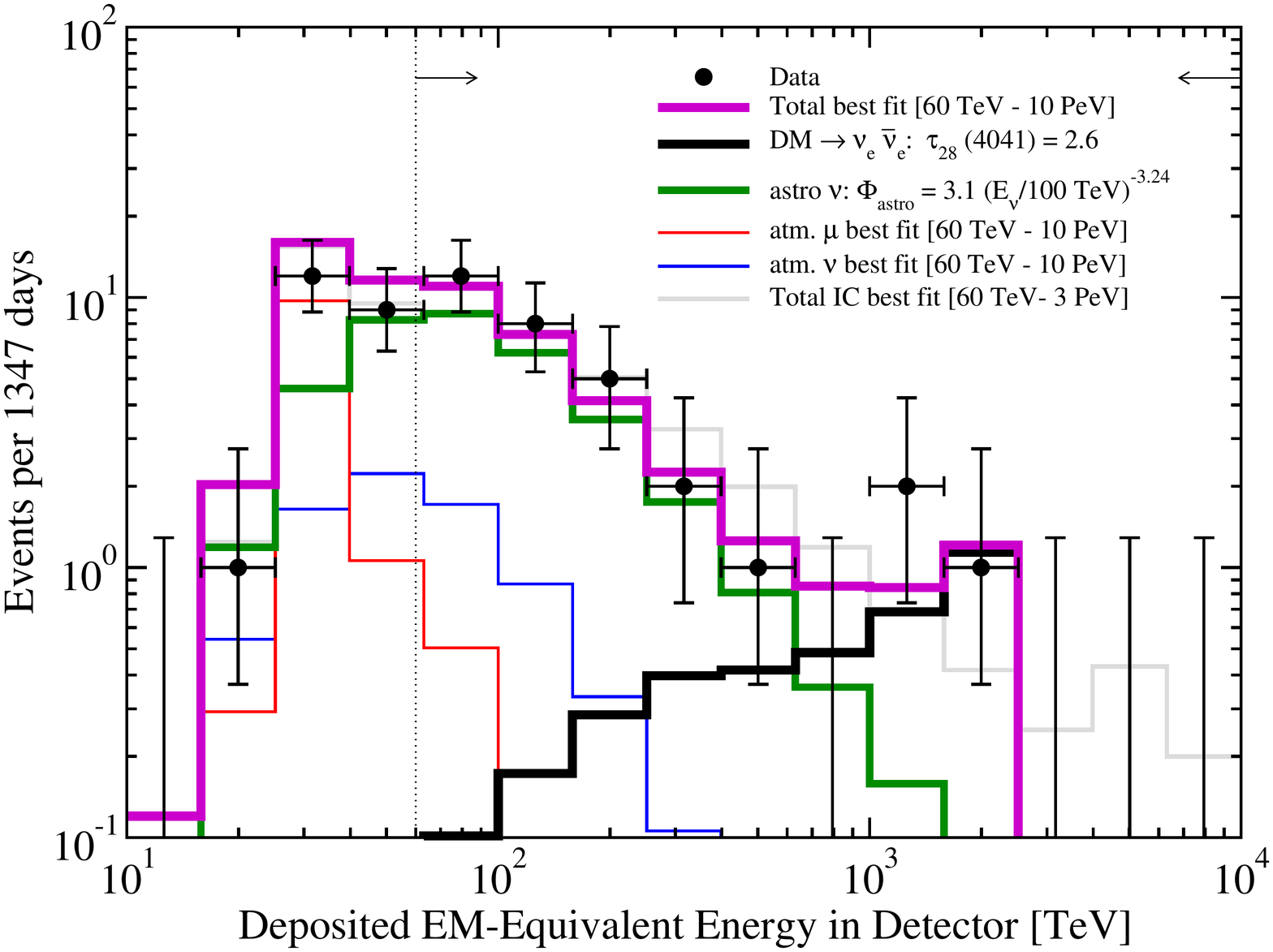}}
	\end{center}
	\caption{\label{fig:events}
		Event spectra in the IceCube detector after 1347 days. We show the results corresponding to the best fits in the EM-equivalent deposited energy interval [60~TeV--10~PeV] for four DM decay channels: DM $\to b \, \bar{b}$ (top-left panel), DM $\to W^{+} \, W^{-}$ (top-right panel), DM $\to \mu^{+} \, \mu^{-}$ (bottom-left panel) and DM $\to  \nu_e \, \, \bar{\nu}_e$ (bottom-right panel). In all panels: atmospheric muon events (red histogram), conventional atmospheric neutrino events (blue histogram), astrophysical neutrino events (green histogram), neutrino events from DM decays (black histogram), and total event spectrum (purple histogram). We indicate the best fit values of the DM lifetime and mass [$\tau_{28} (\mdm)$] in units of $10^{28}$~s and TeV, and the per-flavor normalization of the power-law flux ($\phia$) in units of  $10^{-8} \, {\rm GeV} \, {\rm cm}^{-2} \, {\rm s}^{-1} \, {\rm sr}^{-1}$. We also show the spectrum obtained using the 4-year IceCube best fit in the EM-equivalent deposited energy interval [60~TeV--3~PeV] (gray histogram), $E_\nu^2 \, d\Phi/dE_\nu = 2.2 \times 10^{-8} \, (E_\nu/100 \, {\rm TeV})^{-0.58}  \, {\rm GeV} \, {\rm cm}^{-2} \, {\rm s}^{-1} \, {\rm sr}^{-1}$ (per flavor), and the binned high-energy neutrino event data (black dots)~\cite{Aartsen:2015zva} with Feldman-Cousins errors~\cite{Feldman:1997qc}.
		}
\end{figure}

Here we show the event spectra corresponding to the best-fit parameters for the single-channel DM decays plus astrophysical signal, for four representative DM decay channels. In all cases, we also show the best-fit spectrum obtained by the IceCube collaboration using only a power-law astrophysical flux, in the EM-equivalent deposited energy range [60~TeV--3~PeV]. Given that our results are quite similar for the two energy intervals we consider, we only show results for the [60~TeV--10~PeV] interval. We show our results for the best-fit event rates in Figure~\ref{fig:events} for DM $\to b \, \bar{b}$ (top-left panel), DM $\to W^{+} \, W^{-}$  (top-right panel), DM $\to \mu^{+} \, \mu^{-}$ (bottom-left panel) and DM $\to  \nu_e \, \bar{\nu}_e$ (bottom-right panel). For all cases, the values for the different parameters of the DM decay and astrophysical spectra are indicated in Table~\ref{tab:fits-noprior60}.

For the case of DM decays into quarks, e.g., $b\,\bar{b}$ (top-left panel), we note that DM decays contribute substantially for energies ranging from about 40~TeV up to 200~TeV, with the maximum number of events predicted in the range (60~TeV--200~TeV).  The astrophysical flux is dominant at high energies, while DM decays take over below 200~TeV. For decays to other quarks, we find qualitatively similar results. On the other hand, for the case of DM decays into $W^+  \, W^-$ (top-right panel), the spectrum is harder and one can see that DM decay contributes substantially for a broad range of energies, from about 60~TeV up to 2~PeV, with the maximum number of events in the range from 80~TeV to 800~TeV. The astrophysical flux contributes significantly at lower energies, below 150~TeV and this is why the predicted power-law flux is very soft.

For the case of DM decays into leptons, the final event spectrum is expected to be harder. Our results for DM decays into $\mu^+  \mu^-$ are shown in the bottom-left panel of Figure~\ref{fig:events}. The main contribution from DM decays occurs at energies between $\sim$250~TeV and $\sim$2.5~PeV, while for energies below $\sim$250~TeV, the astrophysical flux component takes over. In the case of DM decays into $\nu_e \, \bar{\nu}_e$, the hardest channel we consider, the main contribution from DM decays lies at energies between $\sim$650~TeV and $\sim$2.5~PeV, while for energies below $\sim$650~TeV, the astrophysical flux component takes over.  We find similar qualitative behavior for other hard channels, i.e., the major contribution from DM decay is at the highest energy events, thus necessitating a very soft astrophysical spectrum to explain the lower energy events.
For the case of DM decays into gauge bosons, the high-energy events are also preferably explained by the DM component, as for hard channels, but there is also a significant DM contribution at lower energies.

When considering the low-energy threshold of 10~TeV, the relative contribution of the atmospheric background increases, but the number of non-atmospheric events also increases. However, for hard channels (and even for DM decays into gauge bosons), the number of DM events is similar to that with a higher low-energy threshold, because they contribute to high energies. Thus, the contribution over the expected background in the interval [10~TeV--60~TeV] must come from the astrophysical power-law flux and therefore, this requires a softer spectrum. For DM decays into quarks, since their contribution to the detected events covers a wider energy range, as can be seen from Figure~\ref{fig:events}, decreasing the low-energy threshold to 10~TeV also results in increasing the number of DM events, as well as the number of astrophysical events. Therefore, there are very small differences between the two analyses for both the contribution from DM decays and the astrophysical flux.

%%%%%%%%%%%%%%%%%%%%%%%%%%%%%%
%%%%%%%%%%%%%%%%%%%%%%%%%%%%%%
\subsection{\label{subsec:param-corr}Parameter correlations and preferred regions}
%%%%%%%%%%%%%%%%%%%%%%%%%%%%%%
%%%%%%%%%%%%%%%%%%%%%%%%%%%%%%

\begin{figure}[t]
	\begin{center}
	  \subfigcapskip=-4pt
		\subfigure[DM $\to b \, \bar{b}$]{\includegraphics[width=0.49\linewidth]{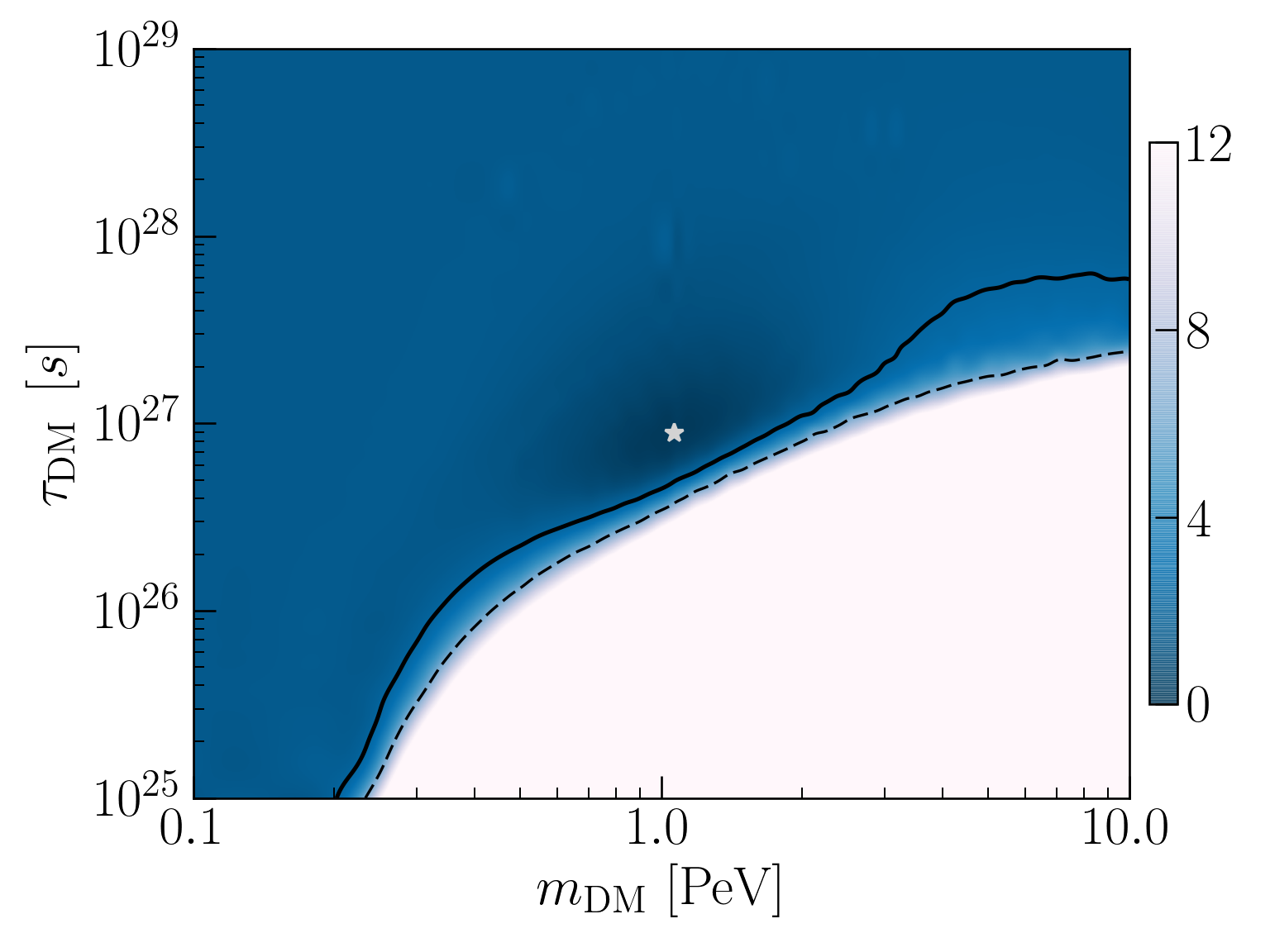}}
		\subfigure[DM $\to W^{+} \, W^{-}$]{\includegraphics[width=0.49\linewidth]{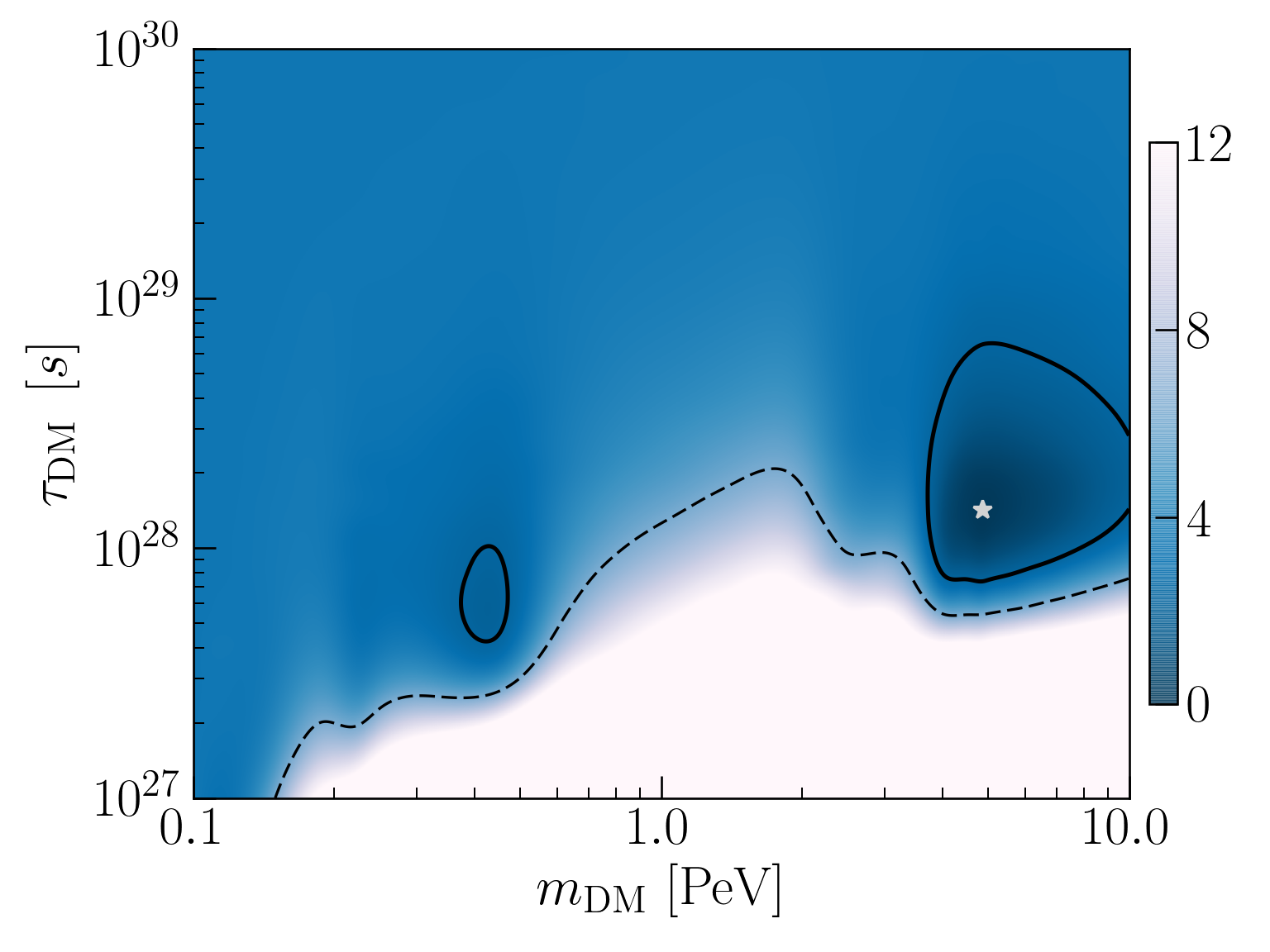}} \\
		\subfigure{\includegraphics[width=0.49\linewidth]{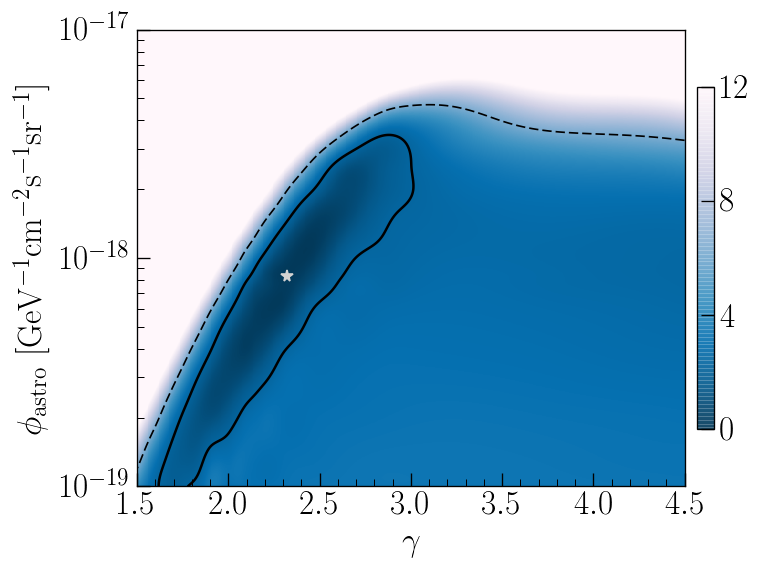}}
		\subfigure{\includegraphics[width=0.49\linewidth]{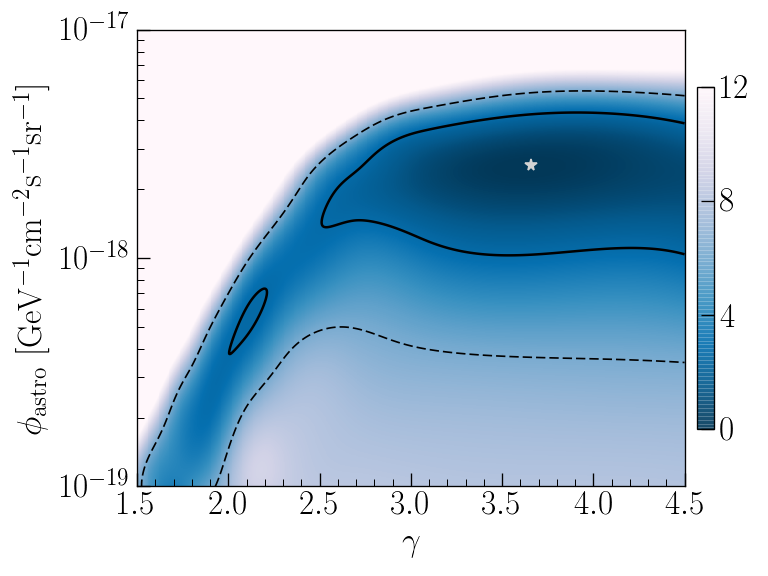}} 			
	\end{center}
	\caption{\label{fig:2DbW}
		DM lifetime-mass (top panels) and astrophysical normalization-spectral index (bottom panels) correlation for $\dm \to b \, \bar{b}$ (left panels) and $\dm \to W^+ \, W^-$ (right panels). The contours indicated by the solid black curves represent the $1\sigma$~CL preferred regions around the best fit (indicated by a white `$\star$' sign), while the corresponding $2\sigma$~CL regions are indicated by the dashed curves. The very different looking $1\sigma$~CL preferred regions between the two channels is representative of the differences between hard-spectrum and soft-spectrum channels.
		}
\end{figure}

\begin{figure}[t]
	\begin{center}
	  \subfigcapskip=-4pt
		\subfigure[$ \dm \to \mu^{+}\mu^{-} $]{\includegraphics[width=0.49\linewidth]{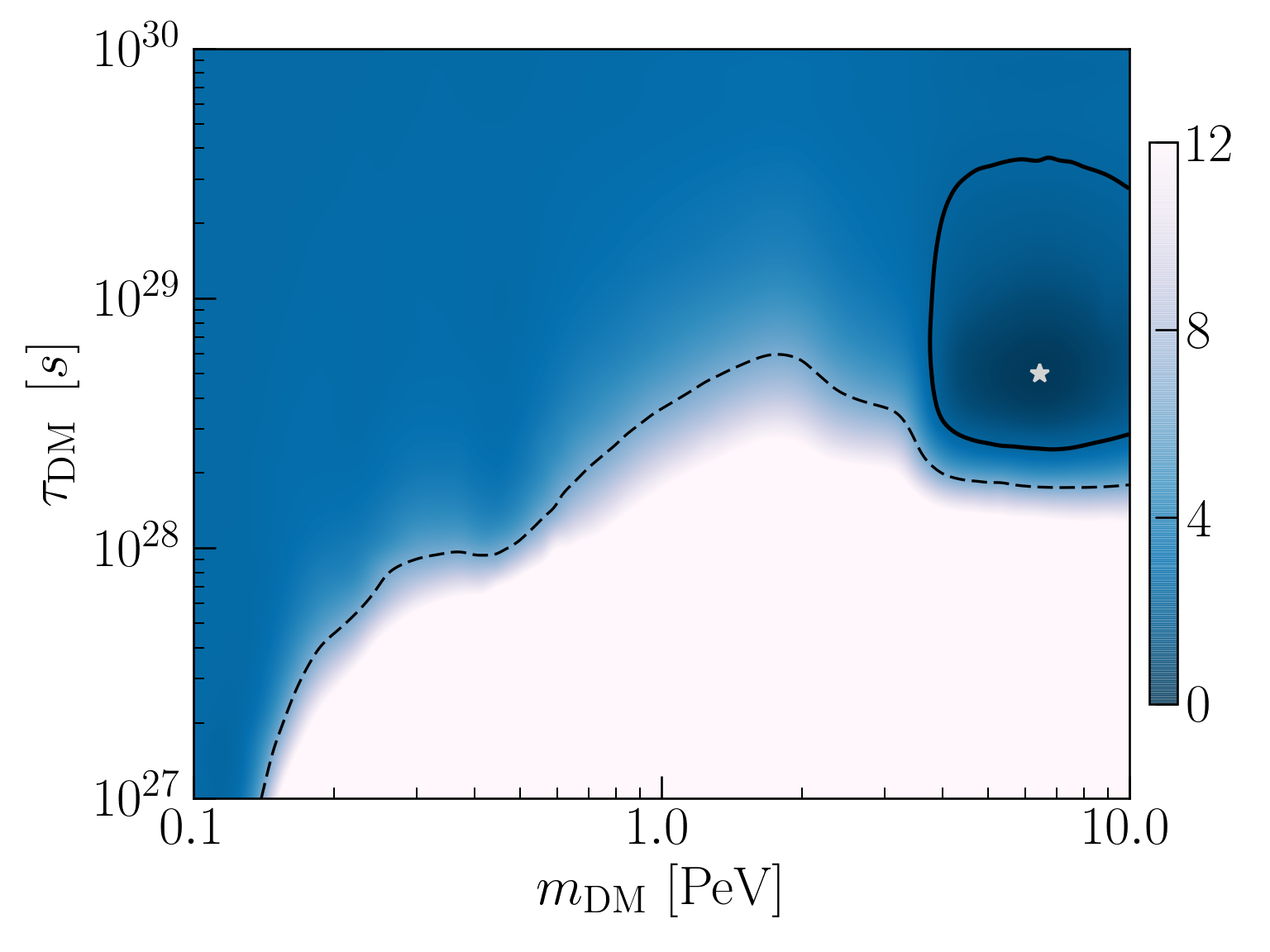}}
		\subfigure[$ \dm \to \nue\bar{\nu}_e $]{\includegraphics[width=0.49\linewidth]{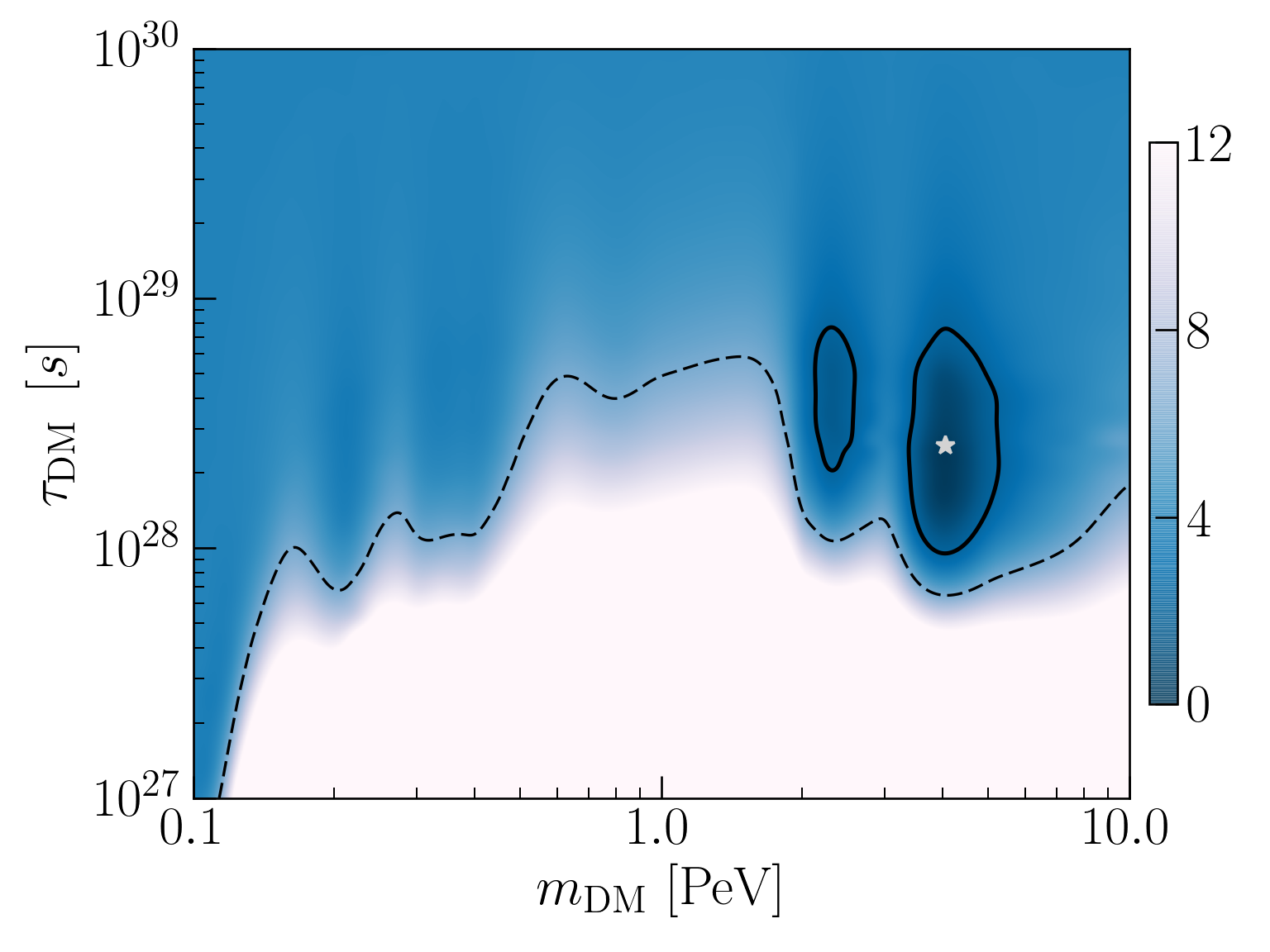}} \\
		\subfigure{\includegraphics[width=0.49\linewidth]{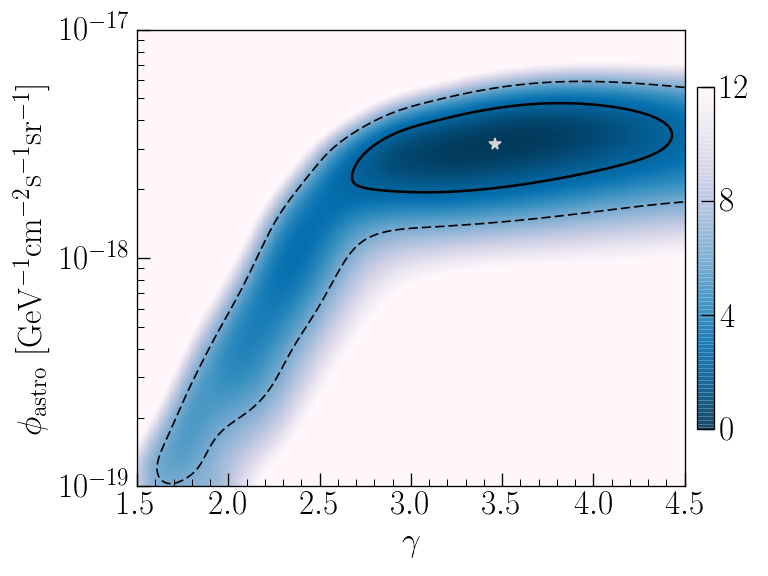}}
		\subfigure{\includegraphics[width=0.49\linewidth]{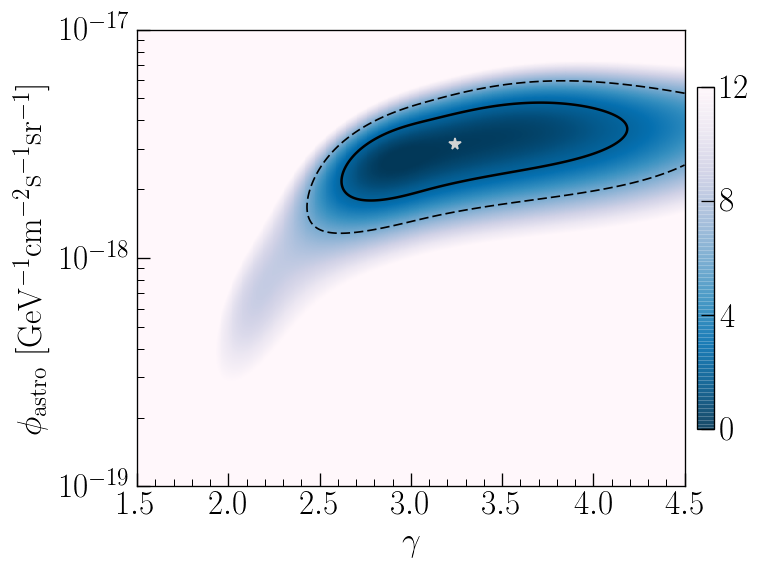}}
	\end{center}
	\caption{\label{fig:2Dmunumu}
		Same as Figure~\ref{fig:2DbW}, but for $\dm \to \mu^+ \, \mu^-$ and $\dm \to \nu_e \, \bar{\nu}_e $.
		}
\end{figure}

In this section, we discuss the correlations between two pairs of parameters from the full set and show, for the same representative channels, the preferred region of parameter space to the statistical confidence levels (CL) of 1$\sigma$ and 2$\sigma$ against the corresponding best-fit point (Figures~\ref{fig:2DbW} and~\ref{fig:2Dmunumu}). Correlations between parameters from the DM and astrophysical components are especially useful, as they demonstrate the interplay between these two disparate origins of neutrinos, and are reflective of the complementarity between the two components that is allowed by current data. On the other hand, correlations between individual parameters of the same component reflect the sensitivity, or lack thereof, of the preferred region to current data. In addition, they also express the significance of the particular component in improving the fit to the data. 

As an example, in the top panels of Figures~\ref{fig:2DbW} and~\ref{fig:2Dmunumu} we show the two-parameter correlation between \tdm\ and $m_{\rm DM}$ for the selected DM decay channels. This illustrates the relative insensitivity to the DM parameters in the case of soft channels such as $\dm \to b \, \bar{b}$, where a large fraction of the scanned \mdm--\tdm\ region is consistent with data even up to $1\sigma$~CL (top-left panel of Figure~\ref{fig:2DbW}). This includes the region where \tdm\ is high enough so that the contribution from DM decays is completely negligible, thereby suggesting that, for the specific channel $\dm \to b \, \bar{b}$, the astrophysical-flux-only fit to the data is statistically as good as that obtained with the DM component added in. This is notably different from the case where the decay spectrum is comparatively harder, e.g., for $\dm \to W^{+} \, W^{-}$ (top-right panel of Figure~\ref{fig:2DbW}). In this case, we see a clearly delineated $1\sigma$-preferred region around the best fit in the \mdm--\tdm\ plane. This suggests that a two-component explanation of the observed IceCube HESE data, with the flux from DM decays explaining the PeV events and the astrophysical power-law component filling-in for events in the $60$--$400$~TeV range, provides a statistically better fit. For the same channel ($\dm \to W^{+} \, W^{-}$), a narrow $1\sigma$-allowed region also opens up at lower $\mdm \sim 400$~TeV, which suggests that a different combination of DM and astrophysical flux components would also be consistent with the best-fit at this significance. In this case, the astrophysical flux would have to be much harder, in order to explain the PeV events. Indeed, this is borne out by a similar, and complementary, $1\sigma$~CL region opening up in the $\gamma\text{--}\phia$ correlation plot (bottom panels of Figures~\ref{fig:2DbW}), preferring low values of $\gamma$.

\begin{figure}[t]
	\centering
	\includegraphics[width=0.75\textwidth]{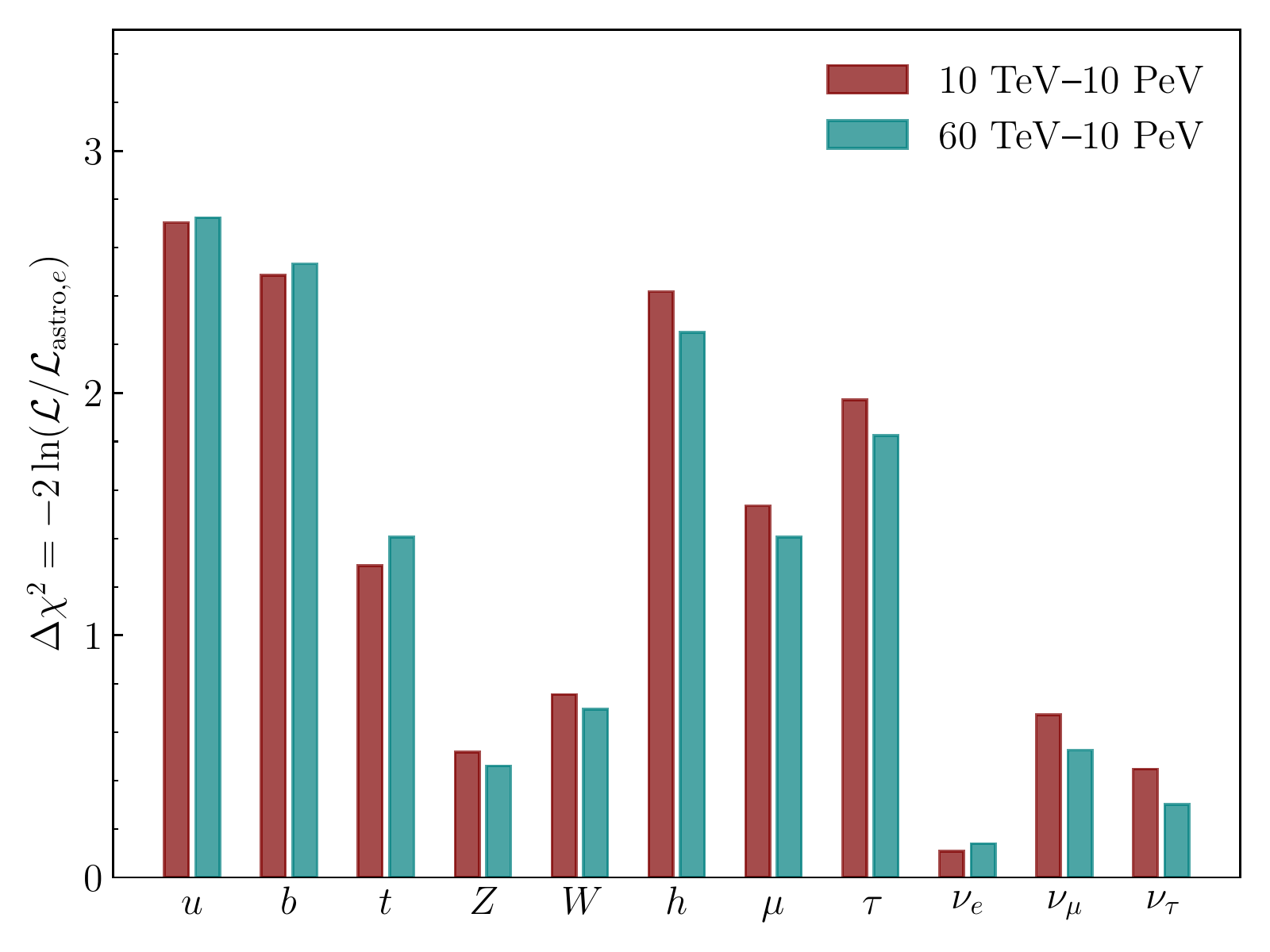}
	\caption{\label{fig:all-ch-like}
		Channel-by-channel comparison of $ \Delta\chisq $ at best fit, computed against the astrophysical flux plus $\dm \to e^+ \, e^-$ channel, which gives the overall best fit. Results for [10~TeV--10~PeV] (brown/left) and [60~TeV--10~PeV] (green/right) are qualitatively similar and indicate the preference for harder DM spectra.
	}
\end{figure}

For channels with even harder spectra, e.g., DM decays into leptons (left panels of Figure~\ref{fig:2Dmunumu}) or neutrinos (right panels of Figure~\ref{fig:2Dmunumu}), the low-mass $1\sigma$~CL preference disappears, while that at high \mdm\ remains qualitatively similar, except for shrinking slightly in extent. The generic shape of the allowed $\gamma$--$\phia$ regions bear out the requirement that a very steep index comes at the cost of lowering the normalization. While, for the soft-spectrum channels, such as $b \, \bar{b}$, the spectral index necessarily has to be on the lower side, the flux normalization rapidly drops as one goes to indexes of $\sim 2$ or lower. For hard channels, which generically provide a better fit to the data, the allowed $1\sigma$~CL region for $\gamma$ extends from around $2.7$ to above $4$, for nearly uniform normalization, thus indicating the necessity of a steeply falling astrophysical flux for these cases. Qualitatively, the more sharply-peaked event-spectrum the flux from DM decays generates, the smaller the preferred region is. Thus, very narrow-width decays directly to neutrinos lead to a more localized $1\sigma$~CL region in the \mdm-\tdm\ plane, whereas for decays to $b \, \bar{b}$, with an event spectrum that is distributed over a wide energy range, the preferred region is much larger. The $2\sigma$~CL regions in all the correlation plots are rather large. Thus, no strong claim can be made on this regard at that level of significance, due to the low statistics of the data under consideration.

When comparing the likelihoods corresponding to the best fits among all the decay channels, the overall best fits come from the channels with hard spectrum, with the high-\mdm\ channels giving the best results (Figure~\ref{fig:all-ch-like}). For instance, of those studied, the overall best fit in terms of likelihoods is obtained for the flux from $\dm \to e^+ \, e^-$, while the neutrino and, to a smaller degree, gauge boson channels provide nearly as good fits. The soft-channel fits are notably poorer, with decays to quarks disfavored. The lack of high-energy tracks also appreciably influences the fit, with the flux from $\dm \to \mu \, \bar{\mu}$ being also slightly disfavored compared to other hard channel cases.

%%%%%%%%%%%%%%%%%%%%%%%%%
%%%%%%%%%%%%%%%%%%%%%%%%%
\subsection{\label{subsec:limits}Limits on the DM lifetime}
%%%%%%%%%%%%%%%%%%%%%%%%%
%%%%%%%%%%%%%%%%%%%%%%%%%

\begin{figure}[t]
	\begin{center}
	  \subfigcapskip=-4pt
		\subfigure[$ \dm \to b\bar{b} $]{\includegraphics[width=0.495\linewidth]{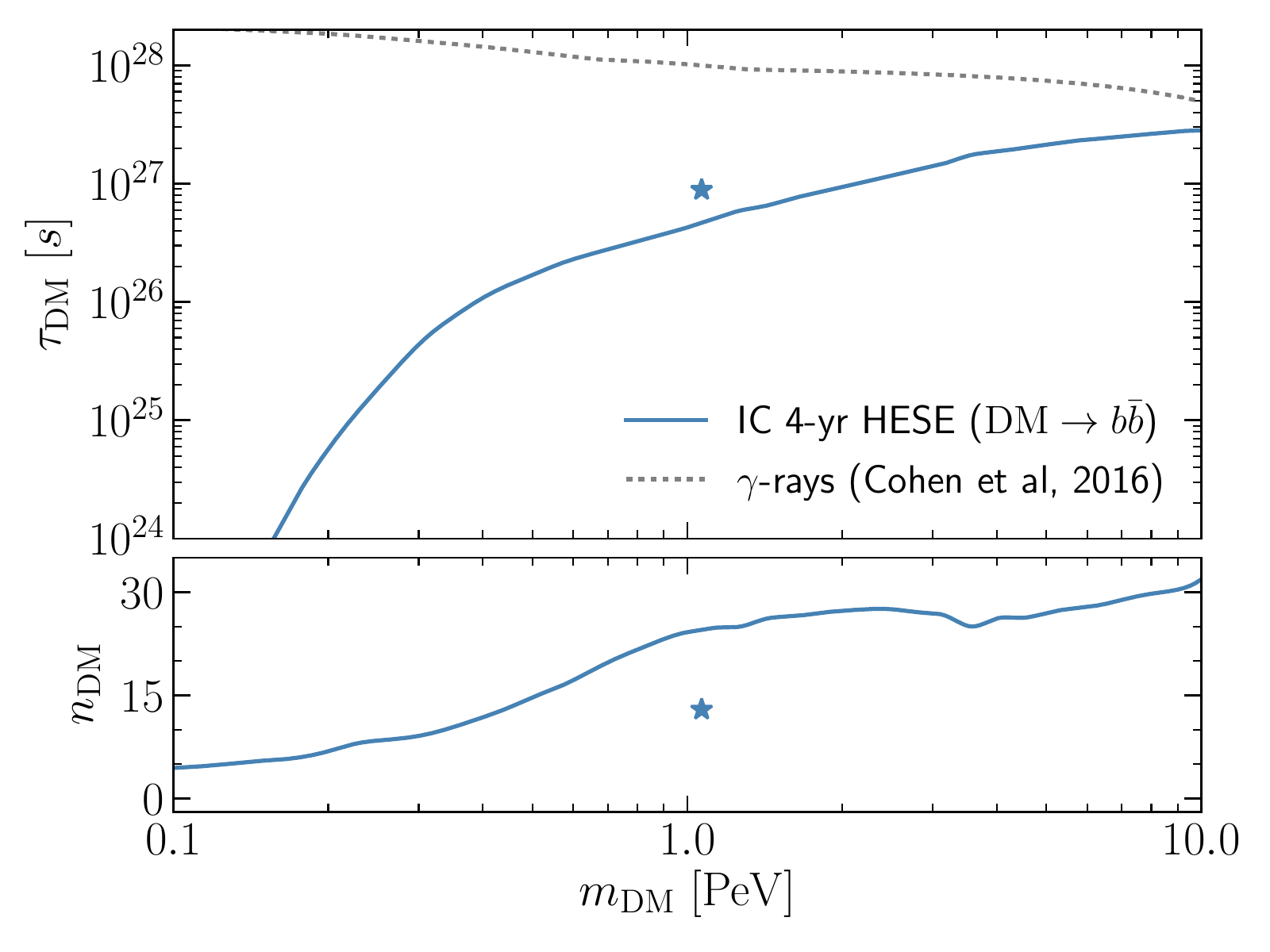}}
		\subfigure[$ \dm \to W^{+} W^{-} $]{\includegraphics[width=0.495\linewidth]{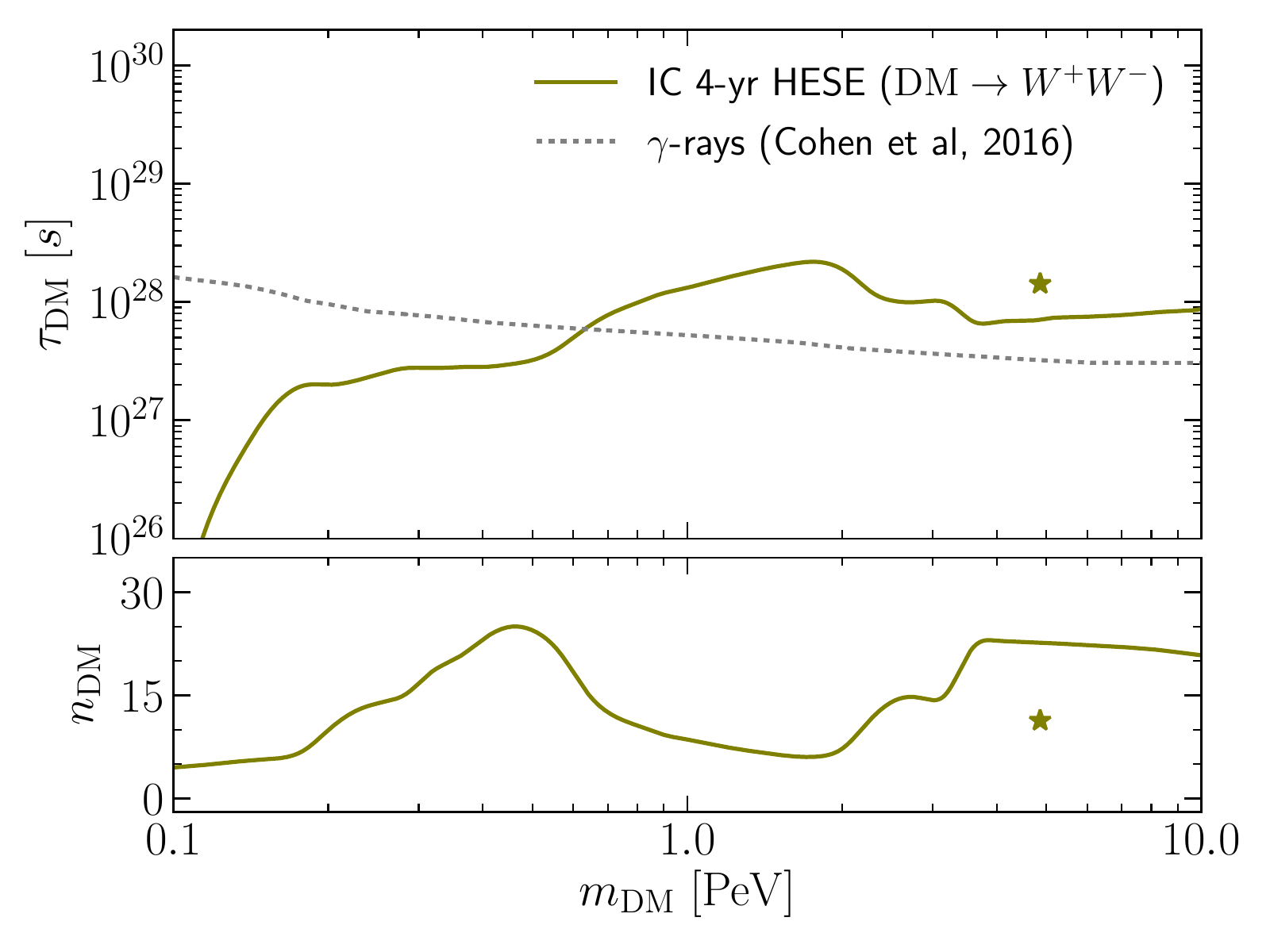}} \\
		\subfigure[$ \dm \to \mu^{+}\mu^{-} $]{\includegraphics[width=0.495\linewidth]{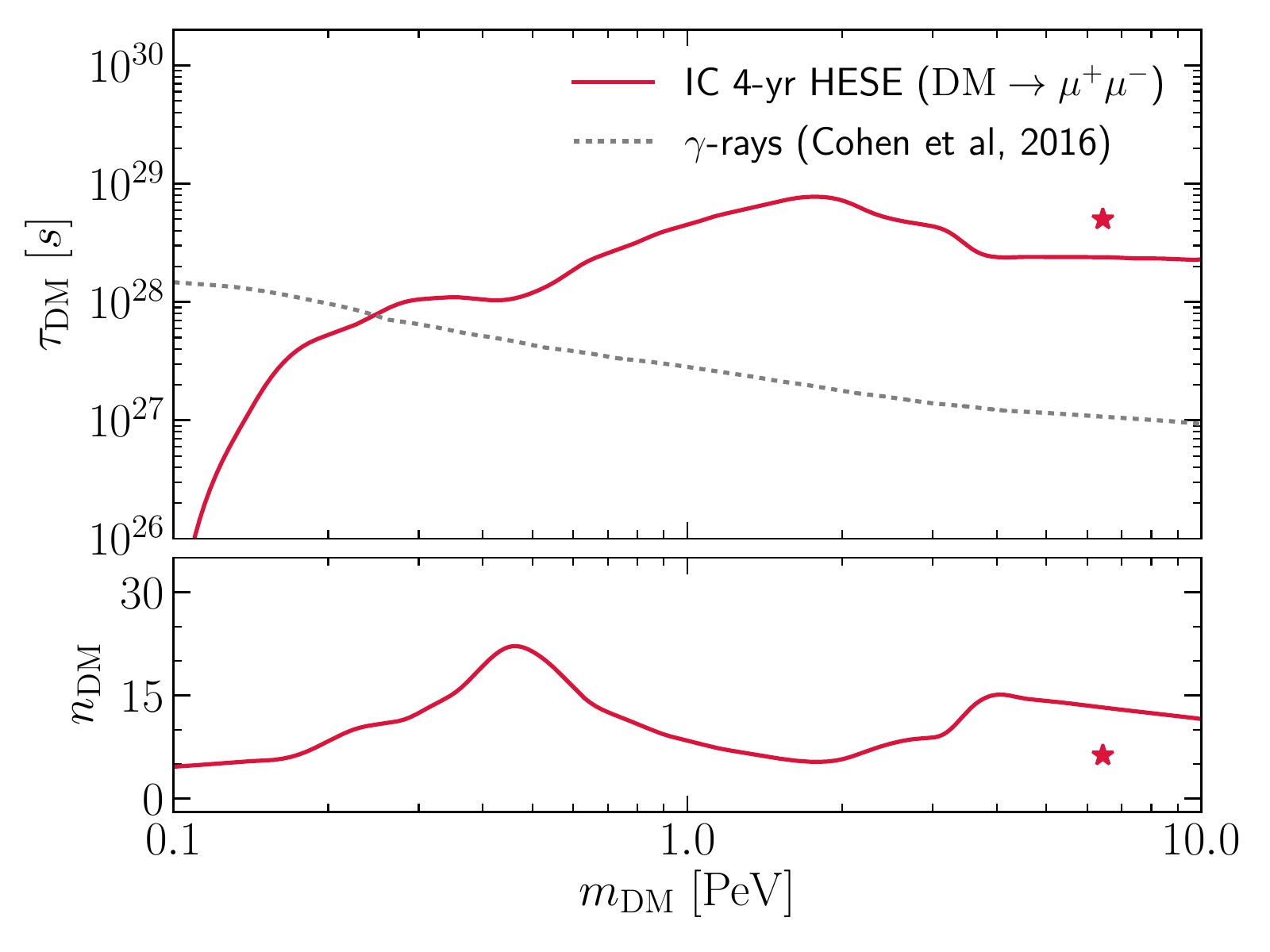}}
		\subfigure[$ \dm \to \nue\bar{\nu}_e $]{\includegraphics[width=0.495\linewidth]{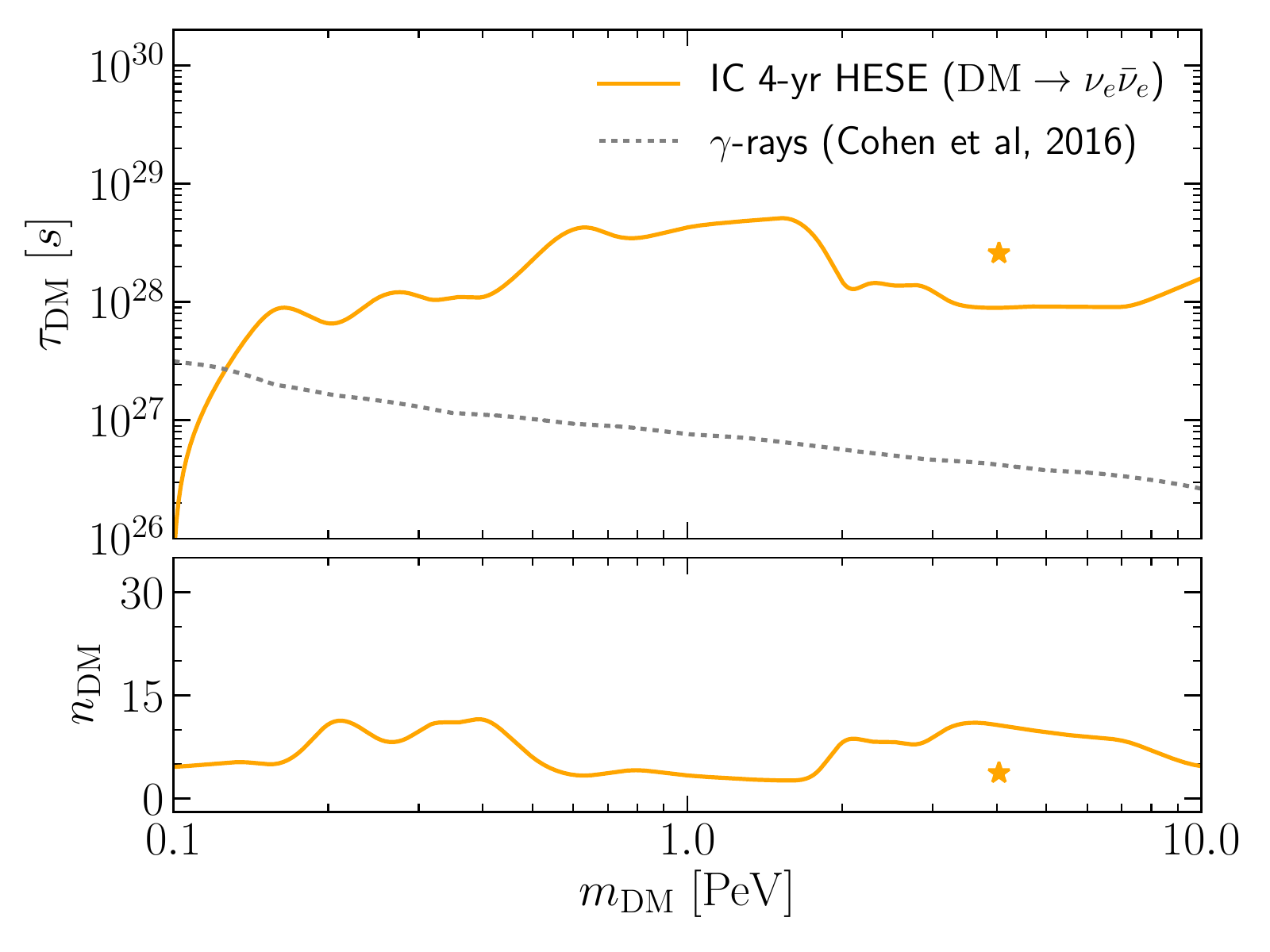}} 			
	\end{center}
	\caption{\label{fig:lt-limits}
		$95\%$~CL limits on \tdm\ and \ndm\ for the four representative channels ($ b \, \bar{b} $, $W^+ \, W^-$, $\mu^+ \, \mu^-$, and $\nu_e \, \bar{\nu}_e $) as a function of \mdm. The best-fit point for each channel is shown by the `$\star$' mark, while the gray dotted curve shows the $ \gamma $-ray constraint on \tdm\ for each channel obtained in Ref.~\cite{Cohen:2016uyg}. These results correspond to the EM-equivalent deposited energy interval [60~TeV--10~PeV].
		}	
\end{figure}

In addition to evaluating how the fit to the HESE spectrum can be improved by adding a DM contribution, it is also interesting to infer the maximum allowed contribution from DM decays and thus, obtain the limit on the DM lifetime.

In Figure~\ref{fig:lt-limits}, we show the 95\%~CL limits for the \tdm\ (and $N_{\rm DM}$) for our benchmark channels as a function of $\mdm$, for the energy interval [60~TeV -- 10~PeV]. In terms of $N_{\rm DM}$, these limits express the 95\%~CL upper limit to number of events from DM decays as a function of \mdm. We compare these limits, obtained by analyzing the IceCube HESE data considering a combined model with contributions from DM decays and from a power-law astrophysical flux, against those obtained from $ \gamma $-ray observations in Ref.~\cite{Cohen:2016uyg}. In the case of the hardest channels (DM decays into leptons, bottom panels), we find our constraints to be stronger than gamma-ray limits, by more than an order of magnitude for $\mdm \gtrsim 100-200$~TeV. For DM decays into gauge bosons (top-right panels), our neutrino limits are more restrictive, by a factor of a few, for $\mdm \gtrsim 600$~TeV. Finally, in the case of soft channels, as DM decays into quarks (top-left panels), our limits are weaker than those obtained from gamma-ray observations throughout the explored range, $\mdm < 10$~PeV. Indeed, the best fits obtained for the contribution from soft channels are in strong tension with gamma-ray limits.
Lifetime limits for all analyzed channels are shown in Appendix~\ref{sec:appB}.

%%%%%%%%%%%%%%%%%%%%%%%%%
%%%%%%%%%%%%%%%%%%%%%%%%%
\section{\label{sec:prior-tracks}Results: Imposing a prior on the astrophysical spectral index}
%%%%%%%%%%%%%%%%%%%%%%%%%
%%%%%%%%%%%%%%%%%%%%%%%%%

IceCube's recent analysis of through-going muon tracks from the northern hemisphere using data accumulated over 6 years~\cite{Aartsen:2016xlq} finds a preference for the best-fit astrophysical flux that is at odds with its best-fit uniform power-law flux obtained with the 4-year HESE data. Using tracks with deposited energies between $\sim 200$~TeV and $\sim 2$~PeV, which originated from interactions of multi-PeV energy neutrinos with nuclei outside the detector's instrumented volume, this analysis finds the unbroken power-law neutrino flux that best fits the data has a harder spectrum, $\PhiAst \propto E^{-2.13 \pm 0.13} $. This is consistent with the expectation from standard Fermi shock acceleration that predicts a spectral index of $\sim 2$, but such a hard spectrum is strongly disfavored by the best fits from the 4-year HESE data at the level of $\sim 4\sigma$~CL.

It is plausible, if future data reinforces this apparent incompatibility, that neutrino fluxes coming from a combination of two or more different origins might prove to be a better fit to the different sets of data than would a simple power-law~\cite{Chen:2014gxa, Aartsen:2015knd, Palladino:2016zoe, Vincent:2016nut}. In this section, we investigate, whether in our scenario, the tension in the predictions of the two different data samples, albeit seen in the same detector and at similar energies, can be alleviated with a heavy sprinkling of neutrinos coming from \dm\ decay.

\begin{figure}[t]
	\begin{center}
		\subfigcapskip=-4pt
		\subfigure[$\dm \to b \, \bar{b}$]{\includegraphics[width=0.49\linewidth]{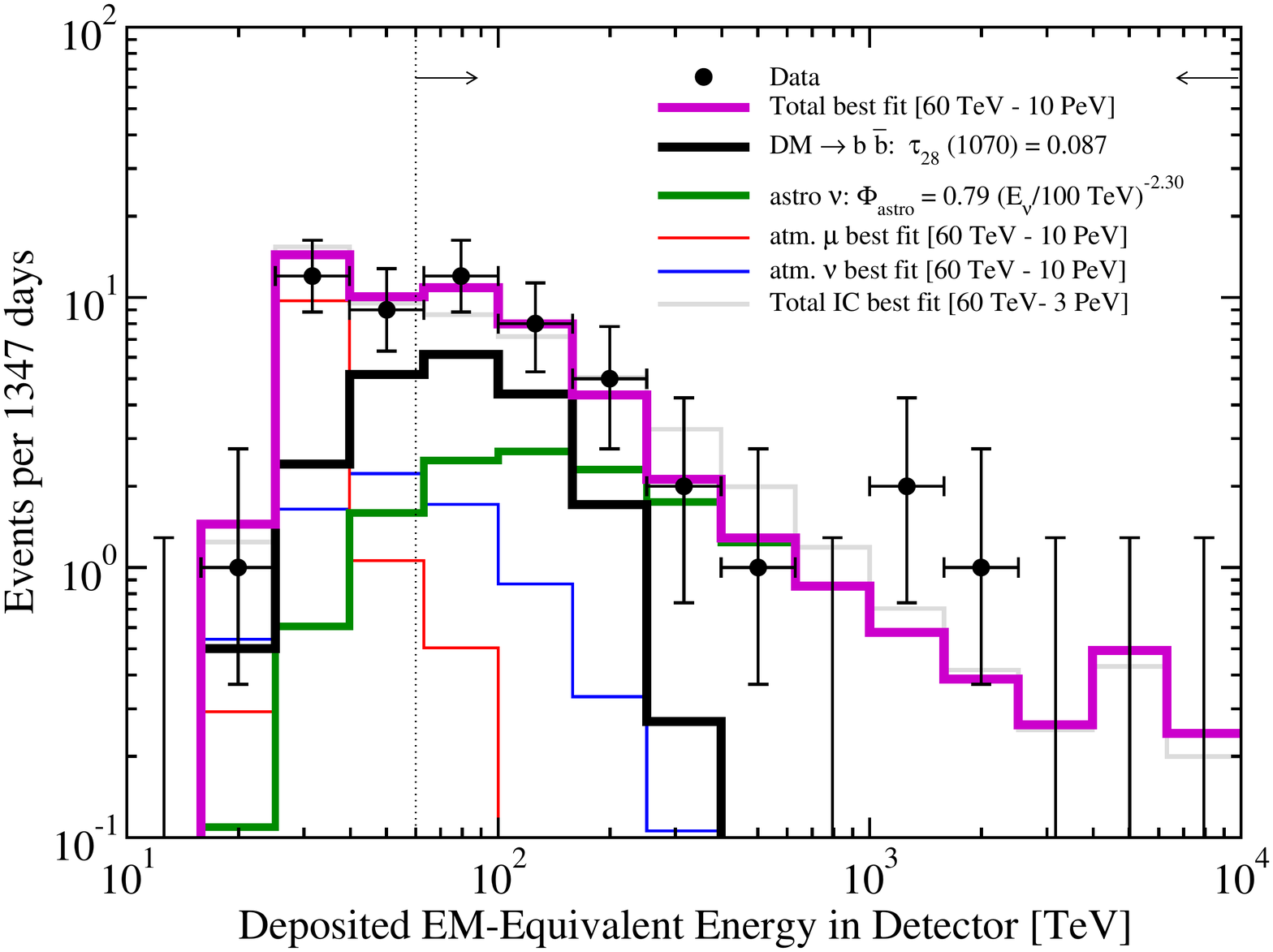}}		
		\subfigure[$\dm \to W^{+} \, W^{-}$]{\includegraphics[width=0.49\linewidth]{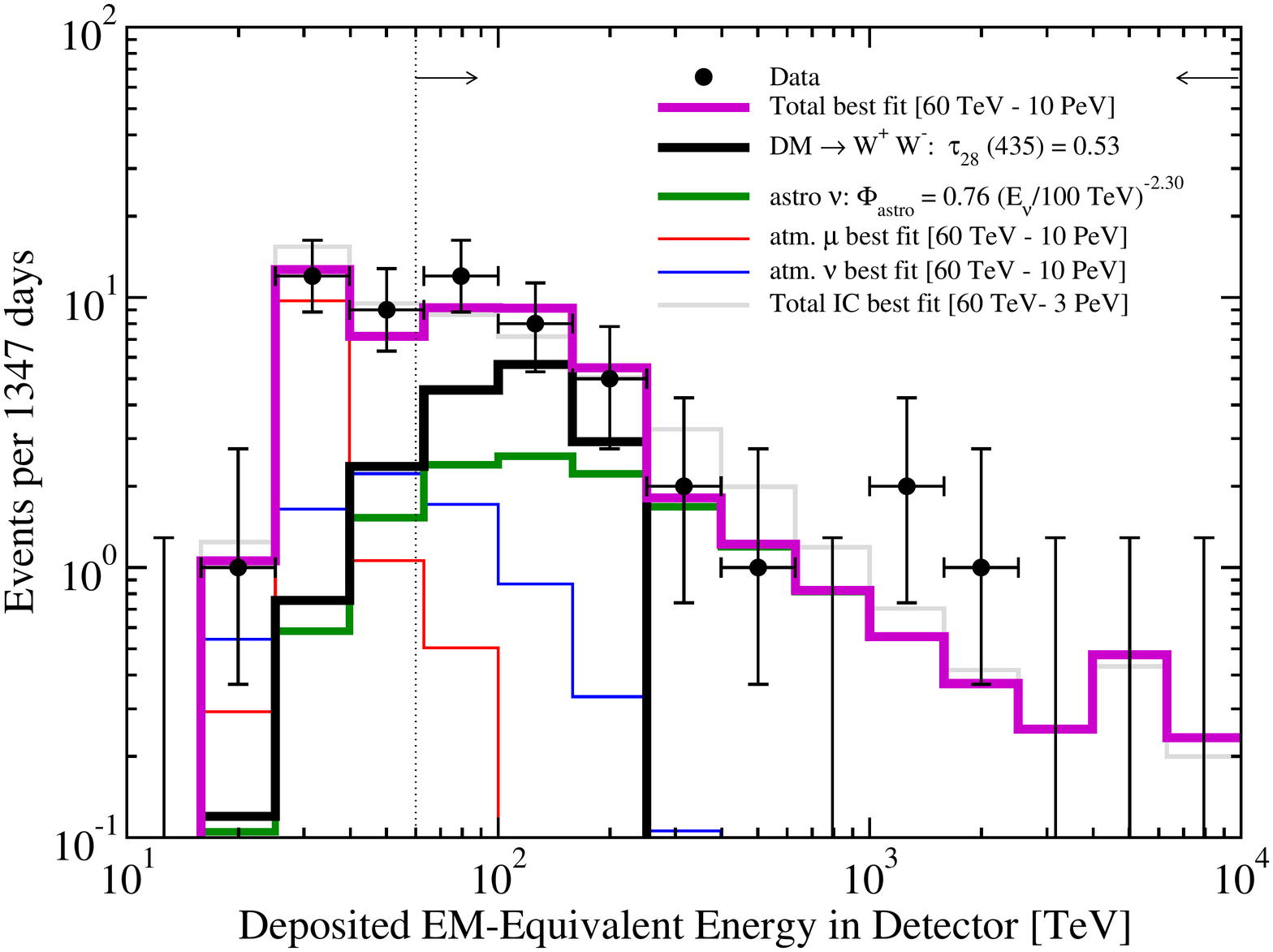}} \\		
		\subfigure[$\dm \to \mu^{+} \, \mu^{-}$]{\includegraphics[width=0.49\linewidth]{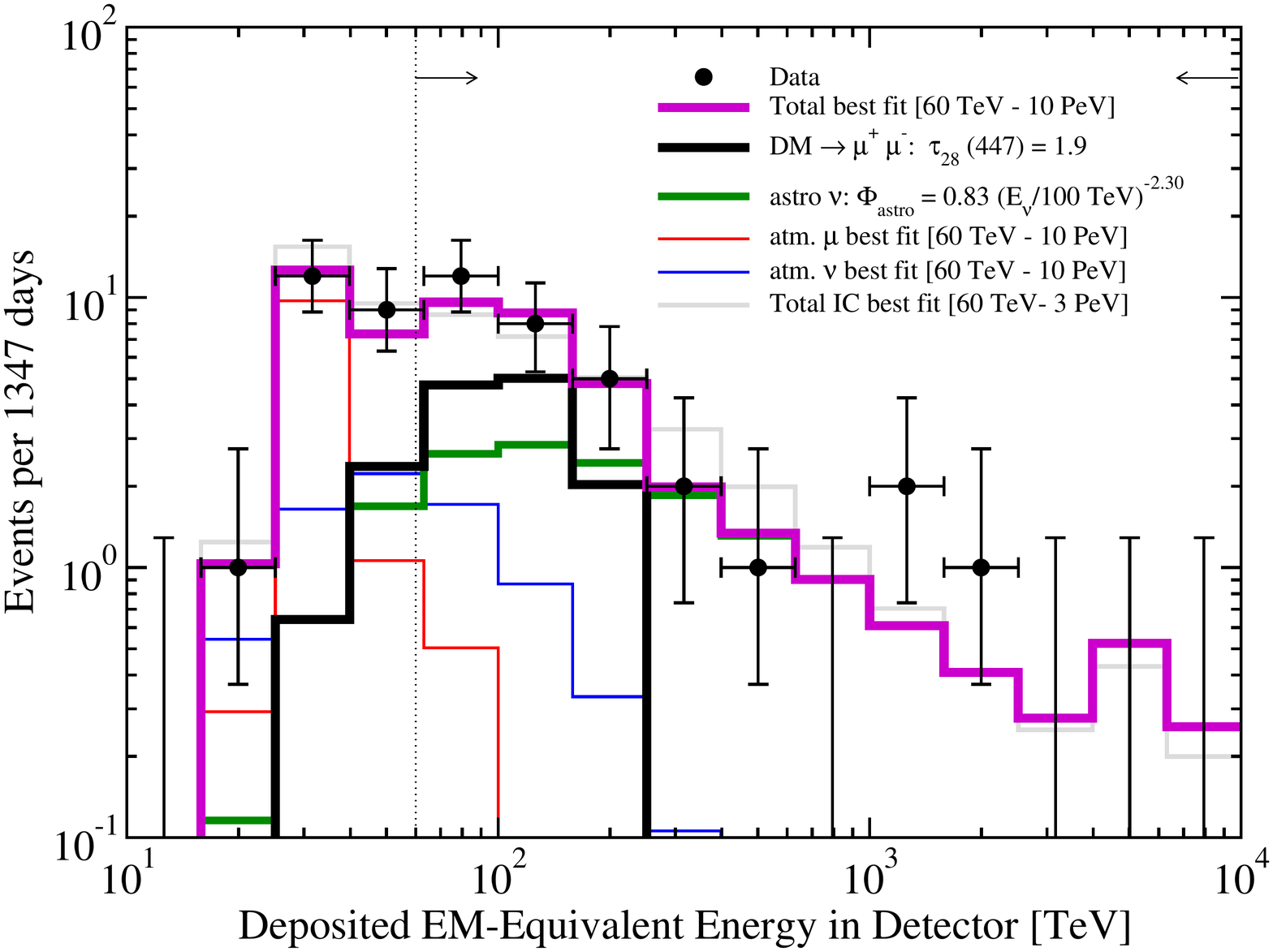}}
		\subfigure[$\dm \to \nu_{\alpha}\bar{\nu}_{\alpha}$]{\includegraphics[width=0.49\linewidth]{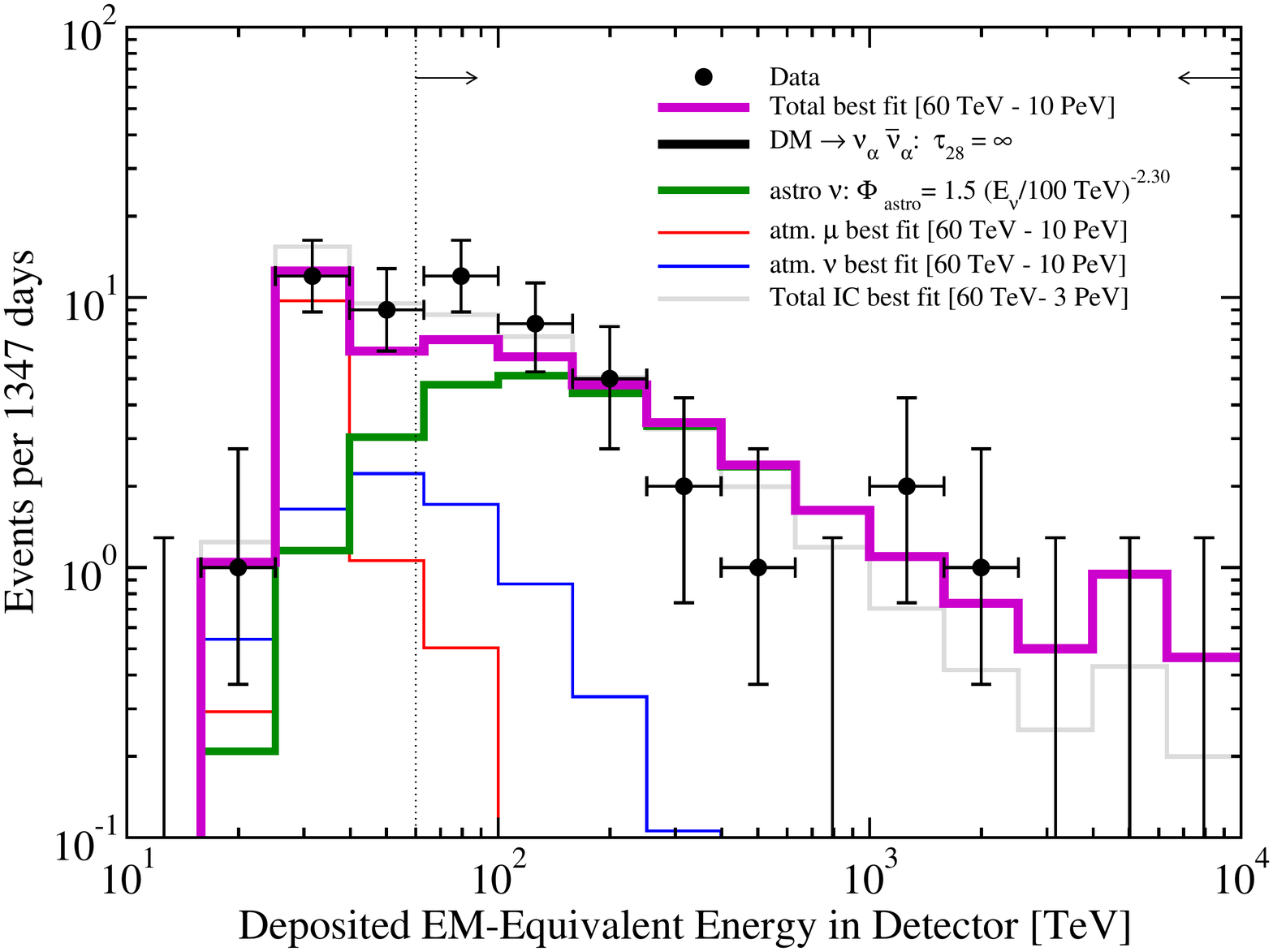}}
	\end{center}
	\caption{\label{fig:events-prior}
		Same event rates as in Figure~\ref{fig:events}, but corresponding the best fits obtained when the astrophysical power-law index is restricted to lie within the range $\gamma = [2.0$--$2.3]$. Note that the bottom-right panel applies to $\alpha = \{e, \mu, \tau\}$.
	}	
\end{figure}

To that end, we restrict the possible values of the astrophysical spectral index to the interval $\gamma = [2.0\text{--}2.3]$. This draws the resulting best-fit spectral index in our analysis closer to the track-events best fit, thus causing the astrophysical flux to flatten out considerably compared to results from the previous section (except from the few channels which already preferred a hard astrophysical flux). As in Tables~\ref{tab:fits-noprior10} and~\ref{tab:fits-noprior60}, the best-fit values for $\boldsymbol{\theta} = \{\ndm (\tau_{\rm DM}), \mdm, \nast (\phia), \gamma\}$, obtained in this way, are now indicated in Tables~\ref{tab:fits-prior10} and~\ref{tab:fits-prior60}. As one could expect, in all the cases, the spectral index $\gamma$ tends to come out to be 2.3. In Figure~\ref{fig:events-prior} we show the results corresponding to the interval $[60~{\rm TeV}\text{--}10~{\rm PeV}]$ for the best-fit event rates for the same channels as in Figure~\ref{fig:events}.

\begin{table}[t]
	\caption{\label{tab:fits-prior10}
		Same as Table~\ref{tab:fits-noprior10}, but restricting the astrophysical power-law index to $\gamma = \left[2.0\text{--}2.3\right]$. The EM-equivalent deposited energy interval is [60~TeV--10~PeV].}
	\begin{center}
		\begin{tabular}{c|cc|cc}
			\hline
			Decay channel & $ \ndm (\tau_{\rm DM} [10^{28}~{\rm s}])$  &  $ \mdm $ [TeV] &
			$\nast (\phia) $  & $\gamma$ \\
			\hline
			$u \, \bar{u}$                &       18.9 (0.035) &        619 &       17.8 (0.90) &      2.30 \\
			$b \, \bar{b}$                &       21.2 (0.082) &       1040 &       14.7 (0.73) &      2.29 \\
			$t \, \bar{t}$                &       19.8 (0.13) &        556 &       15.1 (0.77) &      2.30 \\
			$W^{+} \, W^{-}$              &       17.7 (0.47) &        420 &       15.8 (0.80) &      2.30 \\
			$Z \, Z$                     &       15.5 (0.54) &        346 &       18.2 (0.92) &      2.30 \\
			$h \, h$                     &       20.4 (0.15) &        582 &       14.9 (0.75) &      2.30 \\
			$e^{+} \, e^{-}$              &        9.2 (0.38) &        214 &       23.9 (1.2) &      2.30 \\
			$\mu^{+} \, \mu^{-}$          &       11.4 (1.1) &        230 &       23.5 (1.2) &      2.30 \\
			$\tau^{+} \, \tau^{-}$        &       17.3 (1.1) &        434 &       17.1 (0.86) &      2.30 \\
			$\nu_e \, \bar{\nu}_e$        &        7.7 (1.1) &        208 &       25.0 (1.3) &      2.30 \\
			$\nu_\mu \, \bar{\nu}_\mu$    &       10.3 (0.75) &        209 &       23.7 (1.2) &      2.30 \\
			$\nu_\tau \, \bar{\nu}_\tau$  &       10.8 (0.68) &        210 &       23.3 (1.2) &      2.30 \\
			\hline
		\end{tabular}
	\end{center}
\end{table}

\begin{table}[h!]
	\caption{\label{tab:fits-prior60}
		Same as Table~\ref{tab:fits-prior10}, but for the EM-equivalent deposited energy interval [60~TeV--10~PeV].}
	\begin{center}
		\begin{tabular}{c|cc|cc}
			\hline
			Decay channel & $ \ndm (\tau_{\rm DM} [10^{28}~{\rm s}])$  &  $ \mdm $ [TeV] &
			$\nast (\phia) $  & $\gamma$ \\
			\hline
			$u \, \bar{u}$                &       11.8 (0.018) &        523 &       14.9 (0.87) &      2.30 \\
			$b \, \bar{b}$                &       13.2 (0.087) &       1070 &       13.5 (0.79) &      2.30 \\
			$t \, \bar{t}$                &       13.5 (0.14) &        590 &       13.2 (0.77) &      2.30 \\
			$W^{+} \, W^{-}$              &       13.5 (0.53) &        435 &       13.0 (0.76) &      2.30 \\
			$Z \, Z$                     &       13.3 (0.62) &        433 &       13.2 (0.77) &      2.30 \\
			$h \, h$                     &       13.6 (0.17) &        606 &       13.2 (0.76) &      2.29 \\
			$e^{+} \, e^{-}$              &    0.0 ($\infty$) &        --- &       25.8 (1.5) &      2.30 \\
			$\mu^{+} \, \mu^{-}$          &       12.2 (1.9) &        447 &       14.3 (0.84) &      2.30 \\
			$\tau^{+} \, \tau^{-}$        &       12.7 (1.2) &        470 &       14.1 (0.82) &      2.30 \\
			$\nu_e \, \bar{\nu}_e$        &    0.0 ($\infty$) &        --- &       25.8 (1.5) &      2.30 \\
			$\nu_\mu \, \bar{\nu}_\mu$    &    0.0 ($\infty$) &        --- &       25.8 (1.5) &      2.30 \\
			$\nu_\tau \, \bar{\nu}_\tau$  &    0.0 ($\infty$) &        --- &       25.8 (1.5) &      2.30 \\
			\hline
		\end{tabular}
	\end{center}
\end{table}

\clearpage

For all channels, except from DM decays into $u \, \bar{u}$ and $b \, \bar{b}$ (and also $h \, h$ for the low-energy threshold of 60~TeV), we note that forcing a harder astrophysical flux substantially alters the other parameters and the look of the fit (Figure~\ref{fig:events-prior}). We find that lower DM masses are preferred, between $\sim 200$~TeV and $\sim 600$~TeV, when the astrophysical flux is forced to be harder, compared to multi-PeV masses in the case of no restriction on the power-law index (see Tables~\ref{tab:fits-noprior10} and~\ref{tab:fits-noprior60}). In this cases, the contribution (if any) from DM decays is favored at lower energies, in the energy range between $\sim 30$~TeV and $\sim 200$~TeV. 

On the other hand, except from the hardest channels, increasing the energy threshold from 10~TeV to 60~TeV does not change the best fits of the parameters significantly. However, in the case of the analysis with the low-energy threshold at 60~TeV, in order to account for the more prominent excess of events over and above the now flatter astrophysical spectrum (mainly just below $\sim 100$~TeV), \dm\ decays into the hardest channels (neutrinos or $e^+ \, e^-$), which were previously contributing to the highest energy events that are now described by the astrophysical flux, cannot have any significant contribution. Thus, the resulting best fit for the number of events from DM decays for these channels is zero. At first, this is a bit surprising, as one would expect that the sharp-peaked spectrum of these channels could contribute dominantly to the excess in the $10^{1.8}$--$10^{2}$~TeV bin. However, a closer look at the observed events in that particular bin, reveals a much higher track-to-cascade ratio $(5:7)$ than expected from the sum of the background and the astrophysical flux with the canonical $(1:1:1)$ flavor composition. To account for the higher number of tracks in that bin, too many extra cascades would be predicted in higher energy bins, thus disfavoring this possibility. On the other hand, when the analysis is performed with the low-energy threshold at 10~TeV, there is another bin ($10^{1.6}$--$10^{1.8}$~TeV) with an excess of events over the expected background and the astrophysical component. In that bin, the observed track-to-cascade ratio is $2:7$ and some contribution from DM decays, even for the hardest channels, is preferred. Note that the best-fit DM mass for these channels and the low-energy threshold at 10~TeV is $\sim 200$~TeV and hence, the DM contribution to the track sample is very suppressed.

For the cases of DM decays into $\mu^+ \, \mu^-$ or $\tau^+ \, \tau^-$, the number of DM events is larger than in the case with no restriction on the power-law index of the astrophysical flux because these channels also contribute significantly at lower energies. In the case of DM decays into gauge and Higgs bosons and quarks, a slightly larger number of events is also preferred.  

From the top-left panel of Figure~\ref{fig:events-prior}, one can see that in case of DM decay into $b \, \bar{b}$, for EM-equivalent deposited energies below 250~TeV, DM decays provide the dominant contribution, while the astrophysical component takes over for higher energies. This is exactly the same result as that found for the analysis without restricting the power-law index. This is expected, as the best fit value for $\gamma$ for that decay channel was very close to 2.3. However, in case of DM decays into gauge bosons (top-right panel of Figure~\ref{fig:events-prior}), forcing the astrophysical index to have a maximum value of 2.3, pushes the best-fit for the DM mass to much lower values than in the unrestricted analysis. Now the low-energy excess has to be explained by events from DM decays and the DM contribution dominates in the range of energies between 40~TeV and 250~TeV, in contrast to the
unrestricted analysis in which the DM contribution was at higher energies (between 250~TeV and 2.5~PeV), but the  astrophysical flux was much softer.  

In case of DM decays into $\mu^+ \, \mu^-$ (bottom-left panel of Figure~\ref{fig:events-prior}), the DM contribution dominates over the astrophysical one only in the energy range between 60~TeV and 150~TeV, and hence, events above $\sim 200$~TeV are attributed to the astrophysical flux only. For the case of DM decays into neutrinos (bottom-right panel of Figure~\ref{fig:events-prior}), as explained above, the best fit for the interval $[60~{\rm TeV}$--$10~{\rm PeV}]$ does not include any contribution from DM decays because for such a relatively hard astrophysical flux, the highest-energy events can be partly explained by the power-law component and any contribution from the very peaked DM spectrum is disfavored at other energies. Note that this no-DM fit is not the best astrophysical-only fit, which corresponds to $\gamma \simeq 2.6$.

%%%%%%%%%%%%%%%%%%%%%%%%%%%%
%%%%%%%%%%%%%%%%%%%%%%%%%%%%
\section{\label{sec:multi-channel}Results: DM decays via multiple channels}
%%%%%%%%%%%%%%%%%%%%%%%%%%%%
%%%%%%%%%%%%%%%%%%%%%%%%%%%%

It is clear from the discussions in the previous sections that a mixed flux comprising a steeply dropping astrophysical power-law and a component from the DM decays improves the overall fit to the observed IceCube HESE data. In addition, a better fit can be obtained when the DM component is hard, and so it can explain the multi-PeV events, while the soft astrophysical flux fills in at sub-PeV energies. When DM decays into soft channels, e.g., into $b \, \bar{b}$, the event-spectrum from the DM component populates sub-PeV energies ($\sim 100$~TeV), while the astrophysical component becomes harder to accommodate the PeV events. The fit for these cases is obviously poorer, but it motivates the possibility that a DM particle that can decay via two or more channels, whereby at least one of the channels produces a soft spectrum and another one leads to a hard spectrum, might be able to explain the entire HESE spectrum on its own, without any need for a power-law astrophysical spectrum. This possibility has been discussed in Refs.~\cite{Esmaili:2013gha, Esmaili:2014rma}, assuming specific decay channels and/or specific models predicting the decay channels with fixed branching ratios (e.g., see Refs.~\cite{Anisimov:2008gg, Higaki:2014dwa, DiBari:2016guw}), concluding that the decaying DM scenario might describe data better  than the astrophysical unbroken power-law flux. In this section, we investigate this possibility with the 4-year HESE dataset and probe the parameter space for the best fit points in this type of scenarios. To keep the analysis simple, and the number of free parameters minimal, we specifically study the case of a DM particle decaying via two channels.

\begin{table}[t]
	\caption{\label{tab:fits-mulch-60}
		Best-fit values for DM-only two-channel decays ($\dm \to p_1 \, \bar{p}_1,\, p_2 \, \bar{p}_2$) defined by $\boldsymbol{\theta}_{2c} = \{\ndm, \mdm, \text{BR}\}$, where $\text{BR} = \Gamma_{\dm \to p_{1} \, \bar{p}_{1}} / \left(\Gamma_{\dm \to p_{1} \, \bar{p}_{1}} + \Gamma_{\dm \to p_{2} \, \bar{p}_{2}}\right)$ is the branching ratio for decays to $p_1 \, \bar{p}_1$. The EM-equivalent deposited energy interval is [60~TeV--10~PeV].}  	  	
	\begin{center}
		\begin{tabular}{c|ccc}
			\hline
			Decay channels                         & $ \ndm\ (\tau_\dm \left[10^{28}\mathrm{\ s}\right]) $   &  $ \mdm $ [TeV] &	 BR \\
			\hline
			$u \, \bar{u}$, $e^{+} \, e^{-}$               &       26.6 (0.22) &       3991 &       0.84 \\
			$u \, \bar{u}$, $\nu_e \, \bar{\nu}_e$         &       26.7 (0.19) &       3902 &       0.92 \\
			$b \, \bar{b}$, $e^{+} \, e^{-}$               &       26.5 (0.22) &       4042 &       0.84 \\
			$b \, \bar{b}$, $\mu^{+} \, \mu^{-}$           &       26.4 (0.25) &       5444 &       0.94 \\
			$b \, \bar{b}$, $\nu_e \, \bar{\nu}_e$         &       26.6 (0.19) &       3933 &       0.92 \\
			$b \, \bar{b}$, $\nu_\mu \, \bar{\nu}_\mu$     &       26.6 (0.20) &       4023 &       0.93 \\
			$b \, \bar{b}$, $\tau^{+} \, \tau^{-}$         &       26.5 (0.25) &       5539 &       0.94 \\
			$t \, \bar{t}$, $\nu_\mu \, \bar{\nu}_\mu$     &       26.1 (0.32) &       8866 &       1.00 \\
			$W^{+} \, W^{-}$, $\mu^{+} \, \mu^{-}$         &       25.3 (0.22) &       4633 &       1.00 \\
			$W^{+} \, W^{-}$, $\nu_\mu \, \bar{\nu}_\mu$   &       25.3 (0.22) &       4633 &       1.00 \\
			$h \, h$, $\mu^{+} \, \mu^{-}$                &       26.3 (0.28) &       7031 &       1.00 \\
			$h \, h$, $\nu_e \, \bar{\nu}_e$              &       26.3 (0.20) &       4103 &       0.92 \\
			\hline
		\end{tabular}
	\end{center}	 	
\end{table}

\begin{figure}[t]
	\centering
	\includegraphics[width=0.75\textwidth]{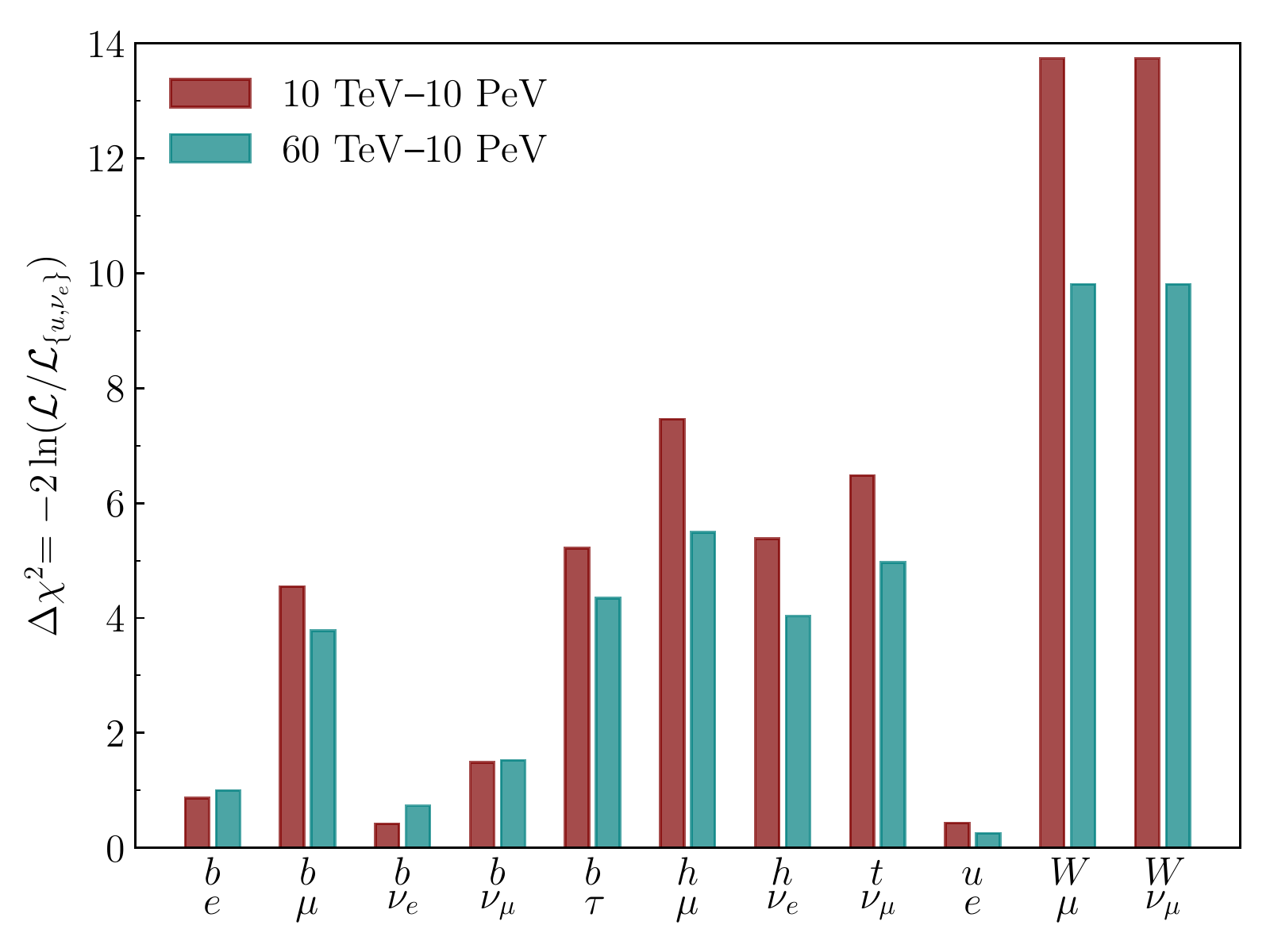}
	\caption{\label{fig:all-mulch-chi}
		Channel-by-channel comparison of $ \Delta\chisq $ at best fit, computed against the $\dm \to \{u \, \bar{u}, \, \nu_e \, \bar{\nu}_e\}$ combination, which gives the overall best fit. Results for [10~TeV--10~PeV] (brown/left) and [10~TeV--10~PeV]  (green/right) are qualitatively similar and indicate the preference for combinations of a soft and a very hard spectra.}
\end{figure}

In the absence of the astrophysical power-law spectrum, the scenario of DM decaying via two distinct channels involves three parameters:
\begin{inparaenum}[\itshape a\upshape)]
  \item the DM mass, \mdm,
  \item the normalization of the DM flux, expressed in terms of expected number of HESE events it leads to, \ndm\ (or equivalently, in terms of the lifetime $\tau_{\rm DM}$), and 
  \item the branching ratio ${\rm BR} = \Gamma_{\dm \to p_{1} \, \bar{p}_{1}} / \left(\Gamma_{\dm \to p_{1} \, \bar{p}_{1}} + \Gamma_{\dm \to p_{2} \, \bar{p}_{2}}\right)$.
\end{inparaenum}
In this scenario, it becomes incumbent upon the \ndm\ to equal the total HESE signal; nonetheless, as done above, we let the fits arrive at the preferred value of \ndm, rather than imposing this as a constraint. It also becomes necessary for the \mdm\ to be at relatively high ($ \geqslant 2 $ PeV) values, in order to be capable of explaining both the PeV and the sub-PeV events.

Guided by the event spectrum from the different channels in Section~\ref{sec:fit-results}, we choose a few representative combinations (Table~\ref{tab:fits-mulch-60}), including DM decays into quark and lepton pairs, gauge boson and lepton pairs, and Higgs and lepton pairs (see Figure~\ref{fig:all-mulch-chi}). We find that the best overall fit comes from DM decays into $u \, \bar{u}$ and $\nu_e \, \bar{\nu}_e$, with $\mdm \sim 4$~PeV, and a branching ratio predominantly in favor of decays into quarks, $\textrm{BR} \sim 0.9$.

Once we can compare several cases of a DM component with two-channel decays (Figure~\ref{fig:all-mulch-chi}) and the results for the cases of only one DM decay channel and a power-law component (Figure~\ref{fig:all-ch-like}), we can ask the next obvious question: Is the overall best-fit in the DM-only scenario with two decay channels better than that obtained when a single-channel DM-decay flux combines with a power-law astrophysical flux component? Despite the complication in making a straight comparison between the two scenarios brought about by the difference in the number of free parameters and of types of models, we find that, at the level of likelihoods corresponding to the best-fit points, $\dm \to \{u \, \bar{u}, \, \nu_e \, \bar{\nu}_e\}$ is a slightly better fit than that for the case of a mixed astrophysical plus $\{\dm \to e^{+} \, e^{-}\}$ scenario, with $\Delta \chi^2 = - 2 \ln (\mathcal{L}_{{\rm astro}, e}/\mathcal{L}_{u,\nu_e}) = 0.61$ and 0.16 for [10~TeV--10~PeV] and [60~TeV--10~PeV], respectively. Given that it is so with one parameter less in the DM-only case with decays into two channels, this suggests that the corresponding fit is indeed better than that resulting one from the combination of a power-law and a DM contribution from a single decay channel. Note, however, that for all these soft-hard combinations, the best fit for the DM lifetime is in tension with gamma-ray limits (see Appendix~\ref{sec:appB}). A more detailed analysis and quantification of the statistical comparison of the two scenarios is beyond the scope of this work.

We noted in Section~\ref{sec:fit-results} that, apart from decays into leptons, DM decays into gauge and Higgs bosons and top quarks also prefer multi-PeV events coming from the DM component, with the astrophysical spectrum fitting the remaining sub-PeV events. Thus, in the absence of the soft astrophysical flux, combinations of relatively hard channels obviously do not work well, ending up being strongly disfavored in comparison to the quark-lepton combinations, and with a branching ratio wholly in favor of one of the two channels. We show this for $ \dm \to \{W^{+} \, W^{-}, \, \mu^{+} \, \mu^{-} \}$ and other similar cases.  

\begin{figure}[t]
	\begin{center}
	  \subfigcapskip=-4pt
		\subfigure[DM $\to \{92\% \, b  \, \bar{b}, \, 8\% \, \nu_e \, \bar{\nu}_e\}$]{\includegraphics[width=0.49\linewidth]{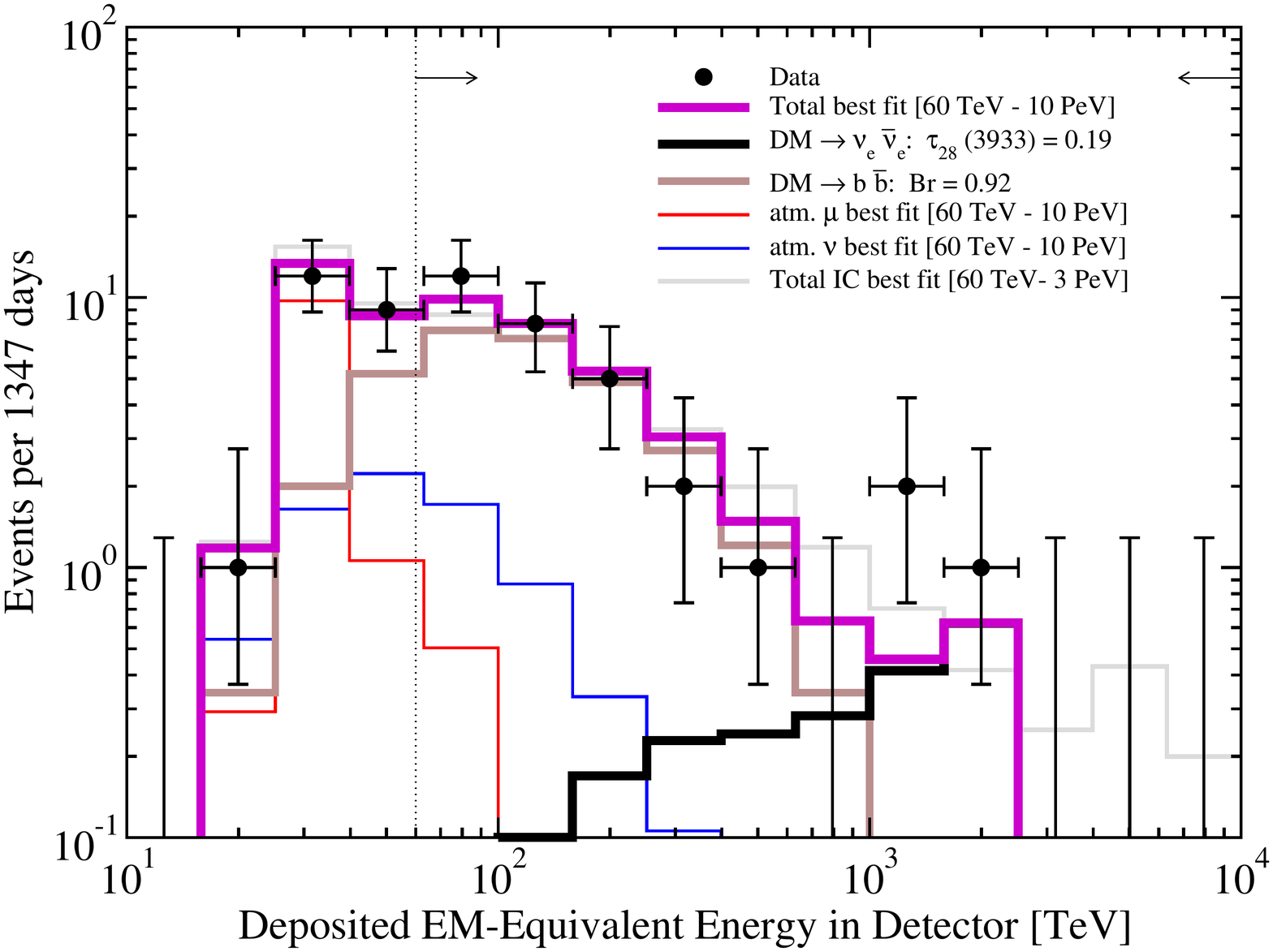}}
		\subfigure[DM $\to \{92\% \, u \, \bar{u}, \, 8\% \, \nu_e \, \bar{\nu}_e\}$]{\includegraphics[width=0.49\linewidth]{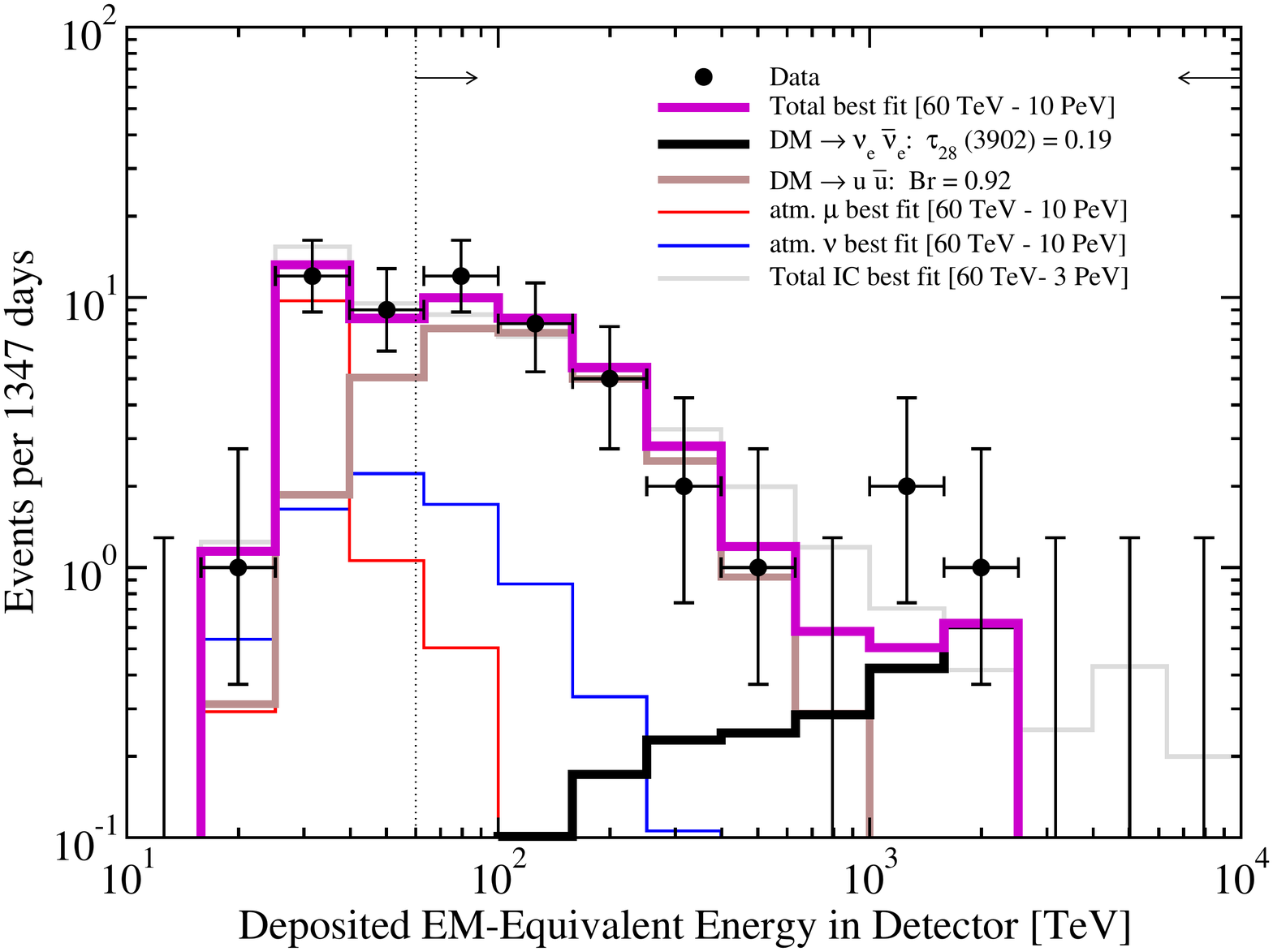}}
	\end{center}
	\caption{\label{fig:events-mulch}
		Event spectra in the IceCube detector after 1347 days. We show the results corresponding to the best fits in the EM-equivalent deposited energy interval [60~TeV--10~PeV] for DM decays into two two-channel combinations: DM $\to \{ b \, \bar{b}, \, \nu_e \, \bar{\nu}_e \}$ (left panel) and DM $\to \lbrace u \, \bar{u}, \, \nu_e \, \bar{\nu}_e \rbrace$ (right panel), with their corresponding branching fractions into the quark channel also indicated.  In both panels: atmospheric muon events (red histogram), conventional atmospheric neutrino events (blue histogram), astrophysical neutrino events (green histogram), neutrino events from DM decays into the quark channel (brown histogram) and into the lepton channel (black histogram), and total event spectrum (purple histogram). We indicate the best fit values of the DM lifetime and mass [$\tau_{28} (\mdm)$] in units of $10^{28}$~s and TeV. We also show the spectrum obtained using the 4-year IceCube best fit in the EM-equivalent  deposited energy interval [60~TeV--3~PeV] (gray histogram), $E_\nu^2 \, d\Phi/dE_\nu = 2.2 \times 10^{-8} \, (E_\nu/100 \, {\rm TeV})^{-0.58}  \, {\rm GeV} \,  {\rm cm}^{-2} \, {\rm s}^{-1} \, {\rm sr}^{-1}$ (per flavor), and the binned high-energy neutrino event data (black dots)~\cite{Aartsen:2015zva} with Feldman-Cousins errors~\cite{Feldman:1997qc}.}
\end{figure}

Finally, in Figure~\ref{fig:events-mulch} we show the event rates obtained in this scenario for two specific combinations of channels:  $\dm \to \{ b \, \bar{b}, \, \nue \, \bar{\nu}_e \}$ and $\dm \to \{ u \, \bar{u}, \, \nu_e \, \bar{\nu}_e \}$, with their corresponding best-fit parameters given in Table~\ref{tab:fits-mulch-60}.

%%%%%%%%%%%%%%%%%%%%%%%%%
%%%%%%%%%%%%%%%%%%%%%%%%%
\section{\label{sec:conc}Discussion and conclusions}
%%%%%%%%%%%%%%%%%%%%%%%%%
%%%%%%%%%%%%%%%%%%%%%%%%%

In view of the increasing incompatibility between the IceCube HESE and through-going muon track data sets if interpreted in terms of a single power-law astrophysical flux, we have considered the possibility of DM decays also contributing to HESE data. We have considered HESE data in the EM-equivalent deposited energy intervals [10~TeV--10~PeV] and [60~TeV--10~PeV], the latter corresponding to the sample analyzed by the IceCube collaboration, as it is less populated by background events. 

In our analyses we have considered simultaneously the topology (shower or track) and energy distributions of the events, as well as the hemisphere where they were originated. In a series of analyses, we have varied four parameters: the astrophysical flux normalization and power-law index, the DM mass, lifetime, and, for multi-channel decays, the branching ratio. For the background events, i.e., atmospheric muons and atmospheric neutrinos, we take the reported values from IceCube analysis~\cite{Aartsen:2014gkd}. We have considered a variety of DM decay channels (see Figure~\ref{fig:flux}), which generally can be classified into soft (decays into quarks or Higgs bosons), intermediate (decays into gauge bosons) and hard channels (decays into leptons). Whereas soft and intermediate channels lead to an event spectrum spanning a broad range of energies, with the intermediate channels contributing to energies closer to half the DM mass, hard channels lead to peak-shaped event spectrum close to $m_{\rm DM}/2$. Our results show that soft channels give a significant contribution to the total number of events, with best-fit DM masses between $\sim 500$~TeV and $\sim 11$~PeV. The events due to DM decays into quarks and Higgs bosons mostly account for the excess of events at energies below $200$~TeV, while the astrophysical flux describes the higher energy data. In the case of DM decays into gauge bosons, the DM contribution is dominant above $200$~TeV and the astrophysical flux accounts for the lower energy events. For the hard channels, when DM decays into leptons, its contribution dominates at higher energies, above $300$~TeV (similar to the gauge boson cases). The best-fit value of the power-law index ranges between $2.4$ (for quark channels) and $3.7$ (for gauge boson channels), except for the top quark channel which gives a softer spectral index $\sim 3.9$. For the hard channels, the best fit for the power-law index is around 3.5, similar to the gauge boson cases. This is, again, because astrophysical flux contributes at lower energies, and is necessarily steep, while DM decays explain the higher energy events where the astrophysical flux is too low. We find very similar best-fit values for both the low, $10$~TeV, and the high, $60$~TeV, low-energy thresholds. However, the background contamination of the data sample in the [60~TeV--10~PeV] interval is much smaller, and as done by the IceCube collaboration, we have focused our attention to this case.

To conclude the discussion of this scenario where the observed data is the result of a combination of events from DM decay and astrophysical power-law spectrum, we provide limits on the DM lifetime and indicate how future IceCube data could pin down the properties of the DM particle.

The IceCube through-going muon track dataset (with energies above $200$~TeV), indicate a much harder energy spectrum than the HESE dataset. In order to incorporate this information and the theoretical expectation of a harder flux, we imposed a prior on the power index: $\gamma = [2.0\text{--}2.3]$. With this prior, only soft channels with DM masses of the order of few hundreds of TeV contribute significantly to the observed HESE data, giving about half of the observed events. These events are all in the low-energy region, below $200$~TeV.  The hard channels do not contribute significantly in this case, because it is now the astrophysical flux that describes highest energy events.  

Finally, motivated by the apparent complementarity of the spectrum from soft-channel and hard-channel decays, we have also investigated the scenario where DM decays via two distinct channels, instead of restricting it to just one. In order to avoid increasing the number of parameters involved in the fit, we have left out the astrophysical flux from this scenario altogether. We have shown that if DM decays to a combination of soft and hard channels, the corresponding events can, by themselves, explain the HESE dataset to a degree as good as, and indeed slightly better than, that obtained from combining single-channel DM decays with an astrophysical power-law flux. As intuitively expected, the best-fit \mdm\ in this case is found to be in the $\sim$ few PeV range, ensuring that DM decays into the hard channel explains the multi-PeV data, while the softer DM decay channel fill up the events at sub-PeV energies. However, we find the branching ratios in these cases to be predominantly in favor of the softer channels, which poses tension with gamma-ray bounds.

We conclude that the current 1347-day IceCube HESE dataset prefers a fit involving multiple-component flux to a single power-law flux. We have investigated this by considering two different scenarios, the first comprising a uniform power-law model for the astrophysical flux and a spectrum from DM decays to a single channel, while in the second case we analyzed DM decays via complementary soft and hard channels, with the astrophysical flux turned off completely. In both cases, we find excellent fits to the data and determine the preferred DM masses and lifetimes consistent with these fits. If future data from IceCube strengthens the trend of disfavoring an astrophysical-only power-law flux, a multi-component fit might likely be required to explain the observations. Various combinations of fluxes from DM decays, whether in collusion with astrophysical fluxes, or by themselves, may provide the best explanation in that case, thereby also providing an indirect evidence for the existence of heavy DM and hints for its particle physics phenomenology.

\section*{Acknowledgments}
AB expresses gratitude to Jean-Ren\'{e} Cudell for helpful discussions and to support from the Fonds de la Recherche Scientifique-FNRS, Belgium, under grant No.~4.4501.15. AB is also thankful to the computational resource provided by Consortium des Équipements de Calcul Intensif (C{\'E}CI), funded by the Fonds de la Recherche Scientifique de Belgique (F.R.S.-FNRS) under Grant No. 2.5020.11 where a part of the computation was carried out. AE thanks the computing resource provided by CCJDR, of IFGW-UNICAMP with resources from FAPESP Multi-user Project 09/54213-0. AE thanks the partial support by the CNPq fellowship No.~310052/2016-5. SPR is supported by a Ram\'on y Cajal contract, by the Spanish MINECO under grants FPA2014-54459-P and SEV-2014-0398, by the Generalitat Valenciana under grant PROMETEOII/2014/049 and by the European Union's Horizon 2020 research and innovation program under the Marie Sk\l odowska-Curie grant agreements No. 690575 and 674896. SPR is also partially supported by the Portuguese FCT through the CFTP-FCT Unit 777 (PEst-OE/FIS/UI0777/2013). IS was supported in part by the Department of Energy under Grant DE-FG02-13ER41976 (DE-SC0009913). SPR and IS would like to thank the Aspen Center for Physics, where this work was initiated, for its hospitality and support by the National Science Foundation grant PHY-1066293.

\appendix
\section{Parameter correlations for combined DM decay and astrophysical power-law flux}
\label{sec:appA}

\begin{figure}[h!]
	\centering
	\includegraphics[width=\textwidth]{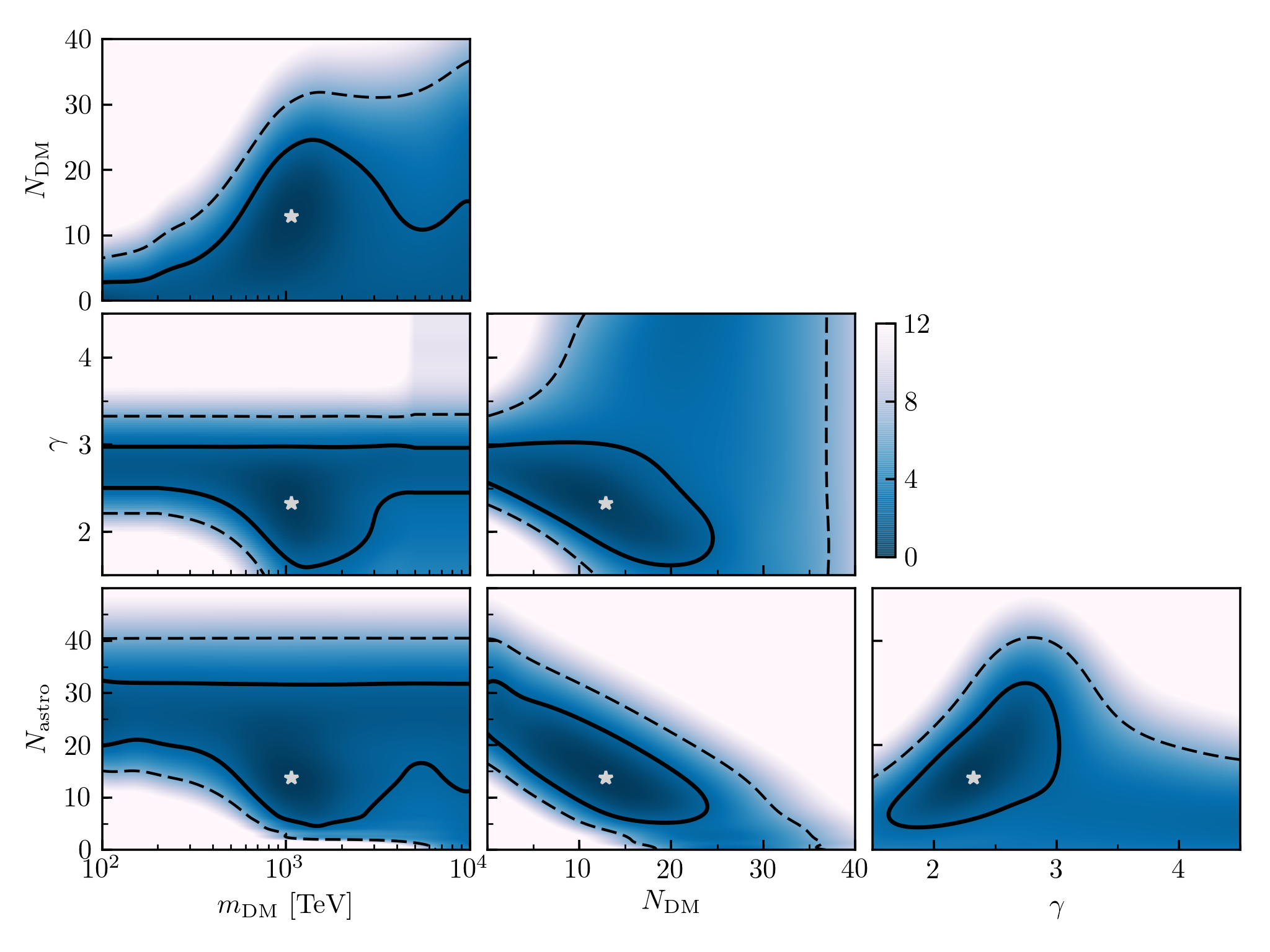}
	\caption{\label{fig:param-corr-b}
		Correlations among all pairs of parameters for $\dm \to b \, \bar{b}$ and power-law flux corresponding to the EM-equivalent deposited energy interval [60 TeV -- 10 PeV].}
\end{figure}

\begin{figure}[t]
	\centering
	\includegraphics[width=\textwidth]{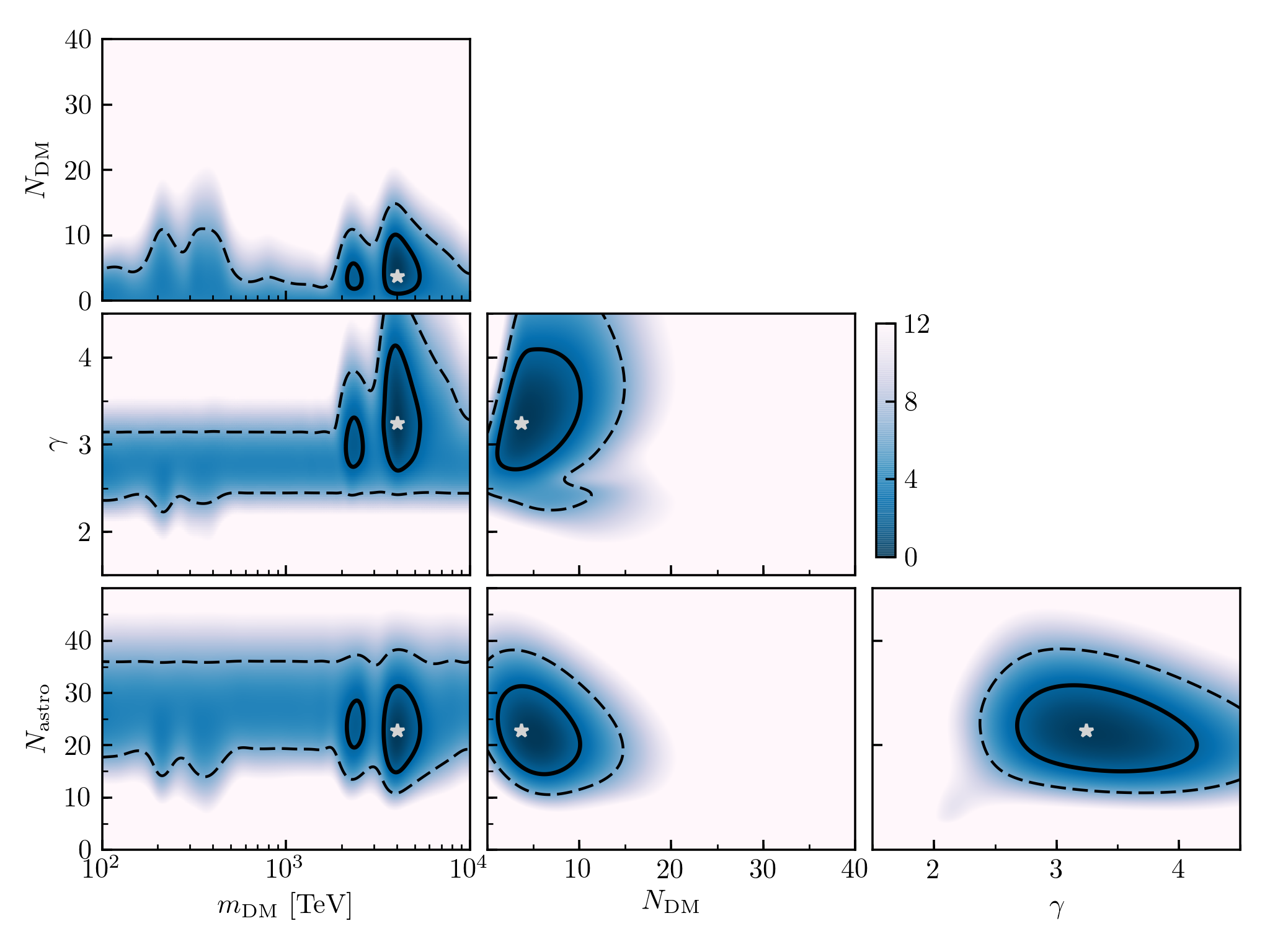}
	\caption{\label{fig:param-corr-nue}
		Same as Figure~\ref{fig:param-corr-b} but for the hard-channel decay $\dm \to \nue \, \bar{\nu}_e$.
		}
\end{figure}

In this section, we show results for all possible two-parameter correlations for the scenario where we fit a combination of a single-channel DM decay and a uniform power-law model for the astrophysical flux, for the EM-equivalent deposited energy interval [60 TeV--10 PeV]. As discussed in Section~\ref{sec:stat}, our fits are parameterized in terms of four variables: $\bm{\theta} = \{\ndm, \mdm, \nast, \gamma \}$, i.e., six pairs of correlations among them. Here we show the $1\sigma$~CL and $2\sigma$~CL contours corresponding to these correlations for a soft-channel decay, $\dm \to b \, \bar{b}$ (Figure~\ref{fig:param-corr-b}), and for a hard-channel decay, $\dm \to \nue \, \bar{\nu}_e$ (Figure~\ref{fig:param-corr-nue}). In each panel, the results have been computed by marginalizing over the other two free parameters. 

\section{Lifetime limits for all DM decay channels}
\label{sec:appB}
	
\begin{figure}[t]
	\subfigure[][]{\includegraphics[width=0.32\textwidth]{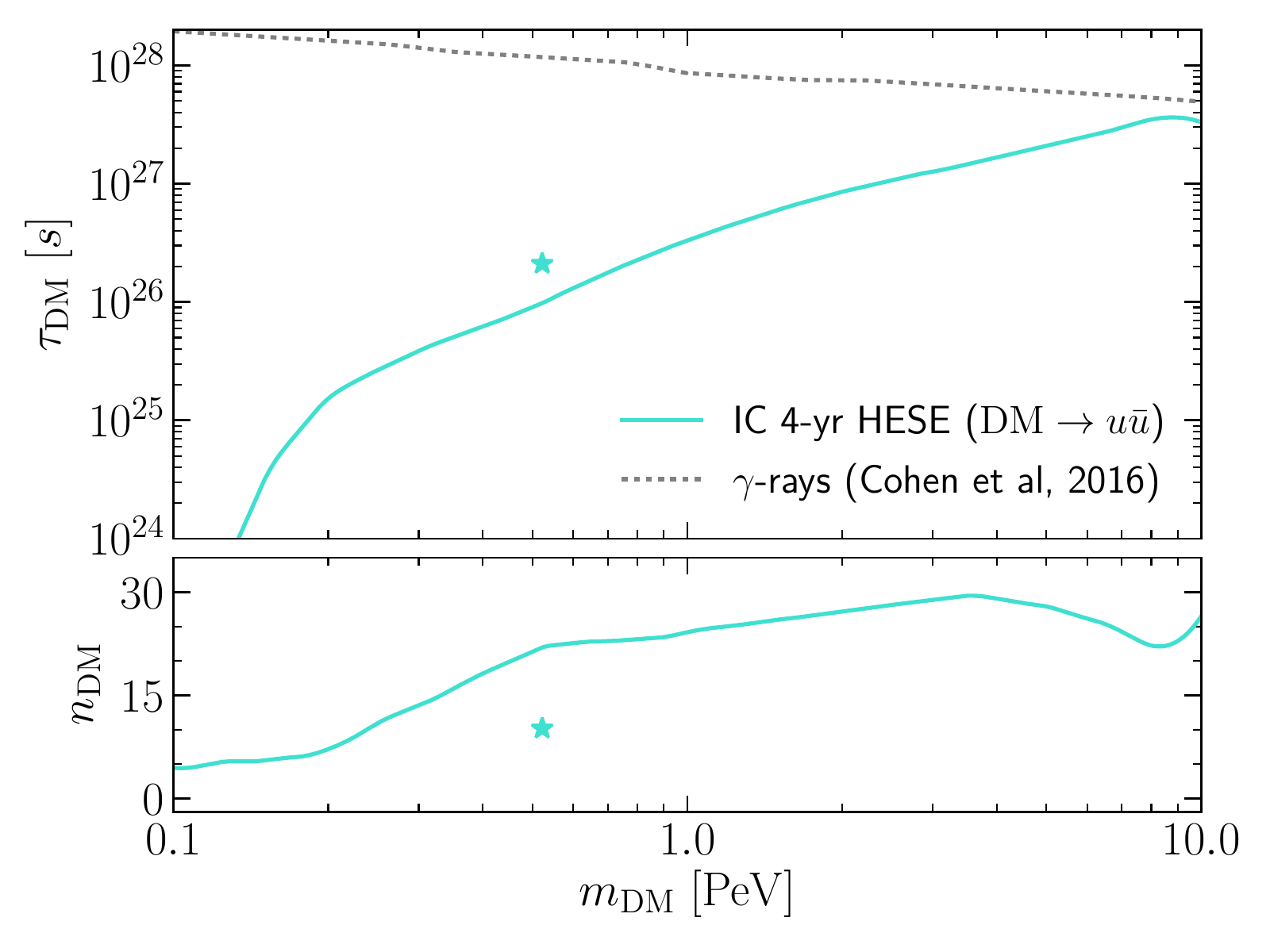}}
	\subfigure[][]{\includegraphics[width=0.32\textwidth]{mDM-LTLims-60to10-b}}
	\subfigure[][]{\includegraphics[width=0.32\textwidth]{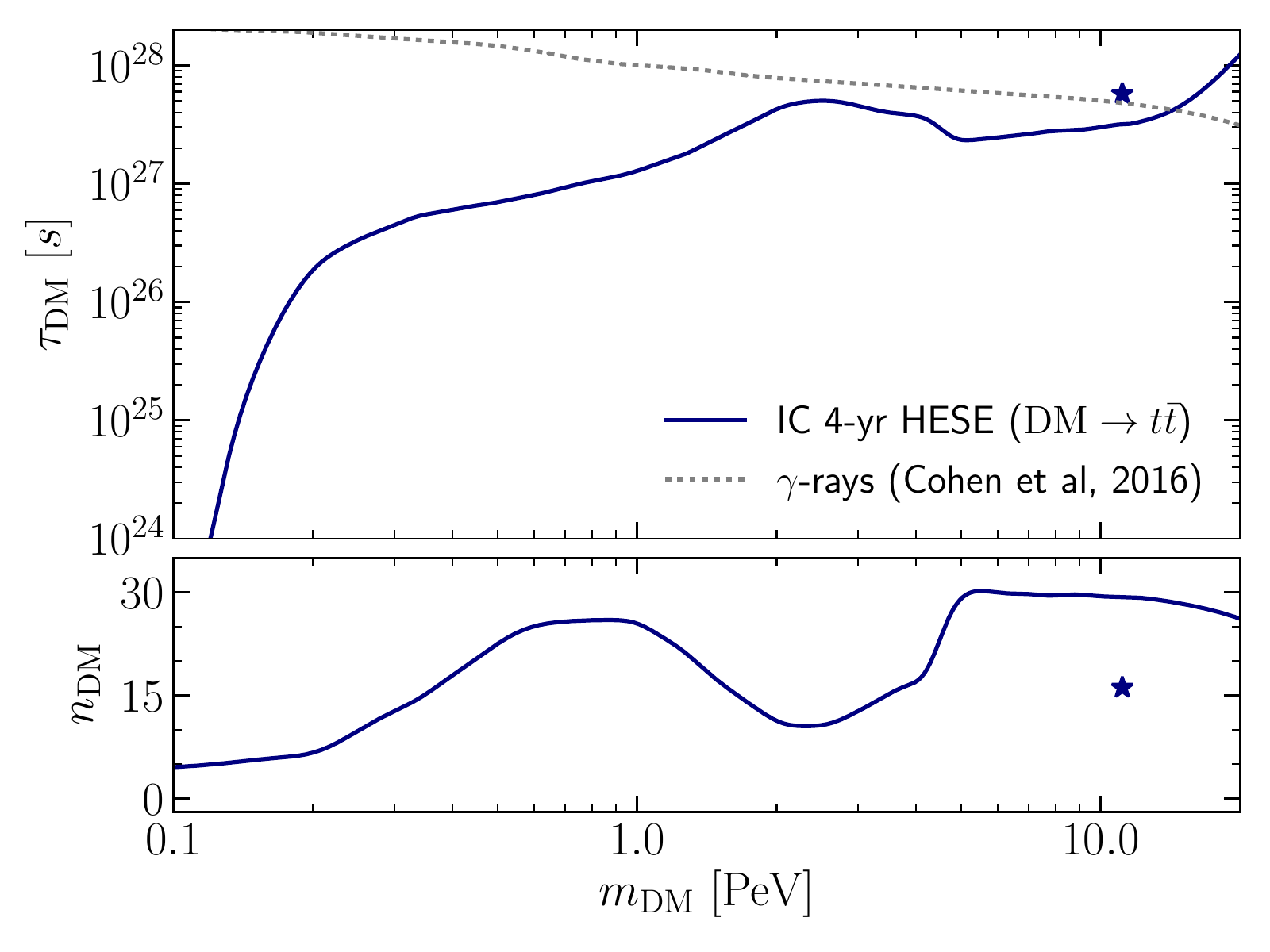}}\\
	\subfigure[][]{\includegraphics[width=0.32\textwidth]{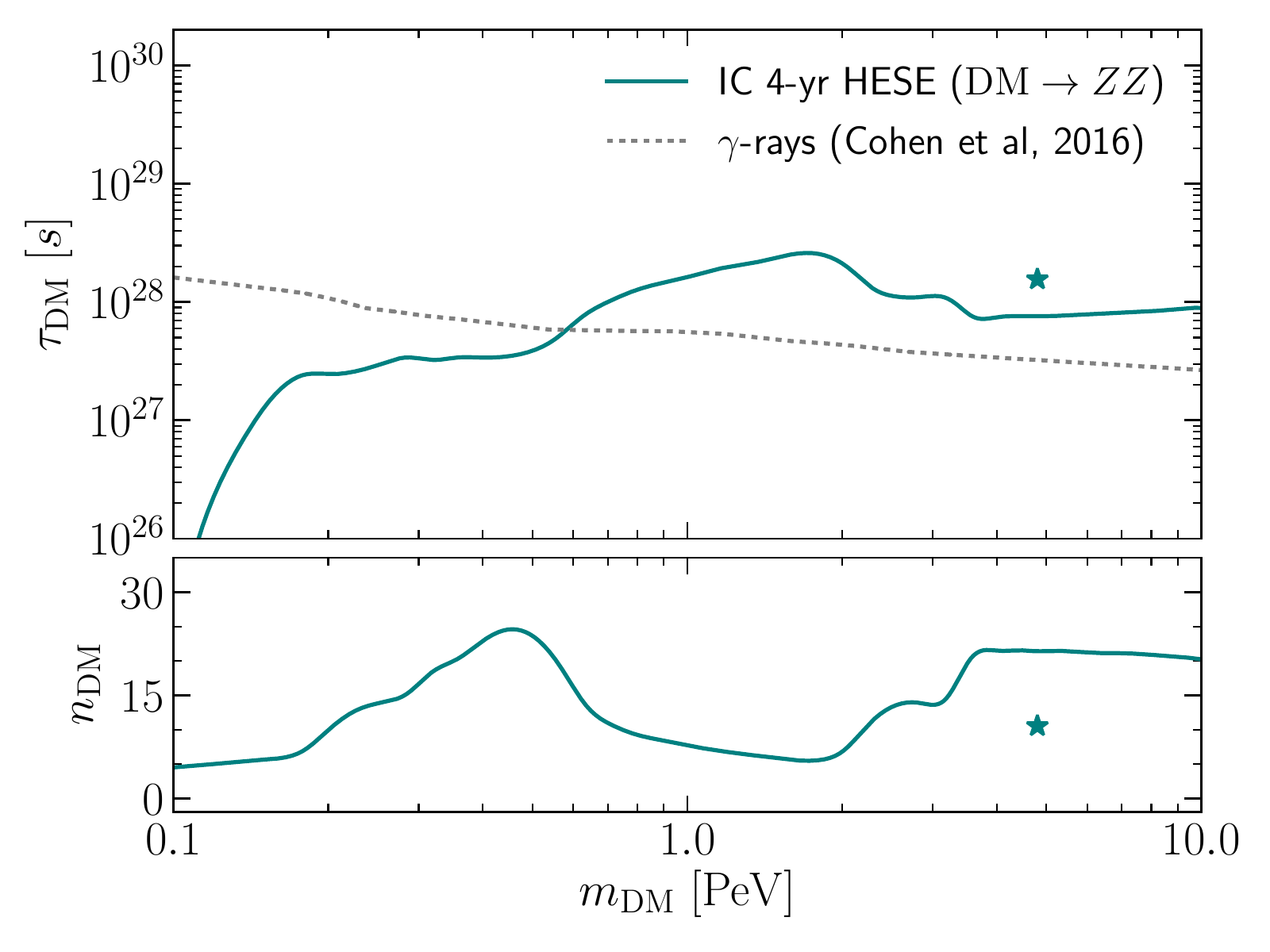}}
	\subfigure[][]{\includegraphics[width=0.32\textwidth]{mDM-LTLims-60to10-W}}
	\subfigure[][]{\includegraphics[width=0.32\textwidth]{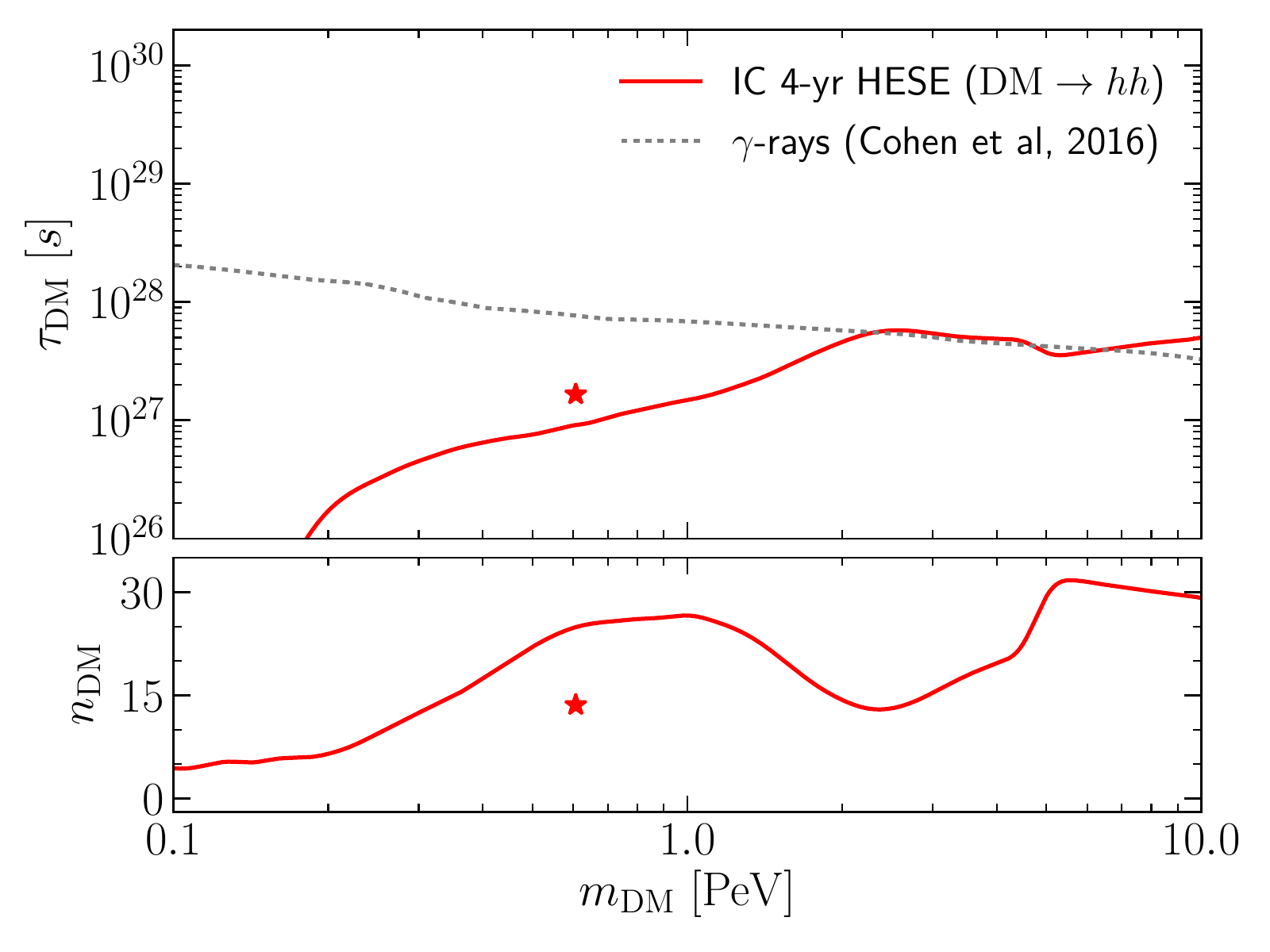}}\\
	\subfigure[][]{\includegraphics[width=0.32\textwidth]{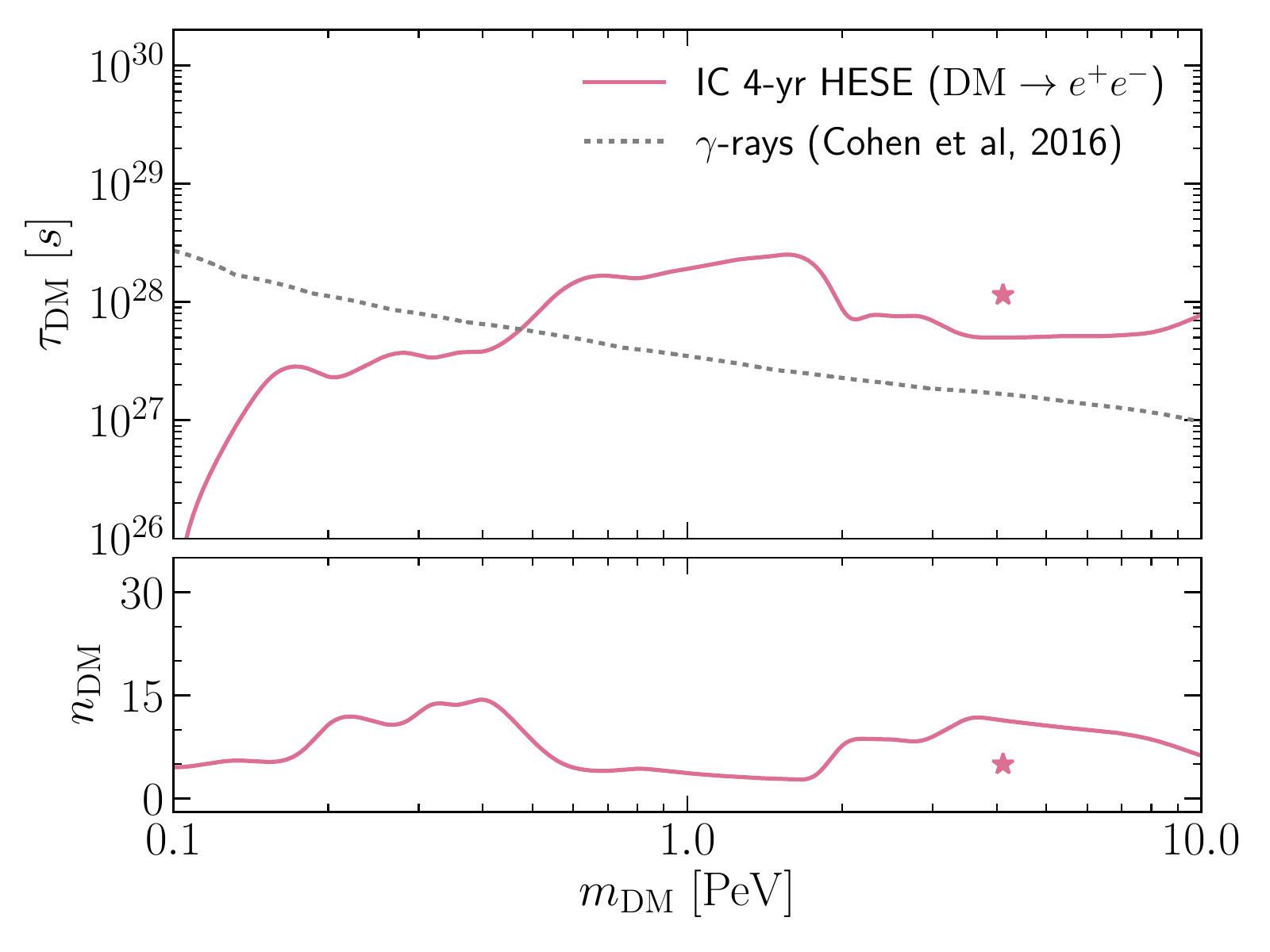}}
	\subfigure[][]{\includegraphics[width=0.32\textwidth]{mDM-LTLims-60to10-mu}}
	\subfigure[][]{\includegraphics[width=0.32\textwidth]{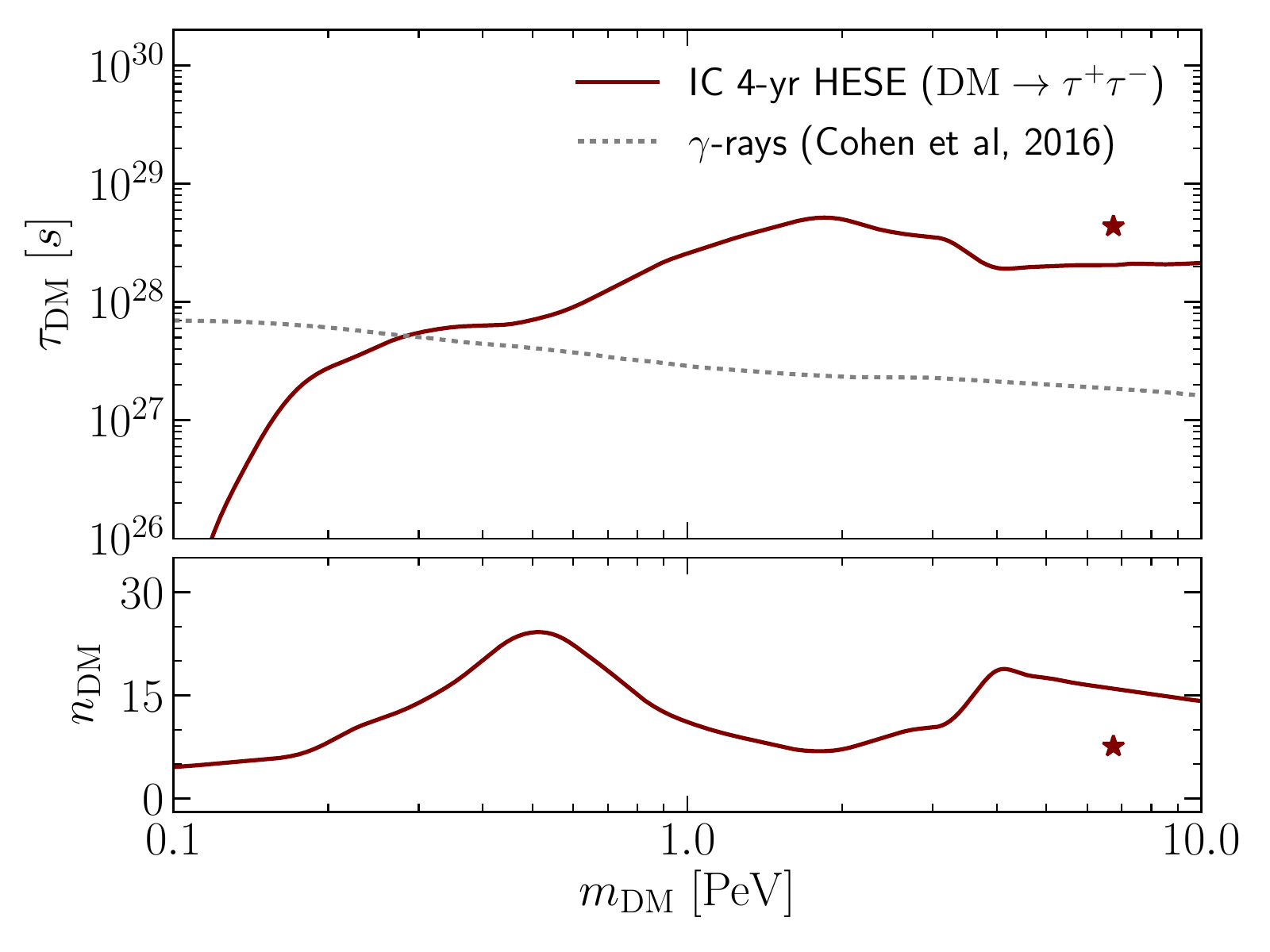}}\\
	\subfigure[][]{\includegraphics[width=0.32\textwidth]{mDM-LTLims-60to10-nue}}
	\subfigure[][]{\includegraphics[width=0.32\textwidth]{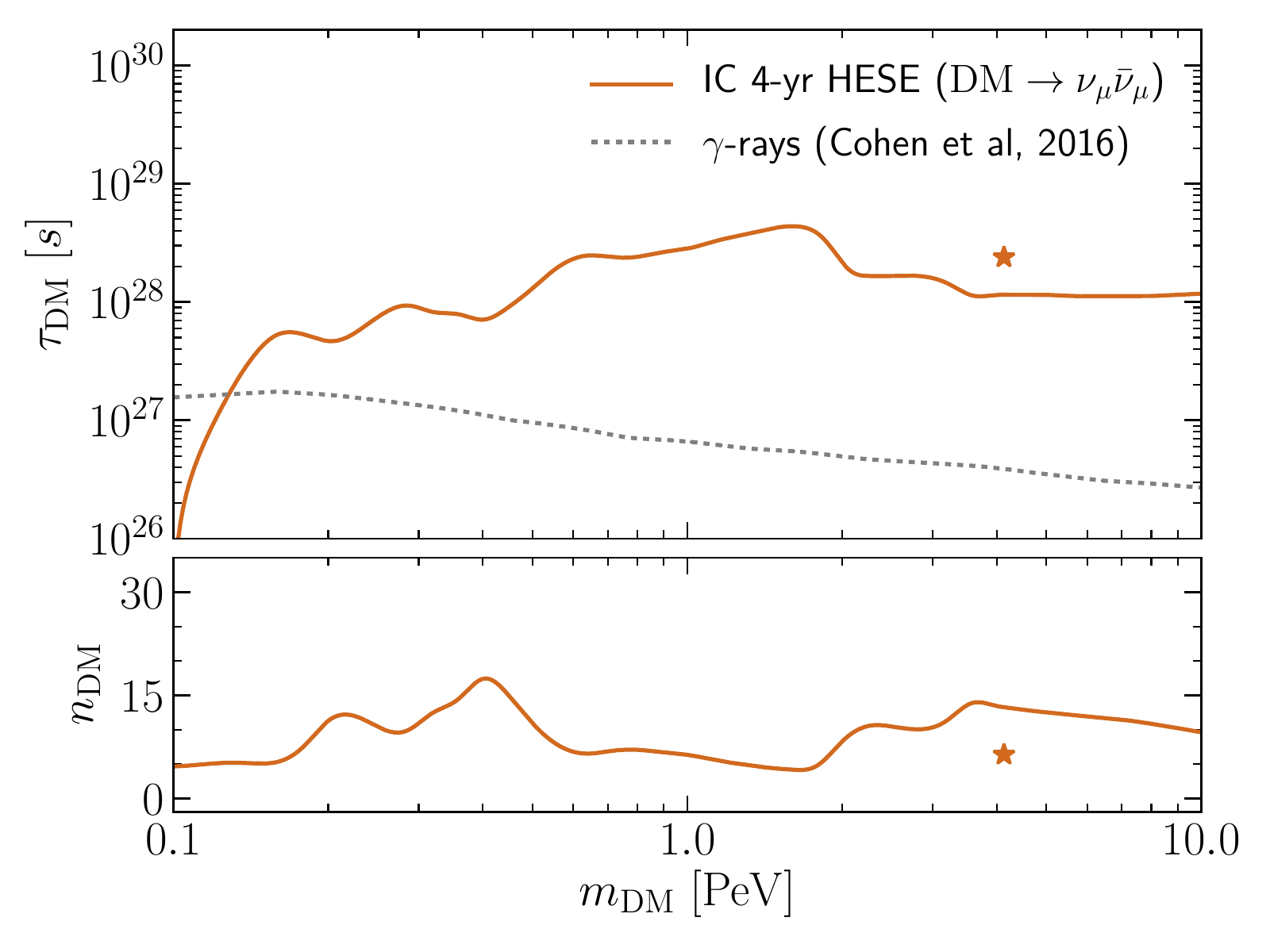}}
	\subfigure[][]{\includegraphics[width=0.32\textwidth]{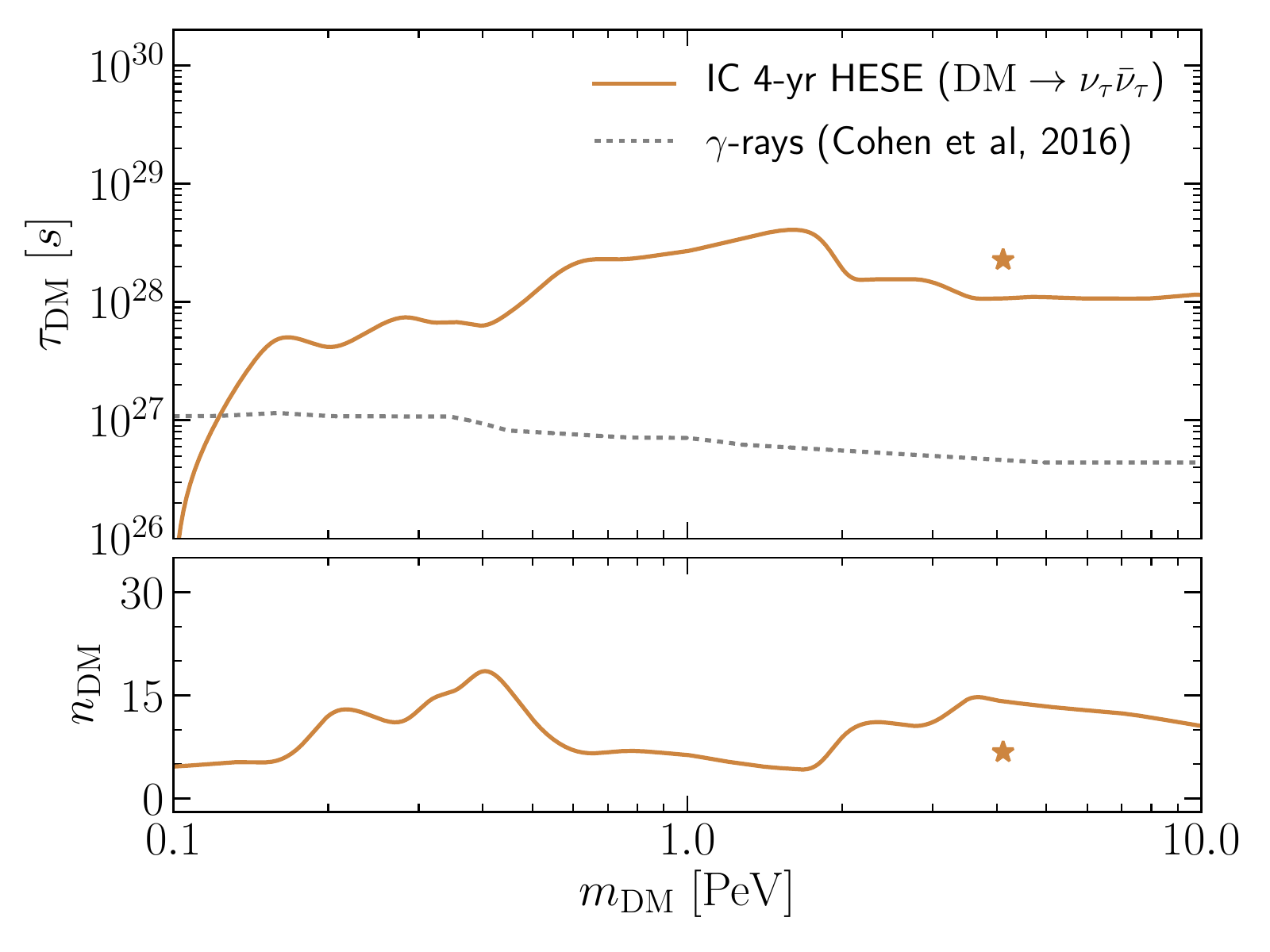}}  
	\caption{\label{fig:ltlims-allch-app}
		Limits on the DM lifetime and $\ndm$ at $95\%$~CL as a function of the DM mass, for all decay channels studied for the single-channel decay and astrophysical flux combination. The best-fit values for $\{ \mdm, \tau_\dm\}$ and $\{\mdm, \ndm \}$ are indicated in each case by the `$\star$' sign. The results correspond to the EM-equivalent deposited energy interval [60 TeV--10 PeV].}
\end{figure}

In this section, we show the DM lifetime limits at $95\%$~CL for all single-channel DM decays as a function of \mdm\, obtained using the 4-year IceCube HESE data in the EM-equivalent deposited energy interval [60 TeV--10 PeV] (Figure~\ref{fig:ltlims-allch-app}). We compare our results with the bounds obtained from $\gamma$--ray observations in Ref.~\cite{Cohen:2016uyg}. Our bounds are palpably stronger at multi-PeV masses, for all but the quark and Higgs decay channels. We also show how our DM lifetime bounds translate into bounds on the observable number of events (\ndm) from DM decays for the 4-year period under consideration.

%\clearpage

\bibliographystyle{JHEP}
\bibliography{dmastrefs}

\end{document}